\documentclass[useAMS,usenatbib]{mnras}

\usepackage{amsmath}
\usepackage{amssymb}
\usepackage{bbm}
\usepackage{graphicx}
{\newif\ifnotend
\notendtrue
\def\veclist{ABCDEFGHIJKLMNOPQRSTUVWXYZabcdefghijklmnopqrstuvwxyz.}
\def\top#1#2.{#1}
\def\tail#1#2.{#2.}
\loop\expandafter\xdef\csname v\expandafter\top\veclist\endcsname%
{{\noexpand\bf\expandafter\top\veclist}}
\edef\veclist{\expandafter\tail\veclist}
\if\veclist.\notendfalse\fi\ifnotend\repeat}
\def\d{{\rm d}}

\def\E{{\cal E}}
\DeclareMathOperator{\pr}{pr}
\def\pc{\hbox{pc}}
\def\kms{\hbox{km\,s}^{-1}}
\def\kpc{\hbox{kpc}}

\makeatletter
\def\@printed{\doi{10.1093/mnras/stz037}}
\makeatother

\title[Dynamical models of the Galaxy's central parsec]{Orbit-superposition models of discrete,
  incomplete stellar kinematics: application to the Galactic centre}
\author[SJ Magorrian]{John Magorrian\\
Rudolf Peierls Centre for Theoretical Physics, Clarendon Laboratory,
Parks Road, Oxford OX1 3PU}
\begin{document}

\date{Accepted 2019 January 1. Received 2018 December 20; in original form 2018 September 16}

\maketitle

\label{firstpage}

\begin{abstract}
  We present a method for fitting orbit-superposition
  models to the kinematics of discrete stellar
  systems when the available stellar sample has
  been filtered by a known selection function.  The fitting method can
  be applied to any model in which the distribution function is
  represented as a linear superposition of basis elements with unknown weights.
  As an example, we
  apply it to Fritz et al.'s kinematics of the innermost regions of
  the Milky Way's nuclear stellar cluster.  Assuming spherical
  symmetry, our models fit a black hole of mass
  $M_\bullet=(3.76\pm0.22)\times10^6\,M_\odot$, surrounded by an
  extended mass $M_\star=(6.57\pm0.54)\times10^6\,M_\odot$ within
  $4\,\pc$.  Within $1\,\pc$ the best-fitting mass models have an approximate
  power-law density cusp $\rho\propto r^{-\gamma}$ 
  with $\gamma=1.3\pm0.3$.  We carry out an extensive investigation of
  how our modelling assumptions might bias these estimates:
  $M_\bullet$ is the most robust parameter and $\gamma$ the least.
  Internally the best-fitting models have broadly isotropic orbit
  distributions, apart from a bias towards circular orbits between 0.1
  and 0.3 parsec.
\end{abstract}


\section{Introduction}

In most stellar systems dynamical times are so long that observations
provide only an instantaneous snapshot of the stars' positions and
velocities.  Any attempt to obtain a dynamical estimate of the
system's mass distribution must therefore rely on assumptions about
its dynamical state.  The most fundamental of these is usually that
the system has settled into an equilibrium configuration, followed by
an assumption about its geometry (e.g., spherical symmetry,
axisymmetry or triaxiality).  The black hole masses deduced in most
galaxies are obtained from models that rely on this symmetric
equilibrium assumption, as are many constraints on the distribution of
dark matter.  For example, models based on the steady-state Jeans
equations make these assumptions, as do models that fit parametrized
distribution functions (hereafter DFs) or those that are based on the
orbit-superposition technique of
\cite{SchwarzschildNumericalmodeltriaxial1979}.

An important exception is the stellar cluster at the centre of our own
Galaxy.  There the dynamical times of certain young stars -- the
so-called S stars -- are so short that, by following their orbits over
time, \cite{GhezMeasuringDistanceProperties2008} and
\cite{GillessenMonitoringStellarOrbits2009} could obtain direct
estimates of the mass of the Galaxy's central black hole (BH) simply
by fitting the stars' orbits.  This neatly avoids the usual
assumptions about the equilibrium of the cluster or its geometry.  The
most recent estimates of the BH mass inferred by such analyses are
$M_\bullet=(4.02\pm0.16\pm0.04)\times10^6\,M_\odot$
\citep{BoehleImprovedDistanceMass2016} and $(4.31\pm0.06\pm0.36)\times10^6\,M_\odot$
\citep{GillessenUpdateMonitoringStellar2017}.

On the other hand, BH mass estimates obtained using the symmetric
equilibrium assumption to fit an instantaneous snapshot of the old
stars that make up the bulk of the Galactic centre stellar cluster
tend to produce significantly lower BH masses.
Most such models have relied on spatial binning of the individual
stellar velocities to construct an estimate of the cluster's velocity
dispersion profile(s) and then use the Jeans equations to deduce the
underlying mass distribution.  For example, using samples of
$\sim10^2$ radial velocities, \cite{GenzelDarkMassConcentration1996}
and \cite{HallerStellarKinematicsBlack1996} fit a BH mass of
$\sim 2.6\times10^6\,M_\odot$, under the assumption that the cluster
is spherical with an isotropic velocity distribution.

Subsequent models relaxed these initial assumptions of
velocity isotropy and spherical symmetry.
\cite{GenzelStellardynamicsGalactic2000} constructed a
sample of several hundred stars within $\sim20$ arcsec
of Sgr A${}^\star$, including 104 with
proper motion measurements and 32 that had both proper motions and
radial velocities.  Their estimates of the BH mass lay in the range
$2.6\times10^6\,M_\odot$ to $3.3\times10^6\,M_\odot$.
Later \cite{SchodelNuclearstarcluster2009} measured proper motions of
$\sim6000$ stars in the same region.  From their anisotropic spherical
Jeans models they inferred a BH mass of
$3.6^{+0.2}_{-0.4}\times10^6\,M_\odot$.

Although the stellar cluster is approximately round in projection
within the innermost parsec or so, deeper observations
\citep[e.g.][]{SchodelSurfacebrightnessprofile2014} show that it becomes
increasingly flattened at larger radii.
\cite{ChatzopoulosOldnuclearstar2015} fit flattened, isotropic Jeans
models to first- and second-order projected velocity moments assembled
from the sample of $\sim10^4$ proper motions and $\sim2500$ radial
velocities measured by \citet{FritzNuclearClusterMilky2016}.  They
found $M_\bullet=(3.86\pm0.14)\times10^{6}\,M_\odot$, with a
systematic uncertainty estimated to be at least a factor of two larger
than the formal uncertainty quoted here.  They also constructed
explicitly the two-integral DF $f(E,L_z)$ that underlies their
best-fitting model and confirmed that its predictions match an
extensive range of binned velocity histograms of the observations.

These results are obtained using the brightest stars as discrete
kinematical tracers.  An alternative is to model the kinematics of the
{\it unresolved} stellar population.
\cite{FeldmeierLargescalekinematics2014} have employed a drift-scan
technique applied to integral field spectroscopy to extract the
kinematics of the innermost few parsecs of the Galaxy.  Applying
anisotropic Jeans models to these data results in
$M_\bullet=1.7^{+1.4}_{-1.1}\times10^6\,M_\odot$, albeit with a very
high reduced $\chi^2$.  Triaxial orbit-superposition models fit to the
same kinematics yield
$M_\bullet=(3.0^{+1.1}_{-1.3})\times10^6\,M_\odot$
\citep{Feldmeier-KrauseTriaxialorbitbasedmodelling2017}.

It is evident that these models -- all constructed using some variant
of the symmetric equilibrium assumption -- struggle to reproduce BH
masses that are as high as the $M_\bullet\gtrsim4\times10^6\,M_\odot$
inferred by following stellar orbits: the symmetric, equilibrium
models produce BH masses that are at best only marginally consistent
with the direct mass measurements.  It is interesting then to try to
identify the dominant sources of systematic error in such models.

The symmetric equilibrium models mentioned so far share the following
shortcomings:
\begin{enumerate}
\item They do not use individual stellar velocities directly, but
  instead fit only to low-order velocity moments estimated by binning
  the data: any constraints on the potential or DF that lurk in the
  details of the joint (position,velocity) distribution are simply
  ignored.
\item They assume a single parameterized form for the stellar number
  density distribution, even though this is difficult to
  measure from discrete data
  \cite[e.g.,][]{MerrittNonparametricestimationdensity1994}.
\item They make strong assumptions about the geometry of the cluster.
  The Galactic centre is a messy place, with at least one tilted disc
  of young stars in addition to the S stars
  \citep{PaumardTwoYoungStar2006,YeldaPropertiesRemnantClockwise2014}. No
  spherical, axisymmetric or triaxial model can treat such a disc
  properly.  If one assumes that the old stellar population is, say,
  axisymmetric, then it would make sense to exclude young stars when
  fitting models, but identifying them requires taking spectra, which
  are expensive to obtain.
\end{enumerate}

Point (iii) is difficult to avoid, but some work has addressed
(i) and (ii), at least partially.  For example,
\cite{DoThreedimensionalStellarKinematics2013} model the cluster's
number density profile as a broken power law whose parameters are fit
simultaneously with the parameters that describe the DF and potential.
Instead of binning the kinematic data, they fit each star's observed
velocity directly under the (implicit, unjustified) assumption that
the cluster's internal three-dimensional velocity distribution is
locally Maxwellian.  They find
$M_\bullet=5.76^{+1.76}_{-1.26}\times10^6\,M_\odot$ for their sample
of 265 stars.

By far the cleanest approach, however, is that adopted by
\cite{ChakrabartyNonparametricEstimateMass2001}.  They assumed only
that the observed stars are drawn from an unknown, spherical,
isotropic phase-space DF $f(E)$, and move in the potential generated
by an unknown spherically symmetric mass distribution $M(r)$.  Making
weak further assumptions about the form of $f(E)$ and $M(r)$, they fit
their model's projected DF directly to the observed stellar velocities
and positions, avoiding any binning of the data.  They fit BH masses
that ranged from $2.2^{+1.6}_{-1.0}\times10^6\,M_\odot$ for models
that fit only the line-of-sight components of stellar velocities to
$1.8^{+0.4}_{-0.3}\times10^6\,M_\odot$ for models that fit all three
components, inconsistent with the results from the S stars.

Our goal in this paper is to understand why analyses that use
the steady-state symmetric assumption to model the kinematics of the
cluster's old stars tend to yield BH masses that differ from those
obtained from the S stars.  We do this by presenting new models that
avoid shortcomings (i) and (ii) above by fitting the underlying DF of
the models directly to the joint (position,velocity) distribution of
the observed stars, sidestepping the need to parametrise the cluster's
number density profile.  We do not evade shortcoming (iii), but we
do investigate some of the biases introduced by fitting symmetric models to
asymmetric galaxies.
The paper is organised as follows.  Section~\ref{sec:data} summarises
the stellar catalogue we use to constrain the models.  Our modelling
procedure is described in Section~\ref{sec:models}.  In
Section~\ref{sec:tests} we test the machinery by applying it to
simulated data drawn from spherical and non-spherical model clusters,
before applying it to the real Galaxy in Section~\ref{sec:app}.
Section~\ref{sec:byebye} summarises and concludes.

Throughout the paper we assume that the distance to the Galactic
centre is $8.3\,\kpc$.

\section{Data}
\label{sec:data}

\begin{figure}
  
\includegraphics[width=\hsize]{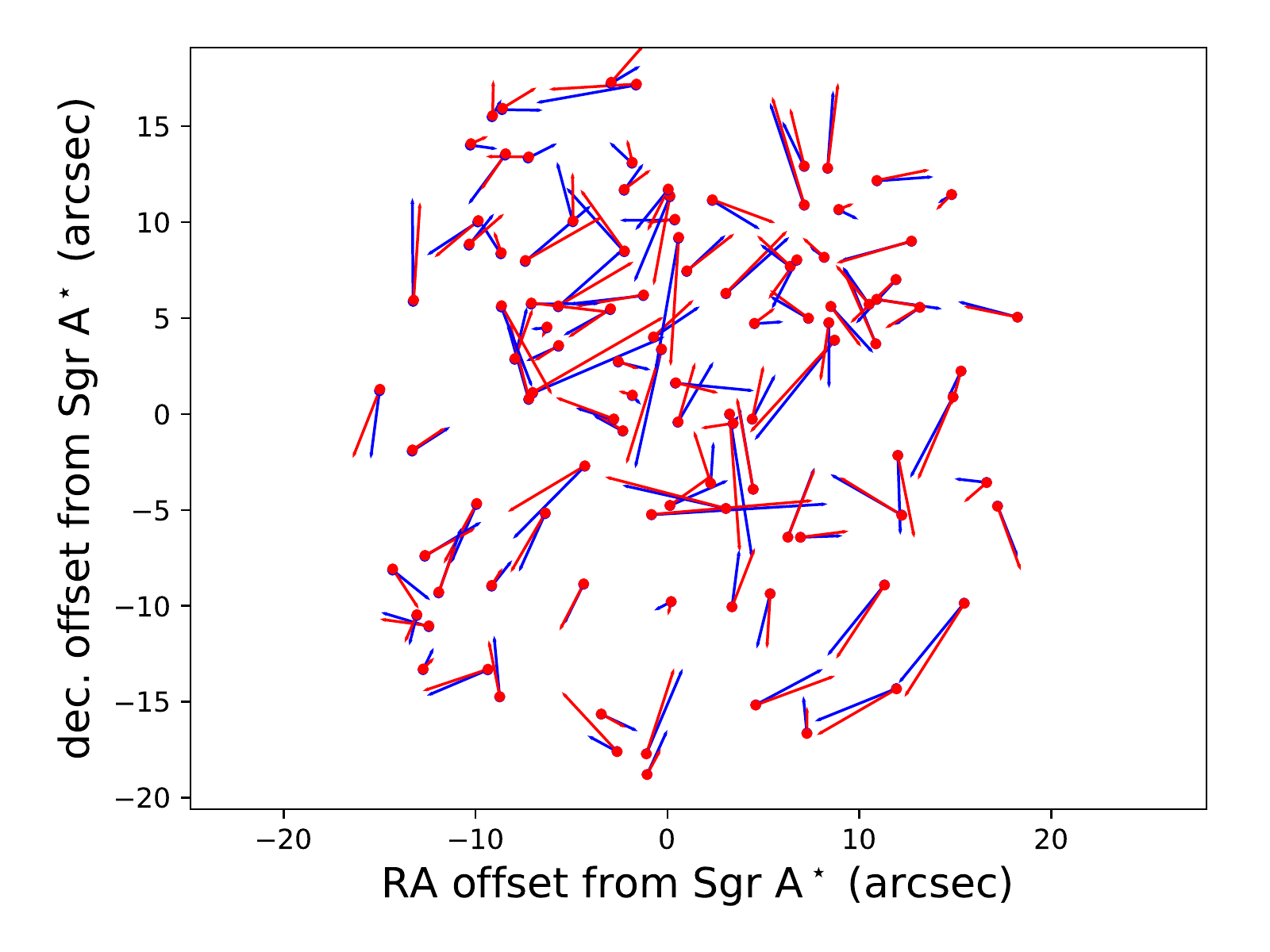}

\caption{Comparison of proper motion measurements near the Galactic
  centre.  Dots plot the positions of every 5th star brighter than
  $K=14$ within 19 arcsec of Sgr A from the sample of Fritz et
  al. (2016), with red streaklines indicating their proper motions.
  For comparison, the blue streaklines show the corresponding proper-motion
  measurements from the Sch\"odel et al. (2009) sample.
}  \label{fig:comparepms}
\end{figure}
\subsection{Discrete stellar velocities}

There are two large recent catalogues of discrete stellar velocities
within 1 pc of Sgr~A${}^\star$.  \cite{SchodelNuclearstarcluster2009}
report proper motions for a sample of more than 6000 stars within an
approximately 40 arcsec-square field of view that includes the
Galactic centre.  \cite{FritzNuclearClusterMilky2016} provide combined
proper motions and/or radial velocities for a sample of over $10^4$
stars that extends beyond 80 arcsec in projection, albeit with
variable coverage in different regions of the sky.  As they have
spectra for many of their stars, they have better early-type star
rejection than~\cite{SchodelNuclearstarcluster2009}.
We use all 4105 stars from \cite{FritzNuclearClusterMilky2016} that
lie within a circle of radius 19 arcsec centred on Sgr A${}^\star$ as
the primary source of kinematic data for our models.  This circle just
touches the boundary of their ``extended field'' sample.  We treat the
\cite{SchodelNuclearstarcluster2009} proper motion sample --
restricted to the same radius and rescaled to our assumed distance
$D=8.3\,\kpc$ -- as an independent catalogue of the same area.  This
secondary catalogue has 5005 stars.

Figure~\ref{fig:comparepms} shows a comparison of these two
catalogues, constructed as follows.  From our primary, 19''-radius
\cite{FritzNuclearClusterMilky2016} sample we extract all stars
brighter than $K=14$.  For each such star, we look for matches from
the \cite{SchodelNuclearstarcluster2009} sample that lie within a 0.2
arcsec-square box centred on the star and take the one that has the
closest match in proper motion.  The rms discrepancy in the on-sky
components of velocity between the two matched samples is 25 km/s,
which is at least a factor four larger than the mean quoted
uncertainty in either sample.  The cause of this discrepancy is
unclear.  We note that \cite{FritzNuclearClusterMilky2016} quote a
median uncertainty of 8 km/s in their measurements of the
line-of-sight components of velocity, which, being obtained from
spectra, are unlikely to be affected by whatever causes the
discrepancy in the proper motion-derived components.  Nevertheless, as
a crude method of dealing with the discrepancy we simply set a
floor of 20 km/s when modelling the uncertainty of {\it any} component
of velocity in either sample.

\subsection{Outliers and contaminants}

\cite{FritzNuclearClusterMilky2016} estimate that as
many as 4\% of the stars beyond 2 arcsec in their sample could be
early type.  There are also some stars with anomalously high
velocities. We make no attempt to identify and exclude these stars,
but instead allow for them in our modelling procedure
(Section~\ref{sec:outliers} below).

\begin{table}
  \centering
  \begin{tabular}{rrrr}
$r_{\rm min}$ & $r_{\rm max}$ & count & RMS l.o.s. vel.\\
(arcsec) & (arcsec) & & (km/s)\\
\noalign{\hrule}
30 & 40 & 1362 & 75\\
40 & 50 & 1487 & 75\\
50 & 60 & 1580 & 75\\
60 & 70 & 1652 & 75\\
70 & 80 & 1710 & 75\\
80 & 100 & 3555 & 75\\
100 & 120 & 3692 & 75\\
120 & 150 & 5726 & 75\\
150 & 200 & 9889 & 75\\
200 & 250 & 10184 & 75\\
250 & 300 & 10391 & 75\\
300 & 500 & 42824& 75\\
  \end{tabular}
  \caption{Adopted zeroth- and second-order moment profiles binned
    into circular annuli beyond 1~pc from the cluster centre.}
  \label{tab:rad}
\end{table}
\subsection{Profile beyond 19''}

Most of the stars in our discrete sample are on orbits that, in three
dimensions, take them well beyond the 0.76 pc radius implied by our
19'' radius cut.  Having some estimate of the outer profile of the
cluster helps constrain the orbits of these more loosely bound stars.
We take the Nuker-law profile fit by
\cite{ChatzopoulosOldnuclearstar2015}, scaled to match the number of
stars in our primary sample between 15 and 17 arcsec, and use that to
predict star counts within the circular annuli listed in
Table~\ref{tab:rad}.  Based on the results presented in
\cite{FeldmeierLargescalekinematics2014} and
\cite{FritzNuclearClusterMilky2016} we assume that the rms
line-of-sight velocity of the stars within each annulus is 75~km/s.
In Section~\ref{sec:testspherical} below we show that the masses fit by our
models are only very weakly dependent on the details of the profiles
adopted in Table~\ref{tab:rad}; these profiles serve more as a weak
prior on the orbit distributions fit by the models.

\section{Modelling procedure}

\label{sec:models}

We would like to learn what constraints these data place on the
cluster's mass and orbit distribution.  The observed stellar catalogue
is taken to be a sample drawn from the cluster's underlying DF.
A flexible way of
representing the orbit distribution is by expanding this DF as
\begin{equation}
  \label{eq:DFbasis}
  f(\vx,\vv)=\sum_{k=1}^Kf_ke_k(\vx,\vv),
\end{equation}
in which $e_k(\vx,\vv)$ are basis functions and $f_k$
are weights that will be constrained by the data.  By Jean's theorem
the $e_k$ should be functions only of the integrals of motion~$\vJ$ in
the assumed potential.  Because the DF should be non-negative
everywhere these $e_k$ are often chosen to be non-negative functions
with compact support
\citep[e.g.,][]{MerrittDynamicalmappinghot1993,KuijkenAxisymmetricDistributionFunction1995,
  PichonNonparametricreconstructiondistribution1998,WuDerivingMassDistribution2006}.
A particularly common choice is to take $e_k=\delta(\vJ-\vJ_k)$ for
some set of representative orbits $\vJ_k$
\citep[e.g.,][]{SchwarzschildNumericalmodeltriaxial1979,McMillanAnalysingsurveysour2012}:
this is the basis for so-called ``Schwarzschild'' models.
The basis we adopt for this paper is given in Section~\ref{sec:DF} below.
Sections \ref{sec:projections} to~\ref{sec:binned} explain how we
calculate the observables for each of our basis elements.
Sections~\ref{sec:mergeblock} and~\ref{sec:dofit} describe how we fit
the coefficients~$f_k$.

The stars are treated as tracer particles that move in the potential
generated by an underlying mass distribution that we would like to
constrain.  Therefore we do not assume that the mass density profile
is proportional to the tracer number density profile: the two are
treated independently.

\subsection{Coordinate system and potential}

\label{sec:coords}
We use a coordinate system whose origin $O$ is at the BH.  The $z$ axis is
parallel to the Galactic rotation axis and the $x$ axis points
towards the sun.  The $y$ axis then points in the direction of
decreasing Galactic longitude~$l$ as viewed from the sun.
We assume that the cluster has a spherically symmetric mass
distribution, with a BH of mass $M_\bullet$ at $r=0$, surrounded by a
cluster with some specified mass-density profile~$\rho(r)$.  For
example, in our most basic models we take a mass density of the form
\citep{DehnenFamilyPotentialDensityPairs1993,Tremainefamilymodelsspherical1994}
\begin{equation}
  \rho(r)\propto
  r^{-\gamma}
  \left(1+\frac{r}{r_{\rm s}}\right)^{-4+\gamma},
\label{eq:rhoprof}
\end{equation}
with scale radius $r_{\rm s}=10\,\pc$, which is well beyond the region
where the discrete kinematics are available.  The latter extend to
$\sim1\,\pc$, which sets a convenient scale at which to normalise the
cluster mass.  Our simplest mass models then have three free parameters: the
black hole mass, $M_\bullet$, the extended mass $M_\star$ enclosed
within 1~pc, and the power-law slope~$\gamma$ of its inner density
distribution.

\subsection{Distribution function}
\label{sec:DF}

Although we assume that the potential is spherical, we allow the
distribution of stars to be axisymmetric about the $Oz$ axis.  We
ignore variations in stellar populations and assume that the the
stellar number density is a function $f(\vx,\vv)$ only of the stars'
phase-space coordinates.  By Jeans' theorem this distribution function
(hereafter DF) can depend only on the binding energy~$\E$ per unit
mass, the total angular momentum~$L$ per unit mass and its
component~$L_z$ projected along the~$z$ axis.  We divide
$(\E,L^2,L_z)$ space into $K$ blocks and parametrize the DF as
\begin{equation}
  \label{eq:dfdefn}
  f(\E,L^2,L_z) = \sum_{k=1}^{K} f_k{\mathbbm{1}}_{V_k}(\E,L^2,L_z),
\end{equation}
where $V_k$ is the physically accessible volume of $(\E,L^2,L_z)$ space
enclosed by the $k^{\rm th}$ block and the indicator function
${\mathbbm{1}}_V(\vw)=1$ if $\vw\in V$ and is zero otherwise.  That is,
the DF
takes on the constant value $f_k$ within the $k^{\rm th}$ block.
Apart from the constraint that the DF must be non-negative, the
parameters $f_k\ge0$ are allowed to vary freely.
%% TODO: say something that this is an NGP method?

\begin{figure}
  \centering
  \includegraphics[width=0.9\hsize]{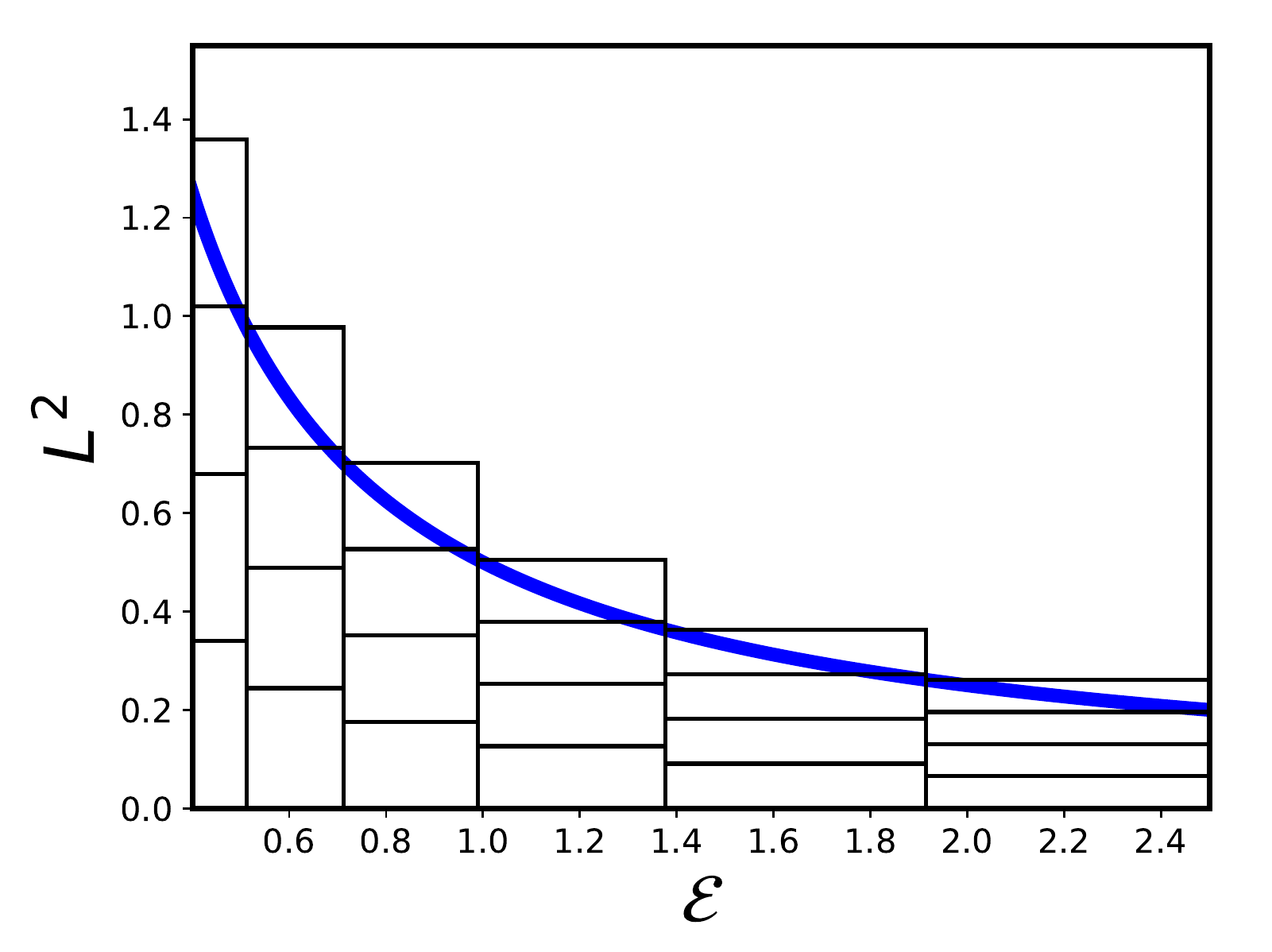}
  \caption{Lindblad diagram illustrating the arrangement of the blocks used
    to represent the cluster DF $f(\E,L^2)$ for the case $n_\E\times
    n_{L^2}=6\times4$.
    Valid orbits have $L^2<L^2_{\rm c}(\E)$ (heavy solid curve).
    The $n_L$ blocks in a given range of $\E$ are equispaced in $L^2$
    with maximum L chosen to just include the boundary of physically
    allowed orbits.
    }
  \label{fig:lindblad}
\end{figure}
Figure~\ref{fig:lindblad} shows an example of how these $K$ blocks are
chosen.  We split $(\E,L^2,L_z)$ integral space into a grid of
$K=n_\E\times n_{L^2}\times n_{L_z}$ (usually either
$200\times10\times1$ or $200\times10\times8$) abutting rectangular
blocks in the following way.  First we first select a range of
energies~$\E_0$,...,$\E_{n_\E}$ to cover by choosing
$\E_i=-\Phi(r_i)$, where the ``apocentre'' radii $r_i$ are spaced
logarithmically between $r_0=0.008\,\pc$ (about 0.2 arcsec) and
$r_{n_\E}=200\,\pc$.  The latter is much larger than any of the
projected radii for which we have data (Section~\ref{sec:data}), but
we find that such extensive sampling is essential when considering
models with extended mass distributions, such as those in
Section~\ref{sec:app} below.
Then for each $(\E_i,\E_{i+1})$ the vertical ($L^2$)
co-ordinates of the block edges are equispaced between 0 and
$L_{\rm c}^2(\E_{i+1})$, where $L_{\rm c}(\E)$ is the angular momentum
of a circular orbit of energy~$\E$.
This scheme ensures that any orbit whose energy lies between $\E_0$
and $\E_{n_\E}$ is included within one of the blocks.
The $L_z$ boundaries (not shown in Figure~\ref{fig:lindblad}) are spaced
linearly between $L_z=-L$ and $L_z=+L$, which allows the models to
have a net sense of rotation about the $z$ axis.
A complete axisymmetric dynamical model then has 16003 parameters: the three parameters
$(M_\bullet,M_\star,\gamma)$ specifying the potential, plus the
parameters $f_1,\ldots,f_{16000}$ describing the DF.

Expressions for the number density distribution $\nu(R,z)$ and
higher-order velocity moments of each orbit block are straightforward to
calculate by hand, but tedious to write out.  From these
we can easily use numerical quadrature to calculate the corresponding
projected moments.

\subsection{Projected DF and discrete likelihoods}
\label{sec:projections}

We assume that the discrete kinematical dataset is a fair sample of
the cluster's DF, modulated by a selection function $S(\vw)$, which
gives the probability that a star at phase-space
location~$\vw=(\vx,\vv)$ would be included in the sample.  For all of
the examples we consider in this paper, this $S$ will depend only on
the star's $(y,z)$ coordinates.  In most cases we adopt the simplest
possible model for~$S$, a step function that returns 1 when the star's
projected radius, $\sqrt{y^2+z^2}$, is less than 19 arcsec, and 0
otherwise.  This is a drastic simplification, which relies on an
assumption that all of the observed stars are approximately at the
same distance and that there are no spatial variations in stellar
populations.  We explore different choices of the selection function
in Section~\ref{sec:app}, including one that models the effects of
dust extinction.

The likelihood of observing a star at phase-space
position $\vw=(\vx,\vv)$ is then simply
\begin{equation}
\pr(\vw|f,\Phi,S)
=\frac1I{
\int\d\vw'\,S(\vw')\pr(\vw|\vw')f(\vw'|\Phi)
}{
},
\end{equation}
in which $\pr(\vw|\vw')$ is the probability that
a star having true phase-space coordinates $\vw'=(\vx',\vv')$ is
observed to be at $\vw=(\vx,\vv)$, $f(\vw'|\Phi)$ is the orbit-block DF~\eqref{eq:dfdefn} and the denominator
\begin{equation}
  I = \int\d\vw \int\d\vw'\,S(\vw')\pr(\vw|\vw')f(\vw'|\Phi)
\end{equation}
ensures that the resulting pdf is correctly normalised.  We ignore
perspective effects and assume that the on-sky coordinates $(y,z)$ of
each star are known perfectly and that every observed star lies
somewhere in the range $|x|<x_{\rm max}=200\,\pc$ along the line of sight.
We assume that each measured component of velocity $v_l$
(with $l=x,y,z$) follows an
independently distributed normal distribution with dispersion
$\Delta v_l$ equal to the
quoted uncertainty or 20 km/s, whichever is larger
(\S\ref{sec:data}).  
That is, we take
\begin{equation}
  \begin{split}
&\pr(\vw|\vw')=
\delta(y-y')\delta(z-z')\pr(\vv|\vv'),
  \end{split}
\end{equation}
with
\begin{equation}
  \begin{split}
&\pr(\vv|\vv')=\\ &\quad
\prod_{l=x,y,z}
\begin{cases}
  \frac1{\sqrt{2\pi}\Delta v_l}\exp\left[-\frac{(v_l-v'_l)^2}{2\Delta v_l^2}\right], & \hbox{if
    $v_l$ measured},\\
1, & \hbox{otherwise}.
\end{cases}
%%\exp\left[-\frac12(\vv-\vv')^{\rm T}C(\vv-\vv')\right],
\label{eq:errorpdf}
  \end{split}
\end{equation}

The likelihood of the discrete kinematical dataset~$D$ is then
\begin{equation}
\pr(D|f,\Phi)=\prod_{n=1}^N\frac{\sum_{k=1}^KP_{nk}f_k}{\sum_{k=1}^KI_kf_k},
\label{eq:discretelik}
\end{equation}
where
\begin{equation}
  P_{nk} = \int_{V_k}\d\vw'_n\,S(\vw_n')\pr(\vw_n|\vw'_n,\Phi)
\end{equation}
is the probability that a star drawn from the $k^{\rm th}$ orbit block
would yield the observed $\vw_n=(\vx_n,\vv_n)$, and
the common normalising factor
\begin{equation}
  I_k=\int\d\vw \int_{V_k}\d\vw'\,S(\vw'u)\pr(\vw|\vw',\Phi)
  \label{eq:I}
\end{equation}
is the probability that a star drawn at random from that block would
be included in the discrete catalogue.

In this work we consider selection functions $S=S(\vx)$ that depend
only on the stars' positions.  Then $I_k$ is
easy to calculate from the three-dimensional density profile of each block.
The matrix elements $P_{nk}$ are more involved.
For each star~$n$ we calculate $P_{n1},...,P_{nK}$
using the following Monte Carlo procedure.  We begin by setting
$P_{n1}=\cdots=P_{nK}=0$.  Then we draw $N_{\rm sample}=10^5$ samples of
the star's unknown true velocity $\vv'$ from a sampling density
\begin{equation}
  \begin{split}
  f_{\rm s}(\vv')&=\pr(\vv_n|\vv',\Phi)\\
&\times
\prod_{l=x,y,z}\begin{cases}
1, & \hbox{if $v_l$ measured},\\
{\cal U}(v_l'|-v_{\rm max},v_{\rm max}), & \hbox{otherwise},
\end{cases}
  \end{split}
\label{eq:fs}
\end{equation}
where the uniform pdf ${\cal U}(x|a,b)=\frac1{b-a}$ if $a<x<b$ and is
zero otherwise.
We follow \citet{McMillanAnalysingsurveysour2013} in using the same
random seed to draw the measured components of each $\vv'$ for each
trial potential: this reduces noise (but not bias) in the resulting
$P_{nk}$ as the potential is varied.
The extent of the sampling
distribution in the unmeasured components of velocity is set by a
conservative bound on the maximum velocity that the star could
possibly have in the assumed potential, namely
$v_{\rm max}^2=-2\Phi(x'=0,y',z')$.  Having $(y',z',v'_x,v'_y,v'_z)$
each position $x'$ along the line of sight belongs to a single DF
cell~$k$.  We then walk along $x'$, identifying the values of $x$ that
mark boundaries between DF cells.  For each cell~$k$ that we encounter
on this walk we increment the corresponding $P_{nk}$ by
$\pr(\vv_n|\vv',\Phi)\Delta x'/N_{\rm sample}f_s(\vv')$, where
$\Delta x'$ is the $x'$-extent of the cell.

\subsection{Interlopers and misidentified stars}
\label{sec:outliers}

Our machinery is designed to model the kinematics of the old stellar
population at the Galactic centre.  Both \cite{SchodelNuclearstarcluster2009} and
\cite{FritzNuclearClusterMilky2016} point out that their catalogues are contaminated by
stars from the less relaxed young population.  Given the
inconsistencies between the two catalogues identified in
Section~\ref{sec:data} it is also conceivable that some stars could
be misidentified when measuring their proper motions.

We account for these possibilities in a simplistic way, replacing
$P_{nk}$ by
\begin{equation}
(1-f_{\rm c})P_{nk}+f_{\rm c}P_{nk}^{\infty},
\label{eq:Pnkc}
\end{equation}
where $f_{\rm c}$ is a contamination fraction
%(which might vary from
%star to star based on the quality of its measurements)
and $P_{nk}^{\infty}$ is the probability that a star drawn from block
$k$ would be observed at on-sky position $(y_n,z_n)$ without any
constraints on its velocity.  That is, we assume that every star's
measured on-sky position is perfectly correct and consistent with the
assumed $S(\vw)$, but we allow for the possibility that its measured
velocity is competely bogus.
%We could in principle assign each star its own
%value of $f_{\rm c}$ depending on the quality of its measurements, but
%we do not do so in the present paper.

Strictly speaking, this is not a true model for interlopers because
it assumes that every star included in the catalogue is bound and
belongs to the old population.  Nevertheless, one might expect that the very
general form of the DF in our models means that they have plenty of
freedom to ``fit around'' any small additional bumps in the projected
$(y,z)$ number-count distribution caused by genuine interlopers, but
that attempting to fit the velocities of such stars would
lead to devastating biases on the fitted potential.

\subsection{Binned projected zeroth and second moments}

\label{sec:binned}
These discrete kinematics are supplemented by
information on the number counts of stars $c_b$ within spatial
bins $b=1,...,B$ and the associated rms line-of-sight velocity
moments~$\sigma^2_b\equiv\langle v_x^2\rangle_b$.  For example, in the
application to the Galactic centre dataset, the binned kinematics
probe projected radii $R>19''$ (e.g., Table~\ref{tab:rad}), beyond the
radius where discrete kinematics are used.
More generally, we assume that the stars in the discrete dataset and within each
spatial bin are independent (i.e., no double counting of stars) and
that each bin contains enough stars that we may approximate the
underlying Poisson distributions by normal distributions.  Then 
the likelihood~\eqref{eq:discretelik} gains an extra factor
to account for the binned data, becoming (dropping an uninteresting
normalisation constant)
\begin{equation}
  \begin{split}
\pr(D|f,\Phi,S)
&=  \left[ \prod_{n=1}^N\frac{\sum_kP_{nk}f_k}{\sum_kI_kf_k} \right]
\exp\left[-\frac12\chi^2\right],
  \end{split}
\label{eq:fulllike}
\end{equation}
in which $\chi^2=\chi_0^2+\chi_2^2$
has contributions
    \begin{equation}
  \begin{split}
    \chi^2_0&=\sum_{b=1}^{B}\left(\frac{c_b-\sum_kQ^{(0)}_{bk}f_k}{\sqrt{c_b}}
    \right)^2\\
  \chi^2_2&=\sum_{b=1}^{B}\left(\frac{c_b\sigma_b^2-\sum_bQ^{(2)}_{bk}f_k}{\sqrt{2c_b}\sigma^2}
    \right)^2,
  \end{split}
\label{eq:chisq}
\end{equation}
from the zeroth- and second-order binned moments, respectively.  Here
the matrix $Q^{(0)}_{bk}$ is the probability that a star drawn from
block~$k$ is found in bin~$b$.  Its elements are calculated by first
writing down the (straightforward but tedious) expression for the
number-density profile $\nu_k(r)$ of orbit block~$k$ and then
integrating numerically along the lines of sight encompassed by
bin~$b$.  Similarly, $Q^{(2)}_{bk}$ is the $k^{\rm th}$ orbit block's
contribution to the number-weighted second moment integrated over the
$b^{\rm th}$ bin and is calculated by numerically integrating the
analytical expression for the second-moment distribution
$\nu\overline{v_x^2}(\vr)$ of the $k^{\rm th}$ orbit block.

In the absence of selection effects (i.e., when all $I_k=1$) the expression~\eqref{eq:fulllike}
reduces to that used by \cite{ChanameConstrainingMassProfiles2008}.

\subsection{Anisotropic and isotropic spherical models}
\label{sec:mergeblock}
Our models assume a spherically symmetric potential $\Phi(r)$, but
when $n_{L_z}>1$ they have the freedom to fit rotating, axisymmetric
DFs $f(\E,L^2,L_z)$ in this potential.  We can enforce spherical
symmetry on the DF by merging the $n_{L_z}$ orbit blocks
for each $\E$ and $L^2$ to produce a DF that is flat in~$L_z$:
having computed the $P_{nk}$, $Q^{(0)}_{nk}$, $Q^{(2)}_{nk}$ and $I_k$
projection matrices for an axisymmetric model, we can construct the
projection matrices for the corresponding spherical model simply by
adding up the $n_{L_z}$ elements corresponding to each $(\E,L^2)$
block.  Similarly, the matrices for spherical isotropic models are
obtained by summing the $n_{L^2}\times n_{L_z}$ elements for each
block in~$\E$.

\subsection{Finding the best-fit model}
\label{sec:dofit}

We follow the usual procedure \citep{RixDynamicalModelingVelocity1997}
of reporting the likelihood of the
best-fit non-negative DF for each assumed mass distribution:
for each potential we find the vector of DF values $(f_1,...,f_K)$
that maximises the likelihood~\eqref{eq:fulllike}, subject only to the
constraint that the DF is everywhere non-negative.  That is, we assign
\begin{equation}
  \label{eq:stdlik}
  \pr(D|\Phi,S) \equiv
  \max_{\{f_k\ge0\}}\pr(D|f,\Phi,S).
\end{equation}
To enforce the non-negativity $\{f_k\ge0\}$ constraint we follow
\cite{ChanameConstrainingMassProfiles2008} in writing $f_k=x_k^2/V_k$
then using a conjugate gradient method to find the set $\{x_k\}$
that maximises the likelihood~\eqref{eq:fulllike}.  We start from an
initial guess of $x_k^2=1/K$, which assigns the same probability mass
$f_kV_k=1/K$ to all orbit blocks.

\subsection{Computational costs}

For the models presented in this paper the dominant expense in
constructing and fitting a single model for an assumed
potential~$\Phi$ is the calculation of the $P_{nk}$ matrix elements
(Section~\ref{sec:projections}).  It takes about 5 CPU seconds per
star to calculate the contribution of all $K$ orbit blocks from
$10^5$ samples of the star's velocity.  The next biggest expense is
typically the calculation of the normalisation coefficients~$I_k$
(equation~\ref{eq:I}).  If $S(\vw)$ varies over an orbit then this
requires numerical integration; the integral over velocities can be
carried out analytically, but the integration over the spatial
coordinates must be carried out numerically.

The likelihood maximisation requires a few thousand
iterations to converge for our spherical anisotropic models with
$n_{\E}\times n_{L^2}\times n_{L_z}=200\times10\times1$ orbit blocks.
This takes about a minute for our test models with 1000 stars
(Section~\ref{sec:tests} below) to about 10 minutes for our models of
the real Galaxy with $4105$ stars (Section~\ref{sec:app}).
Axisymmetric models with $n_{\E}\times n_{L^2}\times
n_{L_z}=200\times10\times8$ take significantly longer to converge.

%%%%%%%%%%%%%%%%%%%%%%%%%%%%%%%%%%%%%%%%%%%%%%%%%% 

\begin{figure*}
  \null\hskip10pt
  \hbox to 0.32\hsize{\hfill all 3 components of $\vv$\hfill}
  \hbox to 0.32\hsize{\hfill proper motions only\hfill}
  \hbox to 0.32\hsize{\hfill line-of-sight only\hfill}
  
  \vspace{-0.3cm}
  \null\leavevmode\raise30pt\hbox{\rotatebox{90}{\hbox{\small
        spherical isotropic}}}
  \includegraphics[width=0.32\hsize]{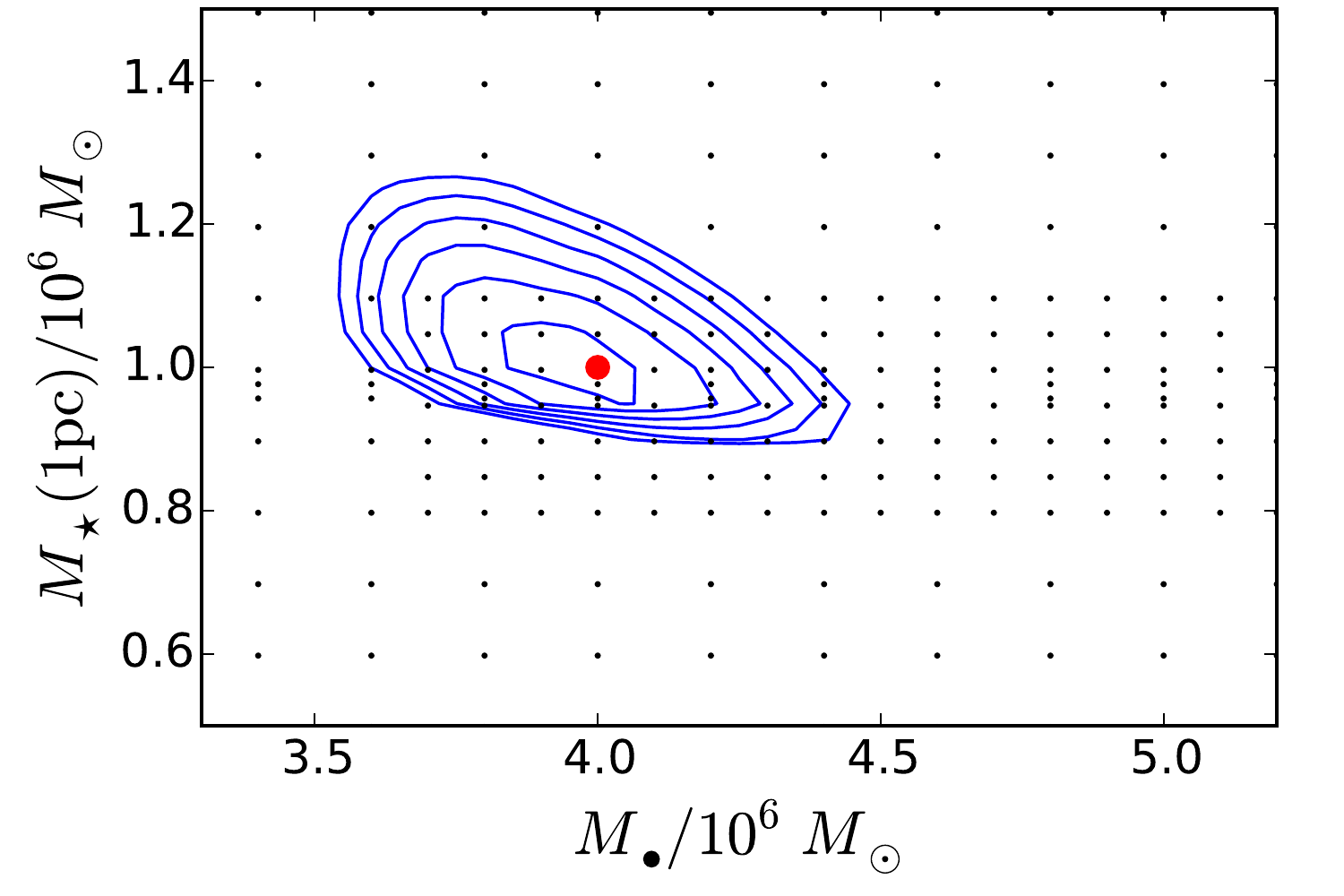}
  \includegraphics[width=0.32\hsize]{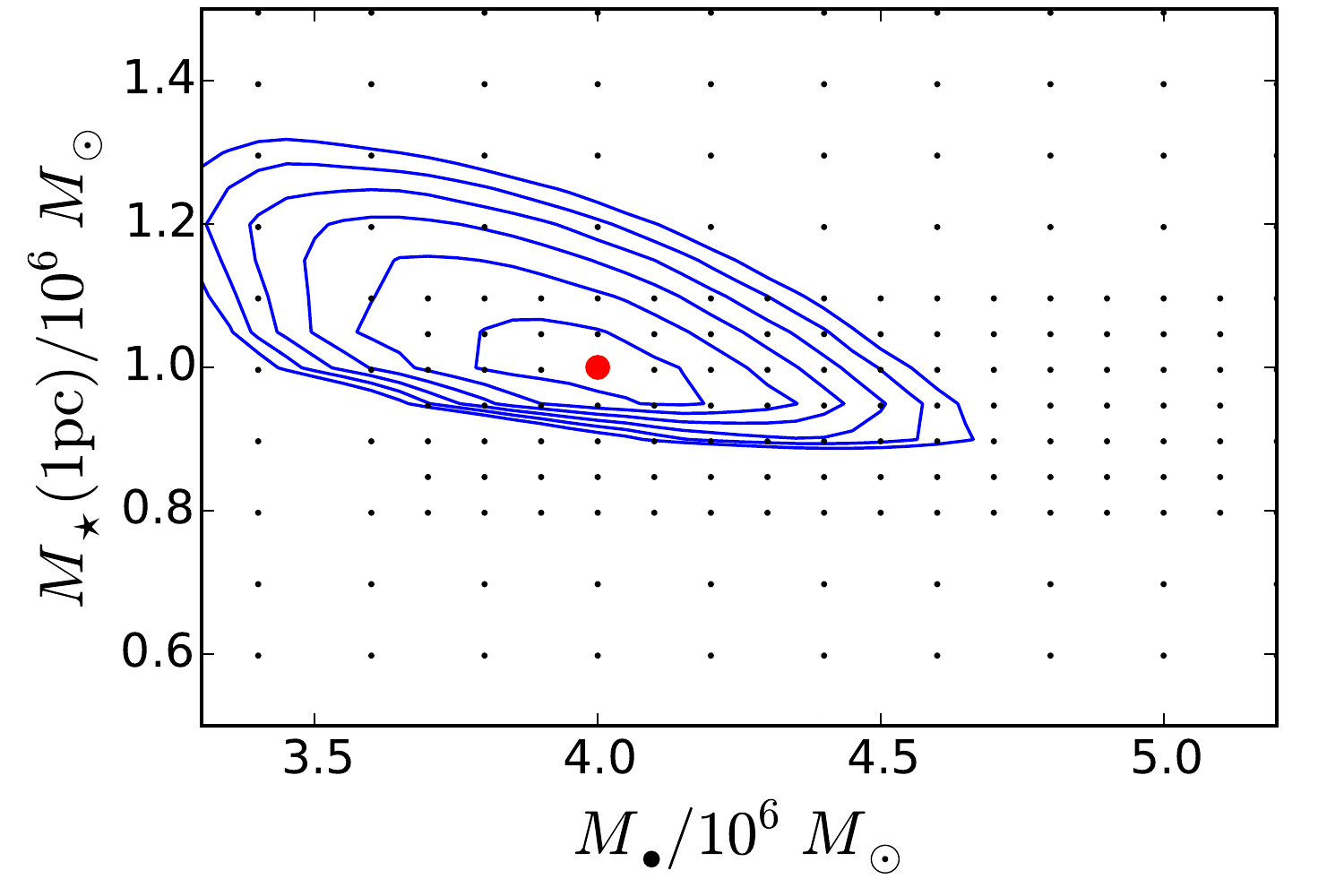}
  \includegraphics[width=0.32\hsize]{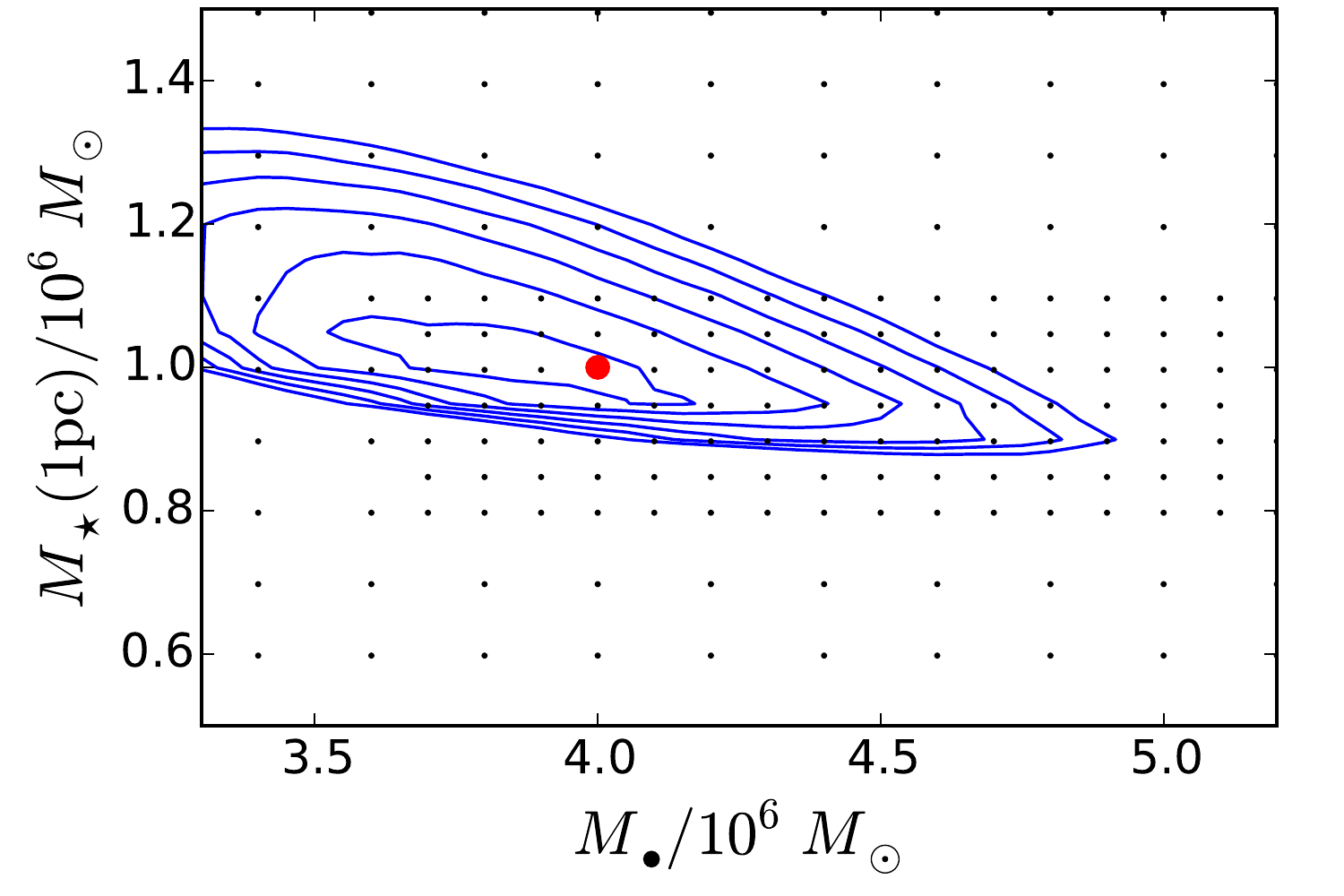}

  \null\leavevmode\raise25pt\hbox{\rotatebox{90}{\hbox{\small
        spherical anisotropic}}}
  \includegraphics[width=0.32\hsize]{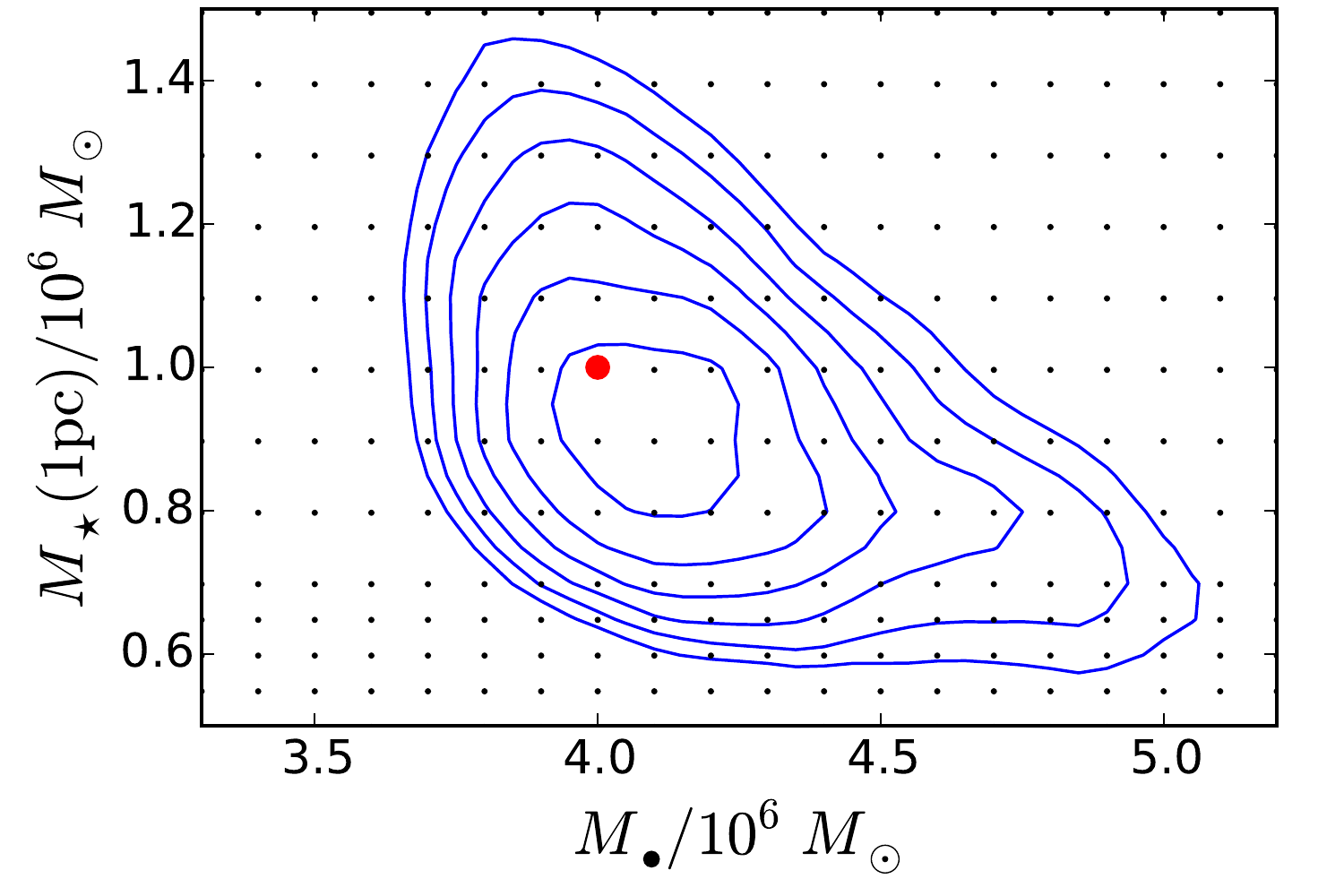}
  \includegraphics[width=0.32\hsize]{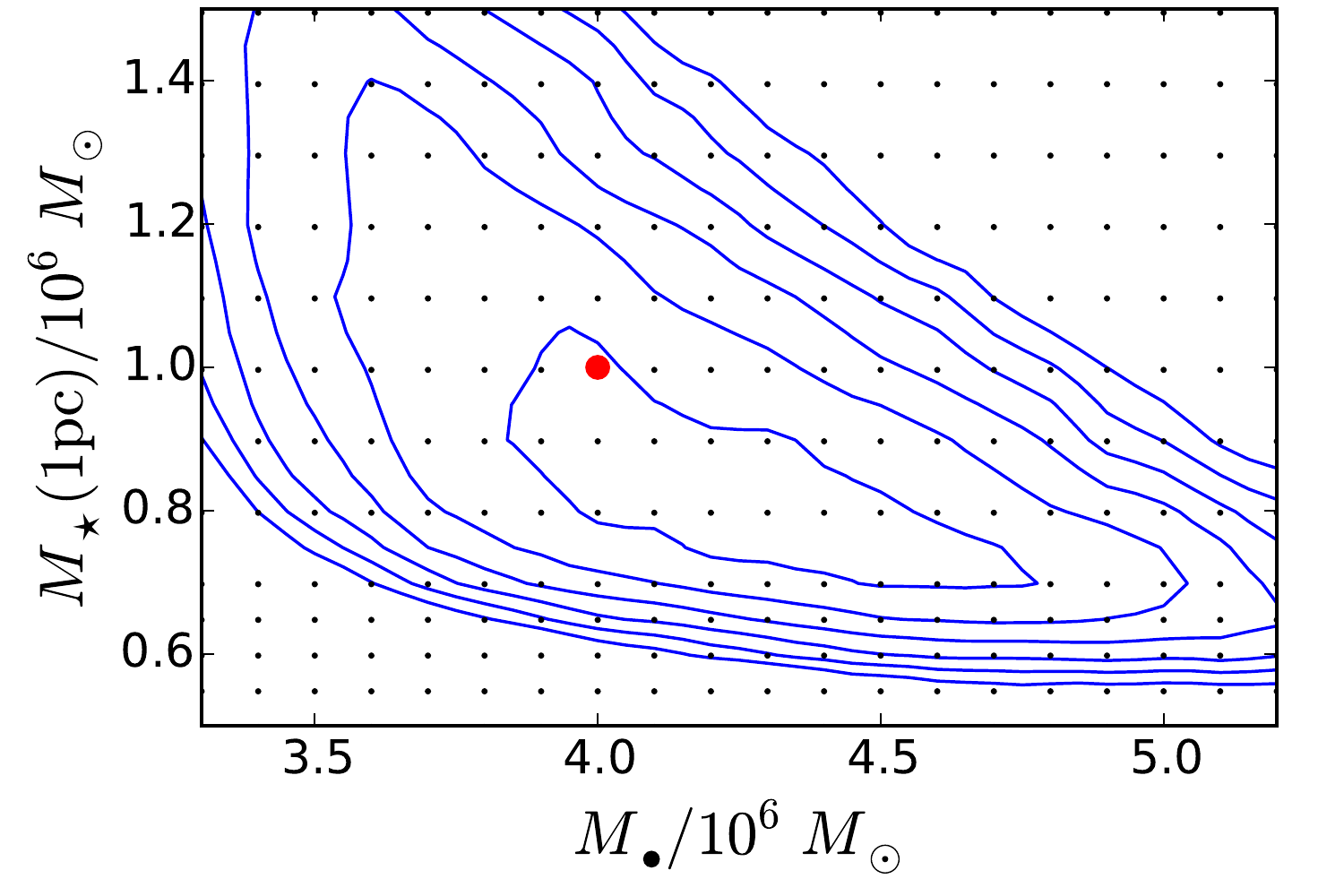}
  \includegraphics[width=0.32\hsize]{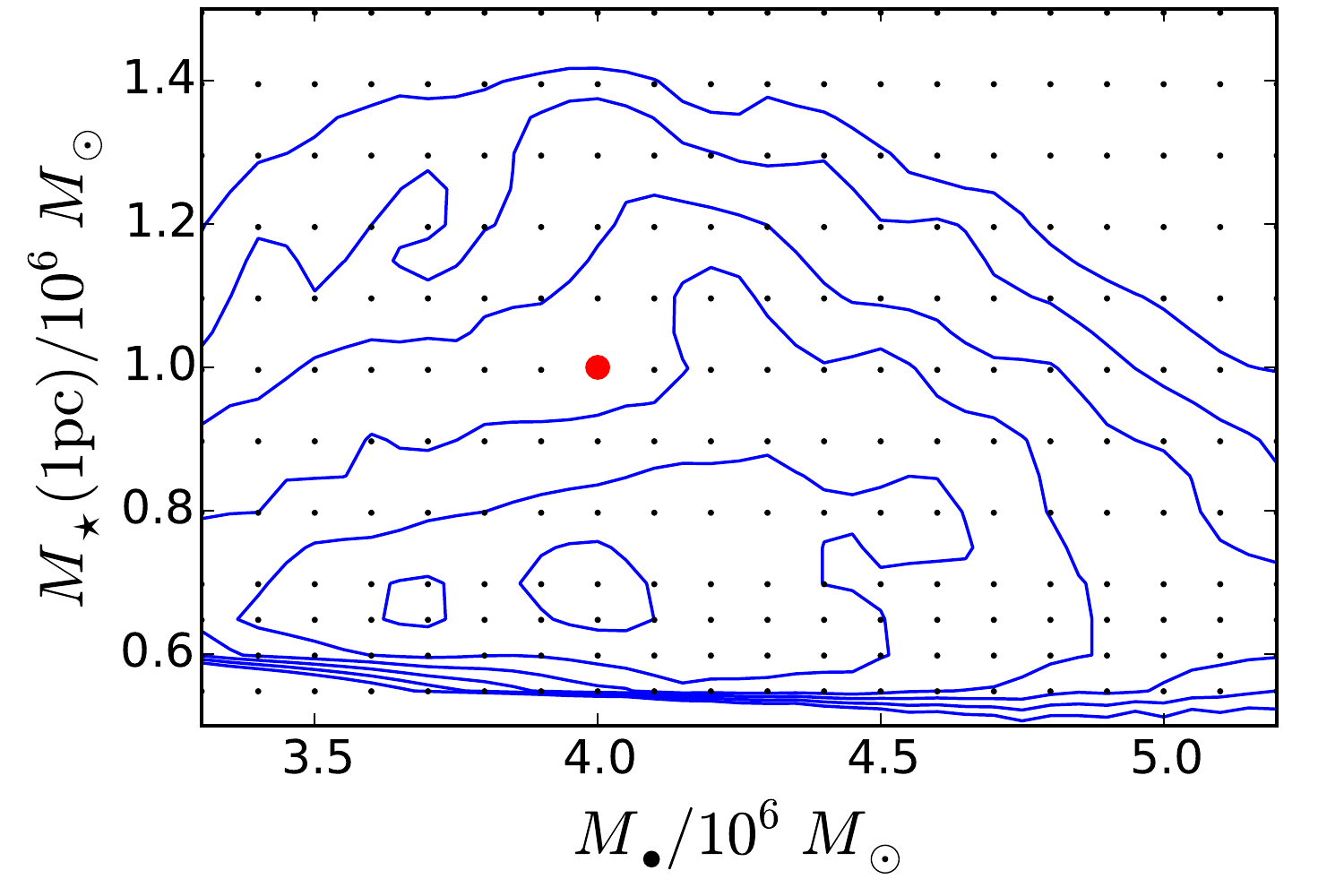}

  \null\leavevmode\raise40pt\hbox{\rotatebox{90}{\hbox{\small axisymmetric}}}
  \includegraphics[width=0.32\hsize]{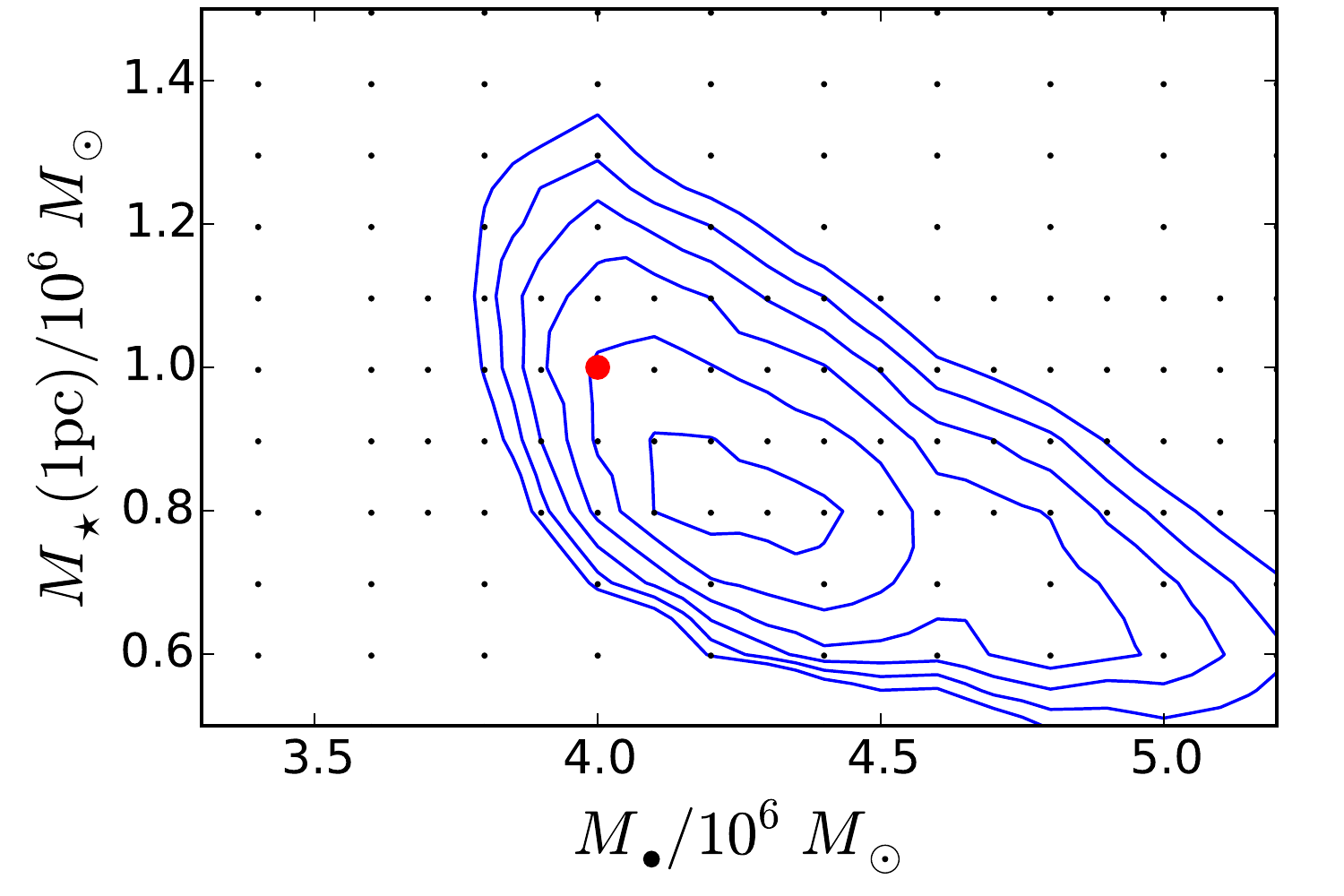}
  \includegraphics[width=0.32\hsize]{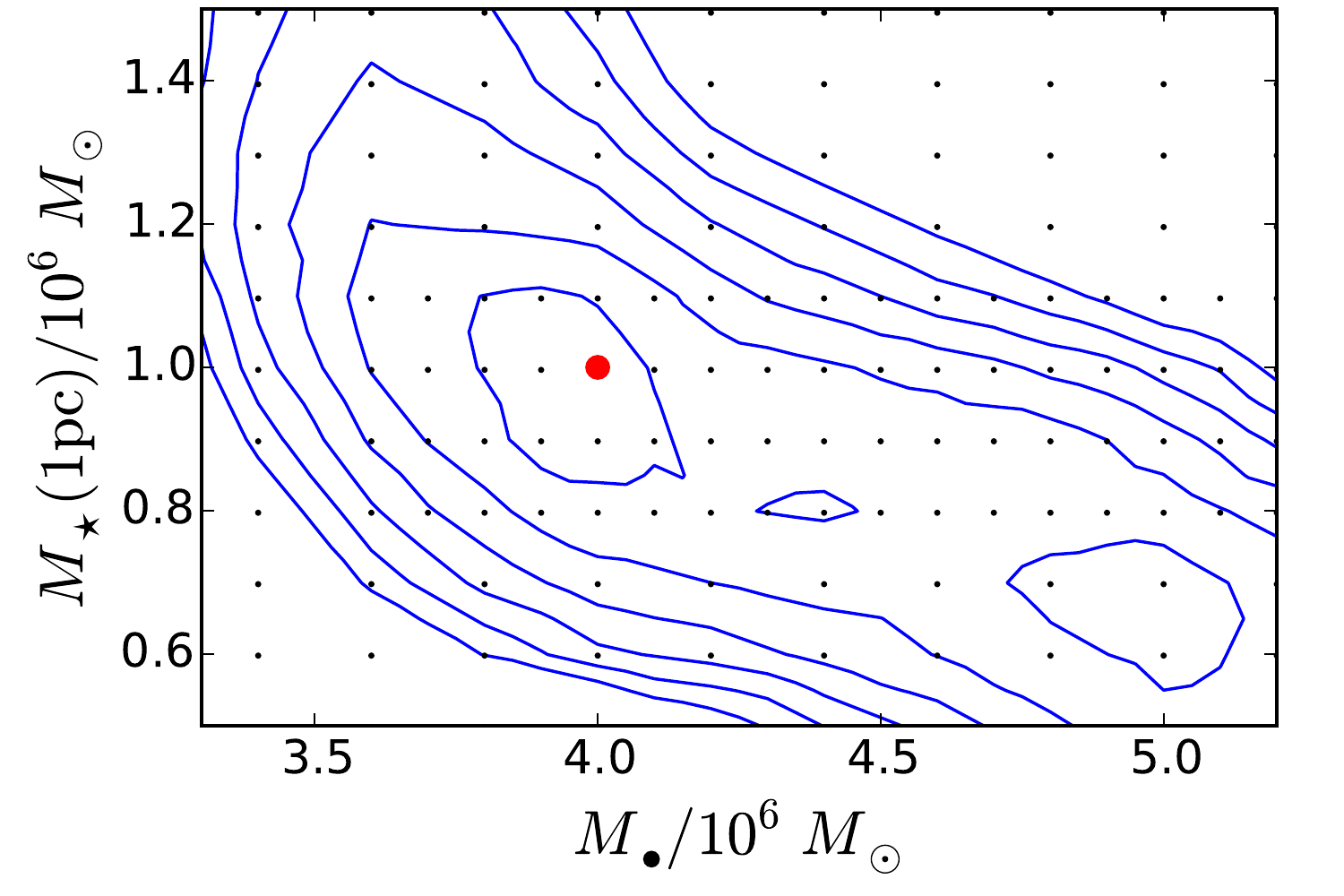}
  \includegraphics[width=0.32\hsize]{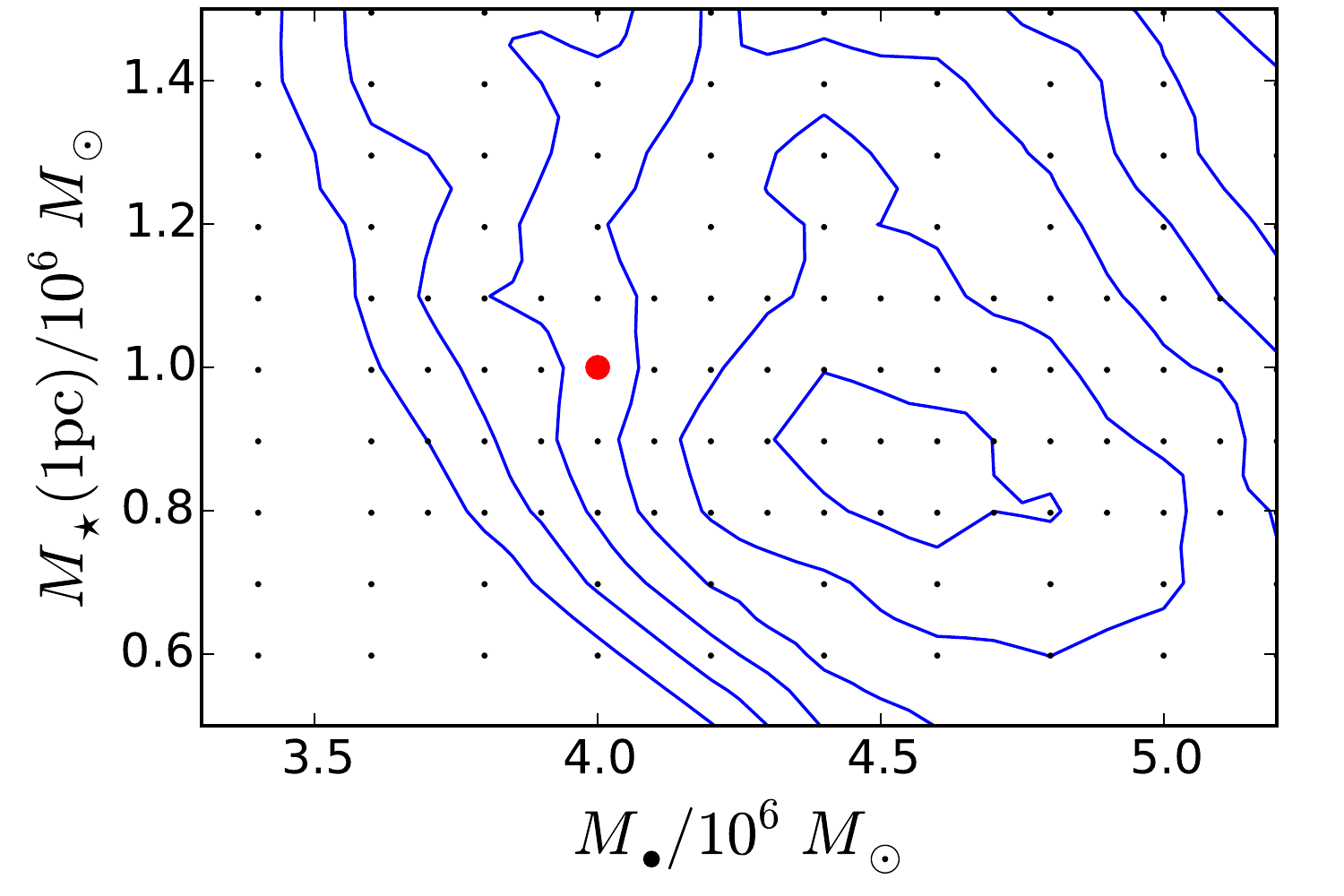}

  \caption{Likelihoods $\pr(D_0|M_\bullet,M_\star,S)$ of models fit to
    data~$D_0$ of the simulated cluster described in Section~\ref{sec:tests}.
    These models assume the correct form for the stellar potential,
    leaving only the black hole mass $M_\bullet$ and cluster mass~$M_\star$ as free parameters.
    From top to bottom the panels show results for spherical
    isotropic, spherical anisotropic and axisymmetric models.
    The left column shows results for the case in which all three
    components of velocity are available for the inner discrete data.
    The middle and right columns show results when only proper motions
    and line-of-sight velocities, respectively, are available.
    Successive contours are spaced by
    $\Delta\log\pr(D_0|M_\bullet,M_\star,S)=1$, with the innermost
    contour starting at $0.5$ below the peak value of $\log\pr(D_0|M_\bullet,M_\star,S)$;
    this innermost contour would then indicate the ``1-sigma'' uncertainty
    on $M_\bullet$ or $M_\star$ if the likelihood were perfectly Gaussian.
    In each panel the red dot indicates the correct value of the
    parameters $(M_\bullet,M_\star)$.  The fainter black dots show the
    locations at which the log likelihoods have been calculated to
    construct the contours.}
  %% TODO: comment that I use moments outside!
  \label{fig:discretebincorrectprof}
\end{figure*}

\section{Tests with mock observations of simulated clusters}

\label{sec:tests}

Before applying our modelling machinery to the real Galactic centre we
first test it against mock data drawn from a variety of simple model
clusters.  The density profiles of these simulated clusters vary, but
each has a central BH of mass $4\times10\,M_\odot$ and an extended mass
of $M_\star=10^6\,M_\odot$ enclosed within $1\,\pc$.

To generate the simulated data for each cluster we solve for the
cluster's DF and from that draw a large number of stellar positions
and velocities, scattering each component of velocity by a random
number drawn from a normal distribution with standard deviation equal
to the assumed observational uncertainty of
$20\,\hbox{km/s}$.
The mock observations consist of a discrete sample of $10^3$ such
stars that lie within a projected radius of 19'' of the BH, plus
stellar number counts and second-order line-of-sight velocity moments summed over
the on-sky annuli given in Table~\ref{tab:rad}.

\subsection{A spherically symmetric cluster}

\label{sec:testspherical}
Our first simulated cluster is spherically symmetric.  Its mass- and
number-density profiles are given by eq.~\eqref{eq:rhoprof} with a scale radius
$r_{\rm s}=10\,\pc$ and inner slope $\gamma=1$
\citep{HernquistAnalyticalmodelspherical1990}.  The cluster has an
isotropic velocity distribution.  To construct our simulated catalogue
we use the
standard Eddington inversion procedure
\citep[e.g.,][]{BinneyGalacticDynamicsSecond2008} to find its DF
$f(\E)$ and then draw stellar positions and velocities from that.

\subsubsection{Recovery of $M_\bullet$ and $M_\star$ assuming 
  the correct (unnormalised) density profile}

\label{sec:testspherical1}
The most basic test of the modelling procedure outlined
in~\S\ref{sec:models} is whether it can recover the parameters of the
mass distribution when the correct underlying functional
form~\eqref{eq:rhoprof} is assumed.  This profile has four free
parameters: $M_\bullet$, $M_\star$, $\gamma$ and $r_{\rm s}$.  A full scan
over all four would be prohibitively expensive.  So, we start by
constructing models that assume the correct shape ($\gamma=1$) and
radial scale ($r_{\rm s}=10\,\pc$) for the mock cluster, leaving only the pair
of mass normalisation parameters $(M_\bullet,M_\star)$ to be constrained.

For each $(M_\bullet,M_\star)$ pair we calculate the potential and set
up a grid of $K=n_\E\times n_{L^2}\times n_{L_z}=200\times10\times8$
orbit blocks with apocentre radii spaced logarithmically between
$0.008\,\pc\simeq0.02''$ and $200\,\pc$.  Then we calculate the $P_{nk}$
matrix for the 1000 discrete stars, the $Q^{(0)}_{bk}$ and
$Q^{(2)}_{bk}$ matrices for the binned moments, and the normalisation
factors $I_k$ assuming the selection function~$S$ equals~1 inside
projected radius $R>19''$ and zero outside.  Having calculated these
matrices we use the maximization procedure of \S\ref{sec:dofit} to
find the likelihood $\pr(D|\Phi,S)$.  To fit spherical anisotropic or
isotropic model to the data we can merge these axisymmetric DF
blocks as described in \S\ref{sec:mergeblock} before carrying out the
likelihood maximization.

Figure~\ref{fig:discretebincorrectprof} shows the 
resulting likelihoods, $\pr(D|M_\bullet,M_\star,S)$, under various
assumptions about the underlying galaxy model.  From top to bottom,
the rows show the results for spherical isotropic models ($n_{L^2}=n_{L_z}=1$),
spherical anisotropic models ($n_{L^2}=10$, $n_{L_z}=1$) and
axisymmetric $f(E,L^2,L_z)$ models ($n_{L^2}=10$, $n_{L_z}=8)$.
The left-most column shows models that fit all three components of each
discrete velocity.  The middle column shows fits to proper motions only
(i.e., the $P_{nk}$ matrix is recalculated assuming complete ignorance
of the line-of-sight component of velocity), while the rightmost column
shows the result of fitting only to the line-of-sight component,
ignoring proper motions.

In all but one case the correct
$(M_\bullet,M_\star)=(4,1)\times10^6\,M_\odot$ is well within the 95\%
credible interval returned by the models (assuming a flat prior on
either $M_\bullet$ and $M_\star$ or on their logarithms).
The spherical isotropic models provide particularly tight constraints
on $(M_\bullet,M_\star)$.  This is not surprising, because such models
have no freedom in their internal dynamics once their radial mass- and
number-density profiles are given.
The number-density distribution is not completely
determined by the discrete observations, however.  Therefore, the greater the
number of components of velocity used for the discrete stellar sample,
the tighter the constraints on the mass parameters from the isotropic
models become.

The spherical anisotropic models have much more freedom, as is evident
from the widening of their likelihood contours compared to the
isotropic ones.  Again, the tightest constraints are obtained when all
three components of velocity are available.
The axisymmetric models have yet more freedom.
On closer inspection of the spherical anisotropic and axisymmetric models
we find that the $\chi^2$ of the fit to the binned outer
data (equation~\ref{eq:chisq}) is typically very small, varying between about
0.1 and 2 (for the 24 datapoints implied by Table~\ref{tab:rad}), provided
$M_\star>0.5\times10^6\,M_\odot$.  Nevertheless, these binned outer
profiles are essential for ruling out the most outlandish
orbit distributions: the likelihood contours of models fit only to the
discrete inner stellar data are much noisier, showing that without the
constraints provided by the outer binned data the anisotropic models are
freer to overfit the details of the discrete stellar distribution.
There is an extremely steep increase in $\chi^2$ as $M_\star$ falls
below the threshold value of $0.5\times10^6\,M_\odot$.  This is the
cause of the sharp cutoff in the
bottom edge of the contours plotted in Figure~\ref{fig:discretebincorrectprof}.

We note that the best-fitting $M_\bullet$ in the anisotropic and
axisymmetric models shown in Figure~\ref{fig:discretebincorrectprof}
tends to be an overprediction of the true value.  There is a
corresponding underprediction of $M_\star$, consistent with the
characteristic mass estimate of the combined BH+cluster system being
better constrained than either $M_\bullet$ or~$M_\star$ individually.
As a quick check of the significance of these systematically high
best-fit $M_\bullet$ values we have generated a number of further
discrete realisations of the cluster model and fit axisymmetric models
to those.  In some cases the resulting best-fit values of $M_\bullet$
are higher than the true one, while in other cases they are lower.
So, although we are unable to prove formally that our implementation
of the DF-superposition method produces unbiased estimates
of~$M_\bullet$ and $M_\star$, the results plotted in
Figure~\ref{fig:discretebincorrectprof} do not constitute evidence for
such a bias.

\begin{figure*}
  \null\hskip10pt
  \hbox to 0.32\hsize{\hfill all 3 components of $\vv$\hfill}
  \hbox to 0.32\hsize{\hfill proper motions only\hfill}
  \hbox to 0.32\hsize{\hfill line-of-sight only\hfill}
  
  \vspace{-0.3cm}
  \null\leavevmode\raise35pt\hbox{\rotatebox{90}{\hbox{\small
        spherical isotropic}}}
  \includegraphics[width=0.32\hsize]{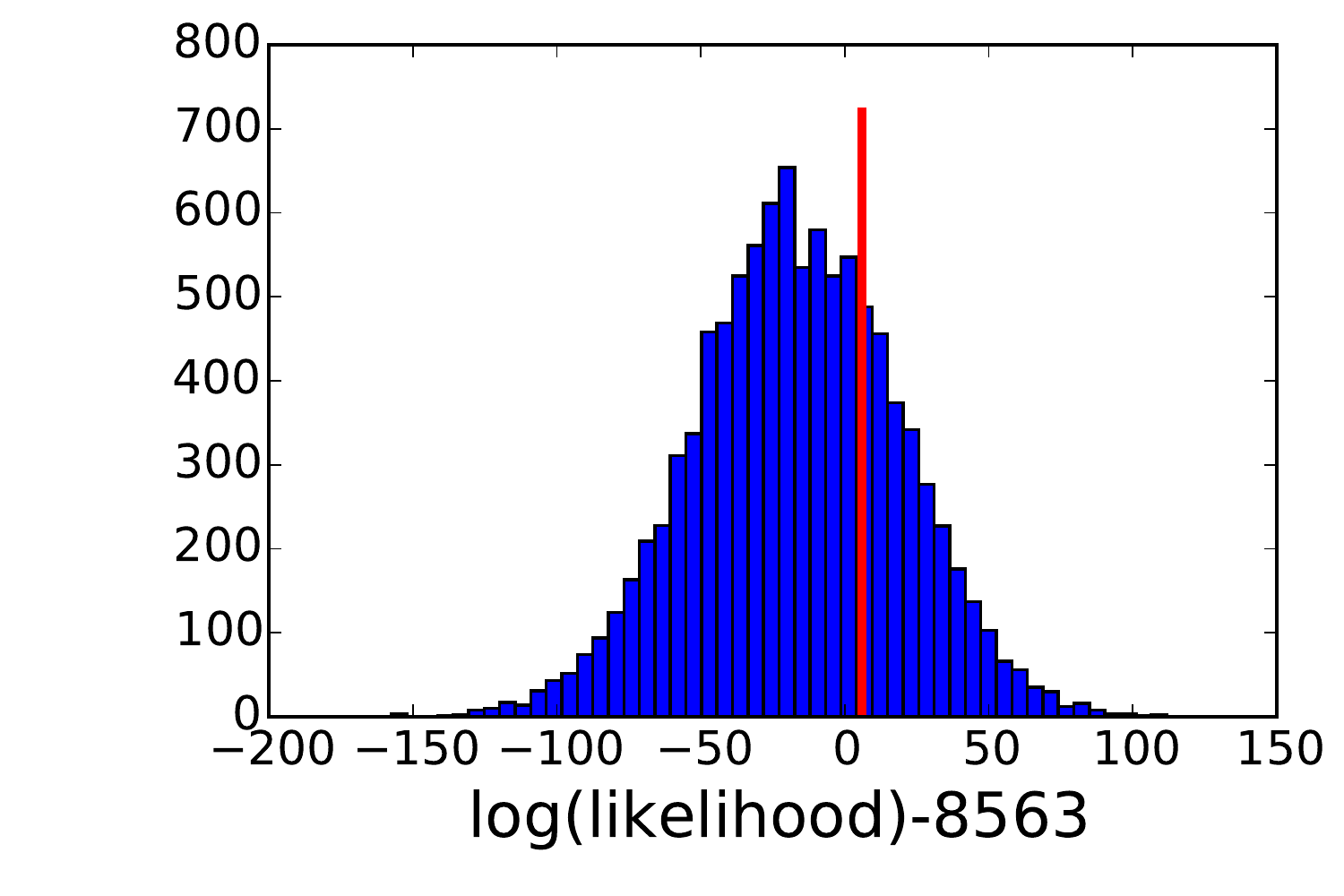}
  \includegraphics[width=0.32\hsize]{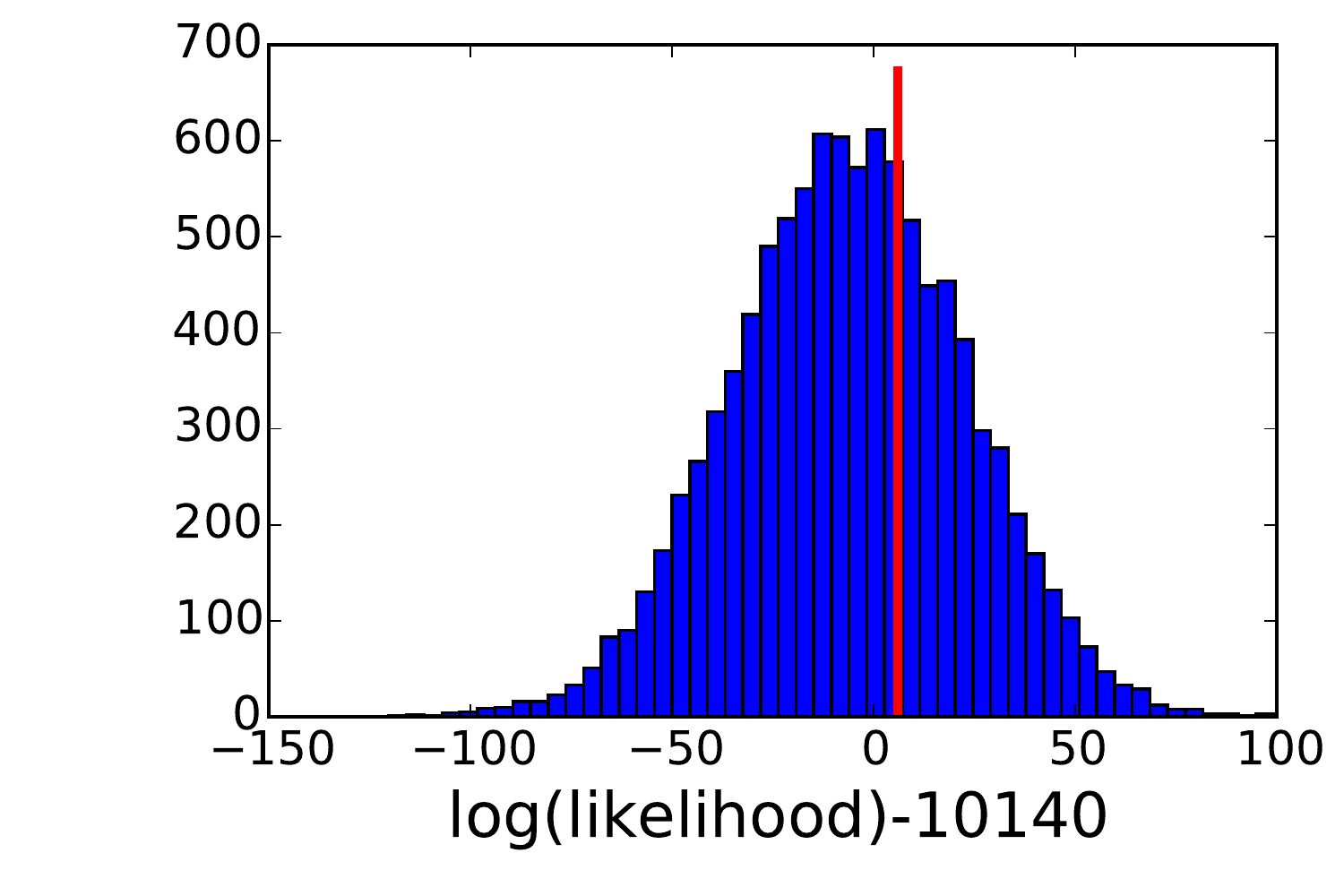}
  \includegraphics[width=0.32\hsize]{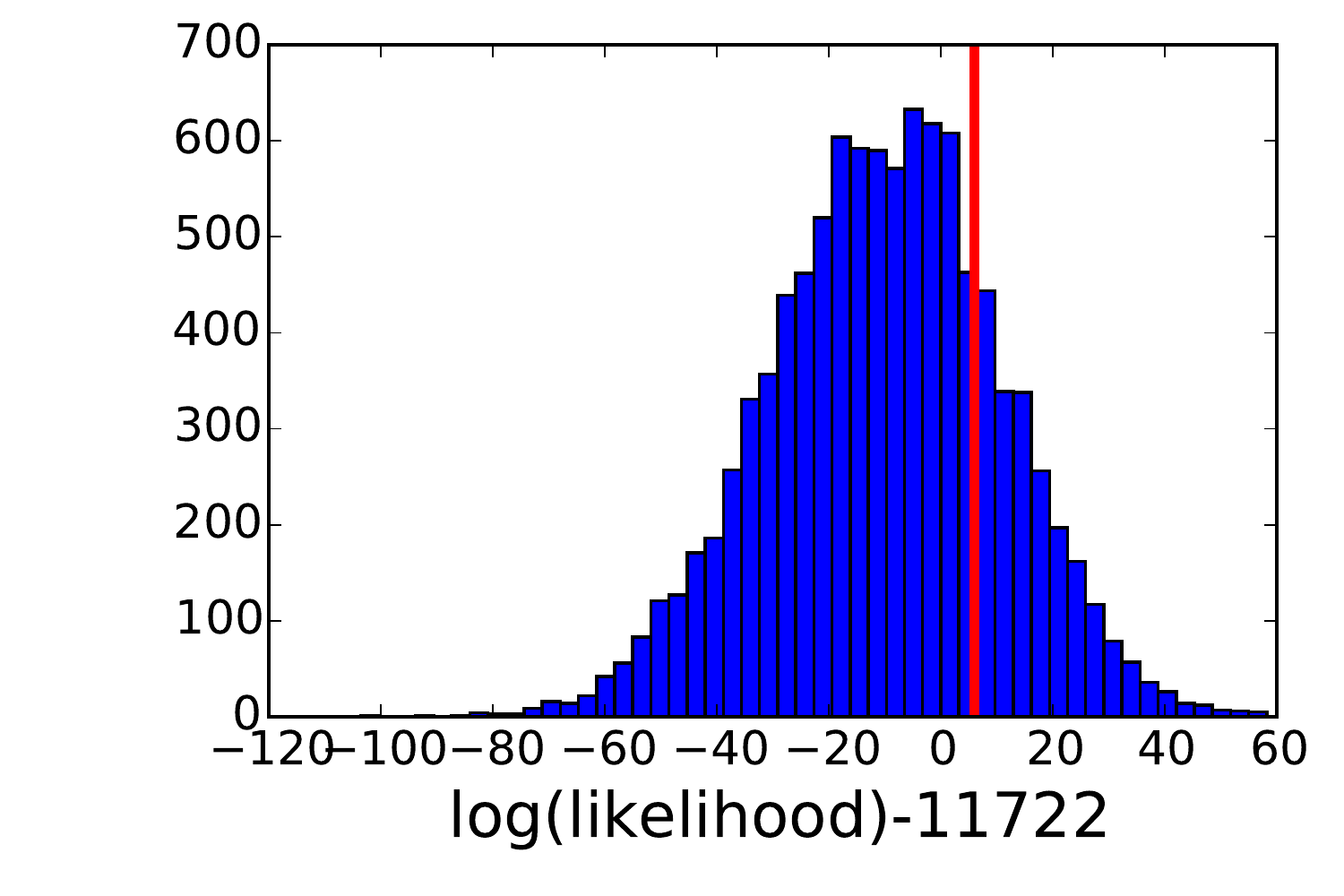}

  \null\leavevmode\raise25pt\hbox{\rotatebox{90}{\hbox{\small
        spherical anisotropic}}}
  \includegraphics[width=0.32\hsize]{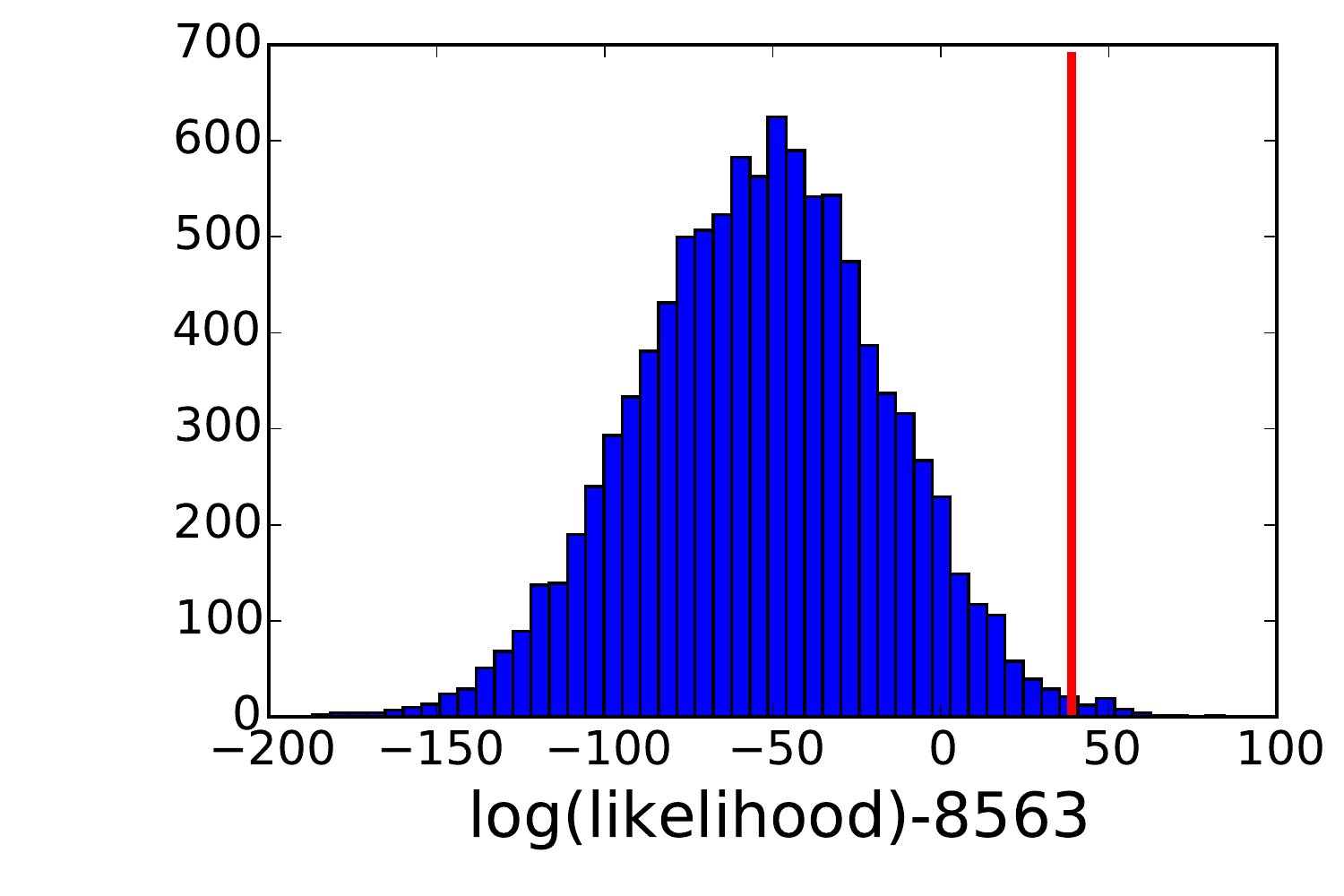}
  \includegraphics[width=0.32\hsize]{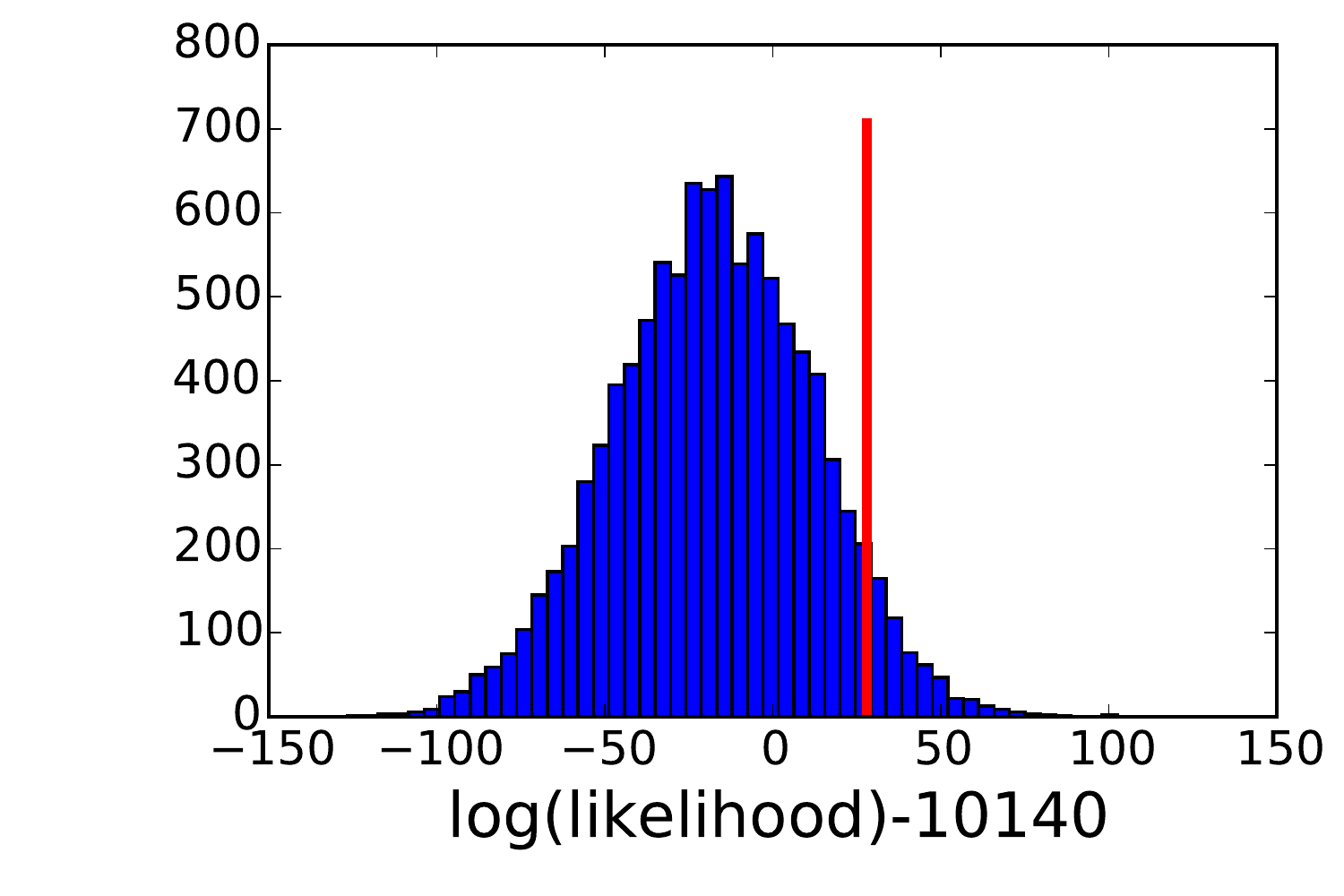}
  \includegraphics[width=0.32\hsize]{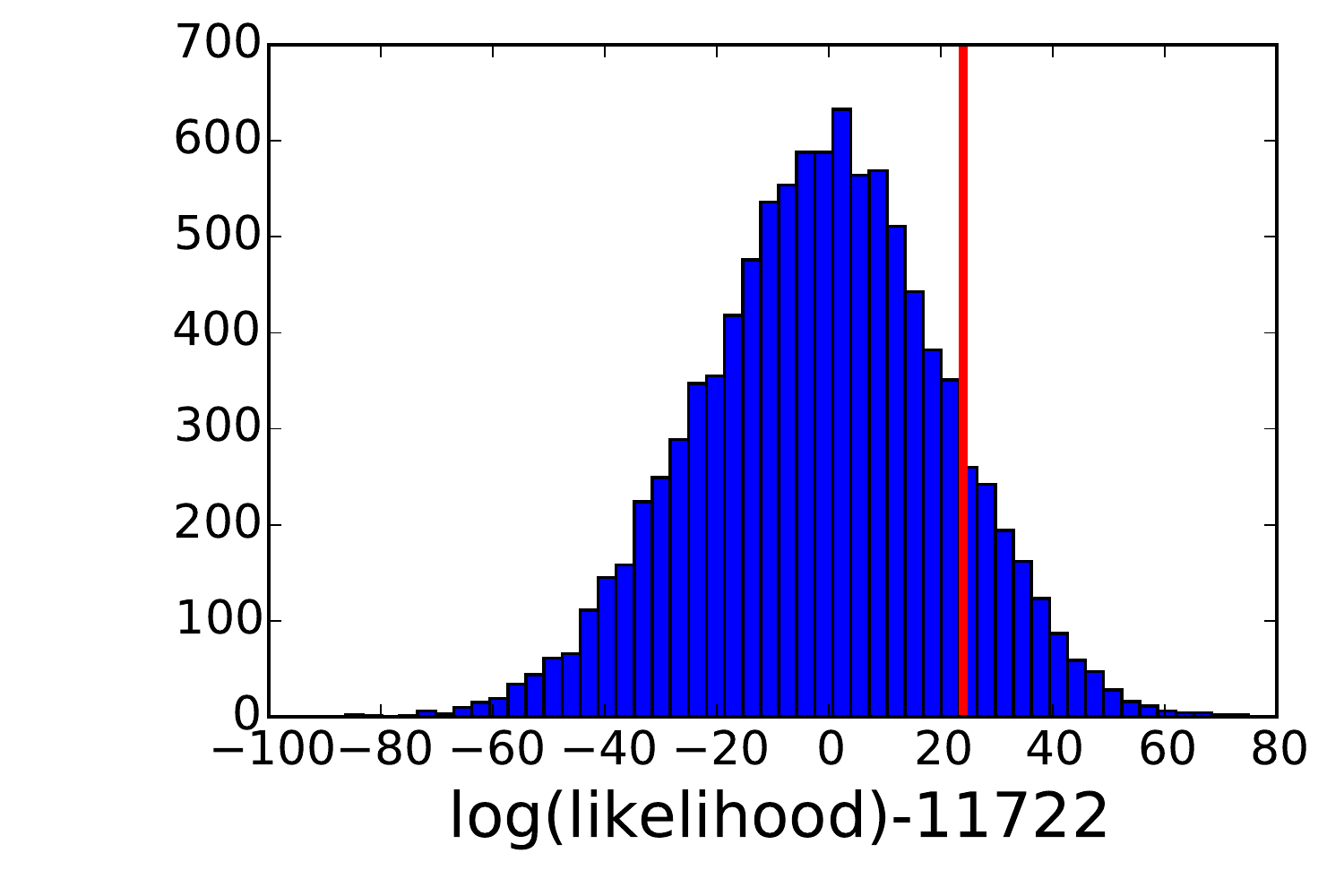}

  \null\leavevmode\raise40pt\hbox{\rotatebox{90}{\hbox{\small axisymmetric}}}
  \includegraphics[width=0.32\hsize]{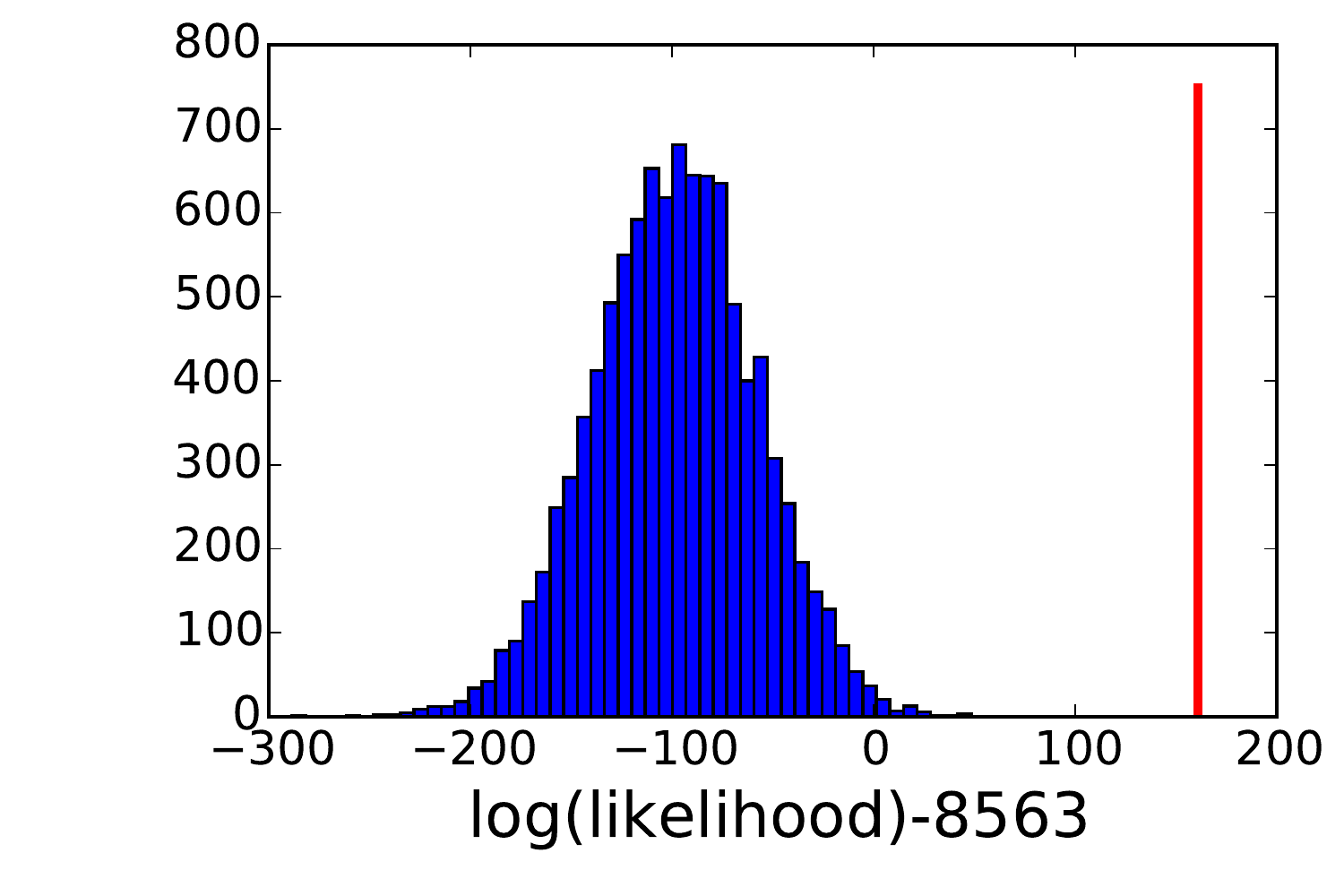}
  \includegraphics[width=0.32\hsize]{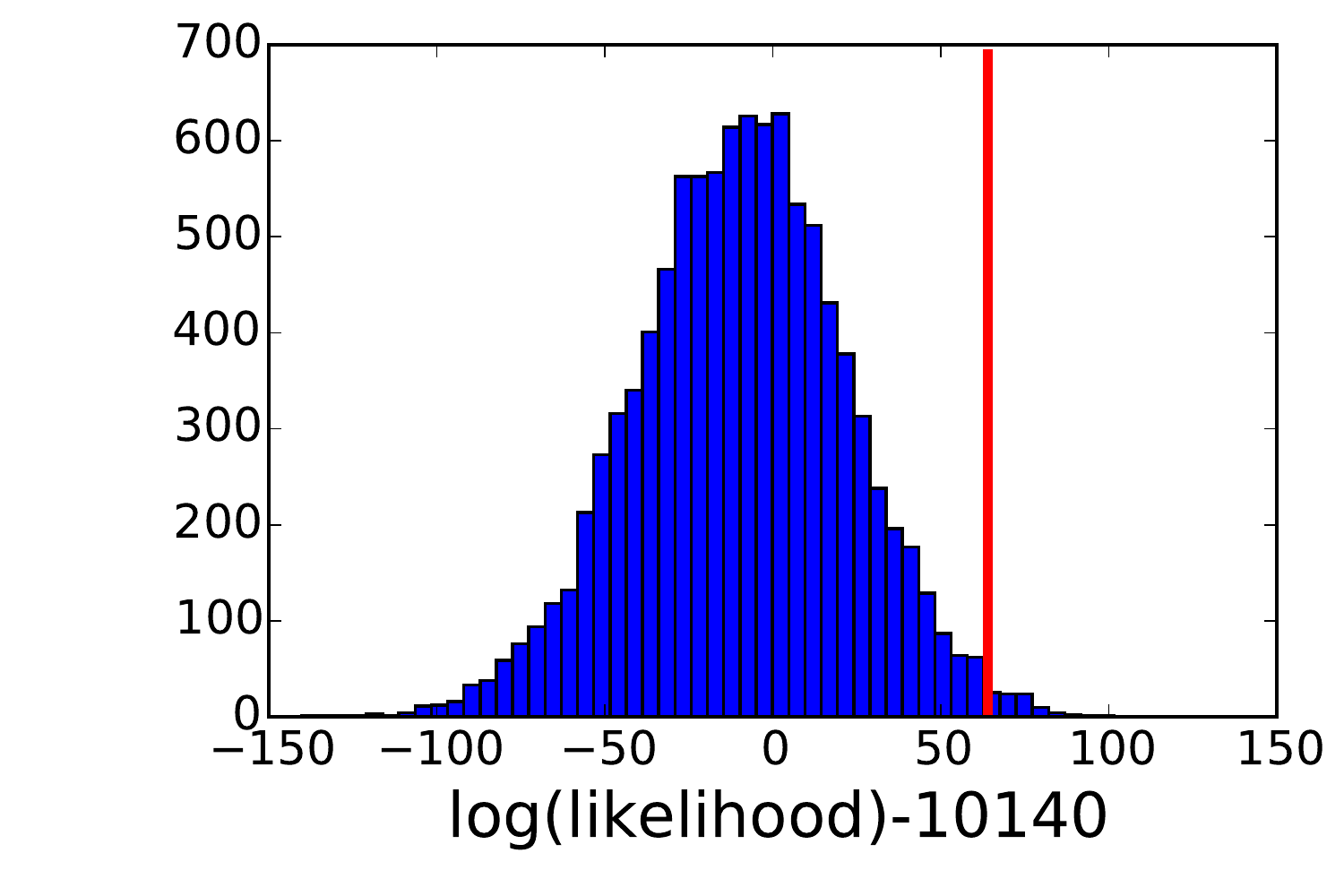}
  \includegraphics[width=0.32\hsize]{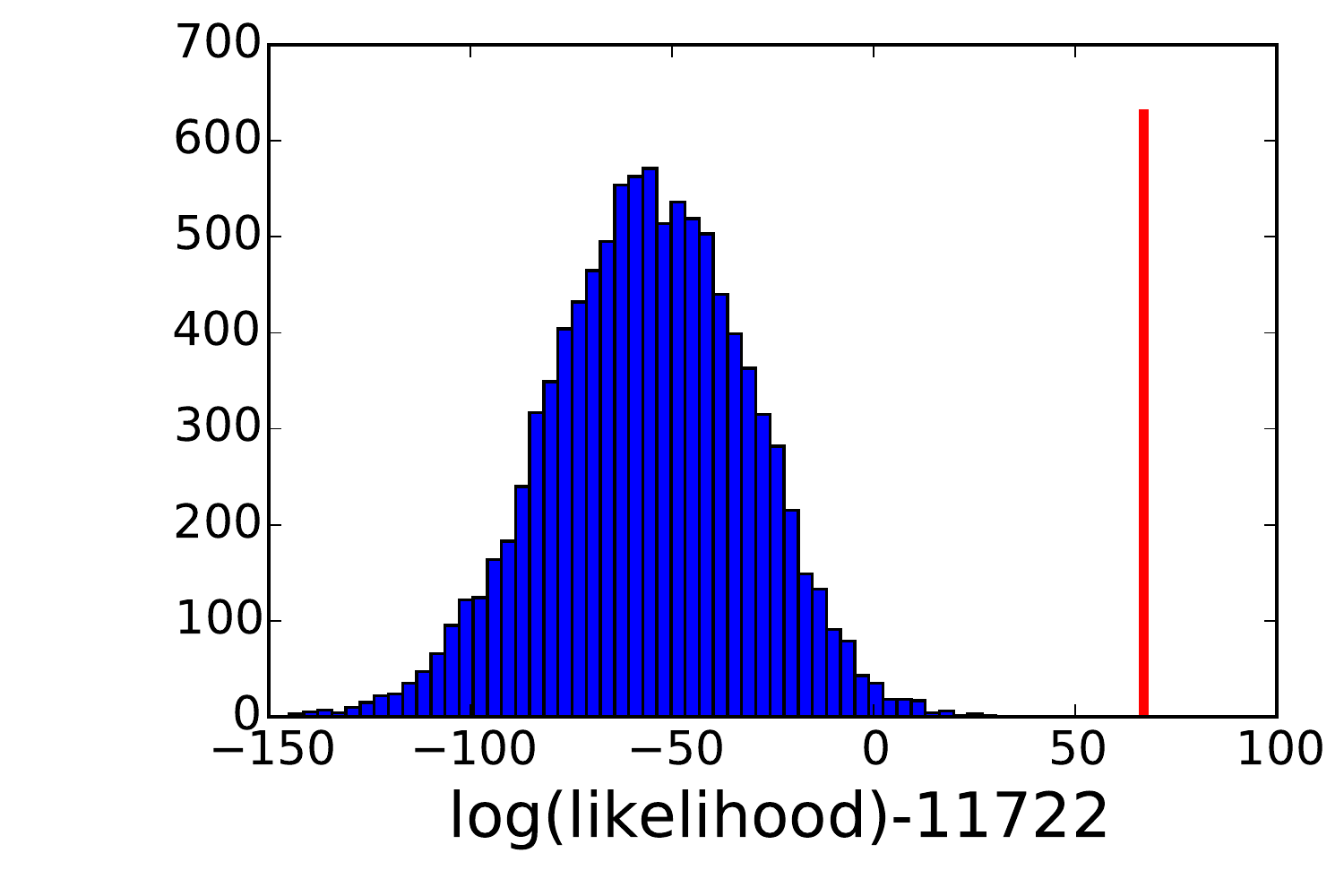}

  \caption{Log likelihoods of various realisations of the models shown
    in Figure~\ref{fig:discretebincorrectprof}.  In each panel the
    zero point of the horizontal axis indicates the log likelihood
    $\log\pr(D_0|f,\Phi,S)$ of the initial dataset $D_0$ drawn from the
    simulated cluster's underlying smooth DF~$f$.  The heavy vertical
    red line indicates the log-likelihood $\log\pr(D_0|f_0,\Phi,S)$ of
    the DF-superpostion model $f_0$ that best fits these data.  Notice that
    this $\log\pr(D_0|f_0,\Phi,S)$ is always at least as large as
    $\log\pr(D_0|f,\Phi,S)$ and becomes larger as $f_0$ becomes more
    flexible (top to bottom rows).  The histograms plot the
    distribution of log likelihoods $\log\pr(D_i|f_0,\Phi,S)$ of an
    ensemble of datasets $D_1$, $D_2$, ..., each of 1000 stars drawn
    from this best-fit DF~$f_0$.  }
  \label{fig:loglikbincorrectprof}
\end{figure*}

\subsubsection{Testing for overfitting}

\label{sec:overfit}
When DF-superposition models are fit to {\it integrated} stellar
kinematics the likelihood becomes a Gaussian,
$\pr(D|\vf,\phi)\propto\exp[-\frac12\chi^2]$, in which $\chi^2[\vf]$ is
some quadratic form in the orbit weights $(f_1,...,f_K)$.
\citet{ValluriDifficultiesRecoveringMasses2004} have pointed out that
the usual ``maximum-likelihood'' procedure for considering only the
very best-fitting orbit weights~$\vf$ for each potential leads
anomalously low values of $\chi^2$: the models overfit the data.
We have just seen that our models tend to produce fits to binned,
outer data that are too good to be true, with $\chi^2<2$.
\citet{MagorrianConstrainingblackhole2006} provided an explanation for
such behaviour, starting from the observation that this $\chi^2$ is a
hugely degenerate quadratic form in the orbit weights.  He argued that
the the correct resolution of the overfitting problem was to
marginalise~$\vf$ after adopting a suitable prior, but nevertheless
found that in an example problem the standard ``maximum-likelihood''
procedure does produce reliable BH masses, albeit with formal
uncertainties on $M_\bullet$ which are slightly too tight.
The situation with {\it discrete} kinematical data is less clear, however,
because the likelihood~\eqref{eq:discretelik} is not a Gaussian.
In the following we carry out some simple checks to test for
overfitting in the discrete case.

Let $D_0$ be the ``observed'' discrete stellar kinematics of our
simulated cluster, and let $f$ the the smooth underlying DF from which
these are drawn.  When we feed this $D_0$ into our DF-superposition
modelling code, it produces some best-fit DF, represented as a
weighted sum of orbit blocks.  We call this fitted DF~$f_0$.  Assuming the
correct potential~$\Phi$ is used, we expect the likelihood
$\pr(D_0|f_0,\Phi)$ of this fitted model to be larger than or equal to
the likelihood $\pr(D_0|f,\Phi)$ of the ``true'' model~$f$, because
the block DF $f_0$ has sufficient flexibility to fit the details of the
``observed'' data $D_0$, while including (an approximate, discretized
version of) the true~$f$ as a special case.

Figure~\ref{fig:loglikbincorrectprof} shows that this is indeed the
case.  The heavy vertical red line in each panel indicates the log
likelihood $\log\pr(D_0|f_0,\Phi,S)$ of the best-fit model for each of
the cases considered in Figure~\ref{fig:discretebincorrectprof}.  The
log-likelihoods of the best-fit spherical, isotropic models (left
column) are within $\sim1$ of the log-likelihoods of the smooth
models~$f$ from which the data were generated.  The best-fit spherical
anisotropic models have $\log\pr(D_0|f_0,\Phi,S)$ larger than
$\log\pr(D_0|f,\Phi,S)$, the difference becoming even larger for more
general axisymmetric models.

From each of these best-fit models~$f_0$ we draw further discrete
realisations $D_1$, $D_2$, ...., each of 1000 stars subject to the
selection function~$S$.  A minimal sanity check of the fit~$f_0$ is
then whether the (log) likelihood of the original dataset,
$\log\pr(D_0|f_0,\Phi,S)$, is typical in the sense that it lies among the
(log) likelihoods $\{\log \pr(D_n|f_0,\Phi,S)\}$ of these resampled
datasets (see also \citet{BinneyModellingMilkyWay2017}, who carry out
a similar test for the models of the MW's globular cluster system).
The distribution of these log likelihoods is plotted as the
histograms in each panel of Figure~\ref{fig:loglikbincorrectprof}.
By the central limit theorem each distribution is approximately
Gaussian: the log-likelihood is a sum of 1000 terms, all drawn from
the same projected PDF.
For our samples
of 1000 stars, the dispersion of this Gaussian ranges from about 20
for models fit only to line-of-sight components of velocity up to
around 50 for models fit to all three components.
It is clear that the spherical isotropic models pass this test,
whereas axisymmetric models do not: for the latter
$\pr(D_n|f_0,\Phi,S)\ll\pr(D_0|f_0,\Phi,S)$, demonstrating that the
fit $f_0$ to $D_0$ is very special indeed in these cases.

Although we do not plot them here, the log likelihoods of samples
$D_n$ drawn from the original smooth DF~$f$ have approximately the
same variance as those drawn from $f_0$, but have means close to the
log likelihood $\log\pr(D_0|f,\Phi,S)$ of our original sample.
Therefore, for the simulated observational setup we consider here, a
condition that is broadly equivalent to the
$\chi^2\simeq N\pm\sqrt{2N}$ plausibility criterion for integrated
stellar kinematics is that
$\log\pr(D|f,\Phi,S)\simeq\log\pr(D_0|f_0,\Phi,S)\pm\Delta\sqrt{N/1000}$,
where $\Delta\simeq 20$ for models that fit only the line-of-sight
components of velocity, up to $\Delta\simeq 50$ for models fit to all
three components.  Our spherical isotropic models pass this test.
Spherical anisotropic models do show a tendency to overfit, but much
less so than the axisymmetric models.
In the remainder of this paper we focus on fitting only spherical anisotropic
models.

\subsubsection{Degeneracies in the DF}

This overfitting is a symptom of the decision to consider only a
single, very special best-fit DF~$f_0$ for each potential~$\Phi$.
This $f_0$ achieves its remarkably good fit by itself becoming
implausibly noisy.  One way of dealing with this \citep[e.g.,][and
references therein]{ValluriDifficultiesRecoveringMasses2004} is by
penalising nonsmooth DFs by maximising a penalized likelihood
$\log\pr(D|\vf,\Phi)+\lambda P[\vf]$, in which a penalty function $P$
is introduced to quantify some measure of ``smoothness'' of the
DF~$f$.  The weight~$\lambda$ given to the penalty function might be
chosen by using experiments such as those in Section~\ref{sec:overfit}
above to ensure that the favoured model's likelihood is within a
plausible range of values.  This is sensible if one wants to pick out
a single representative DF for the assumed potential and believes that
the smoothness conditions imposed by the penalty function are reasonable.

This is less satisfactory if one wants to compare different potentials
or does not want to exclude DFs with sharp features.  An alternative
solution \citep[e.g.,][]{MagorrianConstrainingblackhole2006} is to
note that, although the ``best'' $f_0$ fits the data implausibly well,
there will be vastly more ``nearby'' DFs that produce fits that are
formally slightly worse, but with likelihood values that are in fact
more plausible.  The natural remedy is somehow to take account of
these neighbouring DFs.  In the present case in which we approximate
the DF as a discrete sum of orbit blocks this could be achieved by
marginalising the weights~$f_k$, but carrying out such a calculation
is beyond the scope of this paper.

\begin{figure}

  \includegraphics[width=0.9\hsize]{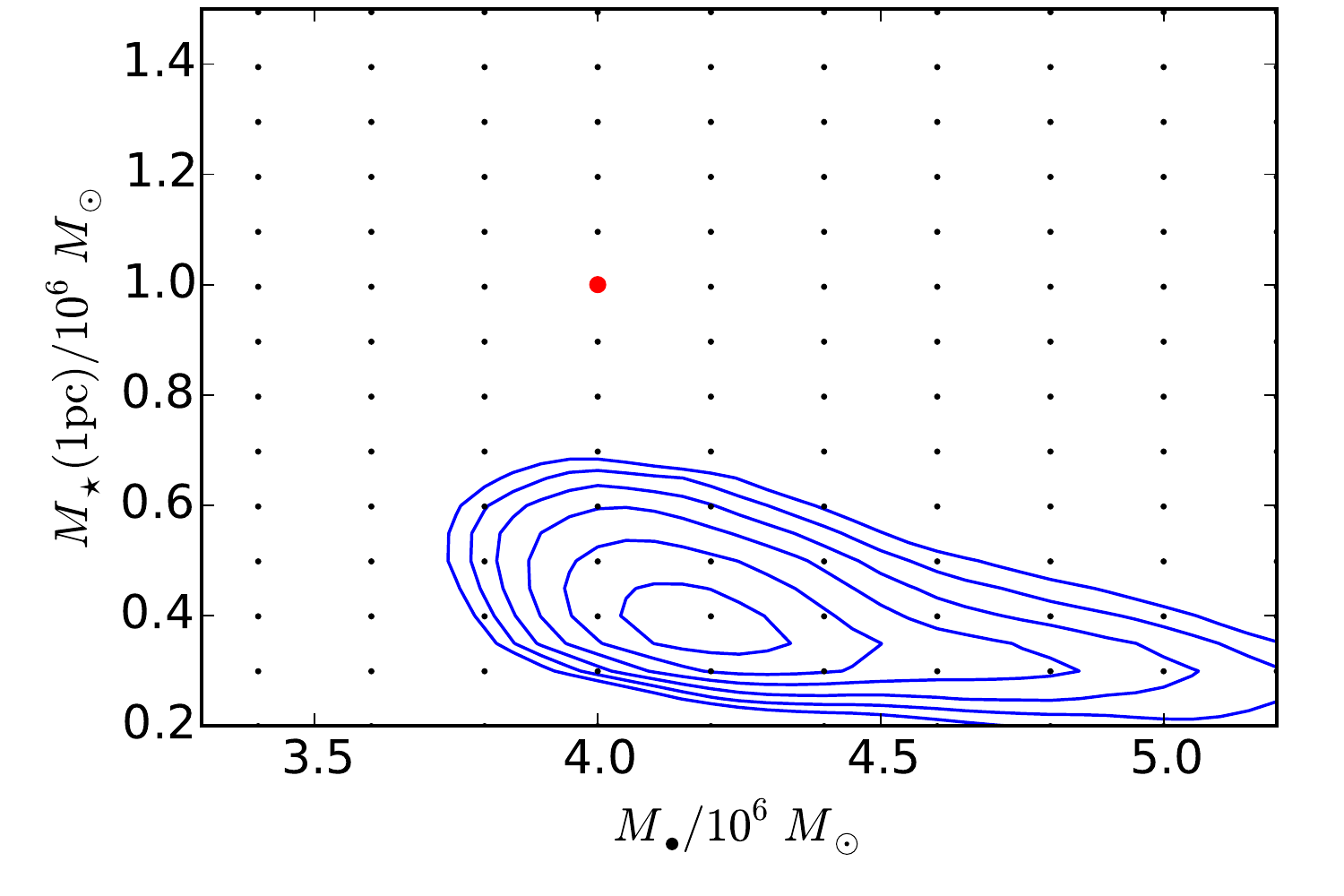}

  \caption{Likelihoods of spherical anisotropic models fit to
    data comprising all three components of velocity from our simulated cluster assuming a mass density~\eqref{eq:rhoprof}
    with the correct slope $\gamma=1$ but with a scale radius $r_{\rm s}=100\,\pc$
    that is a factor of 10 too large.  Fitting only the proper motions
    produces similar results. }
  \label{fig:discretebinbigr0}
\end{figure}

\begin{figure}
  \includegraphics[width=0.9\hsize]{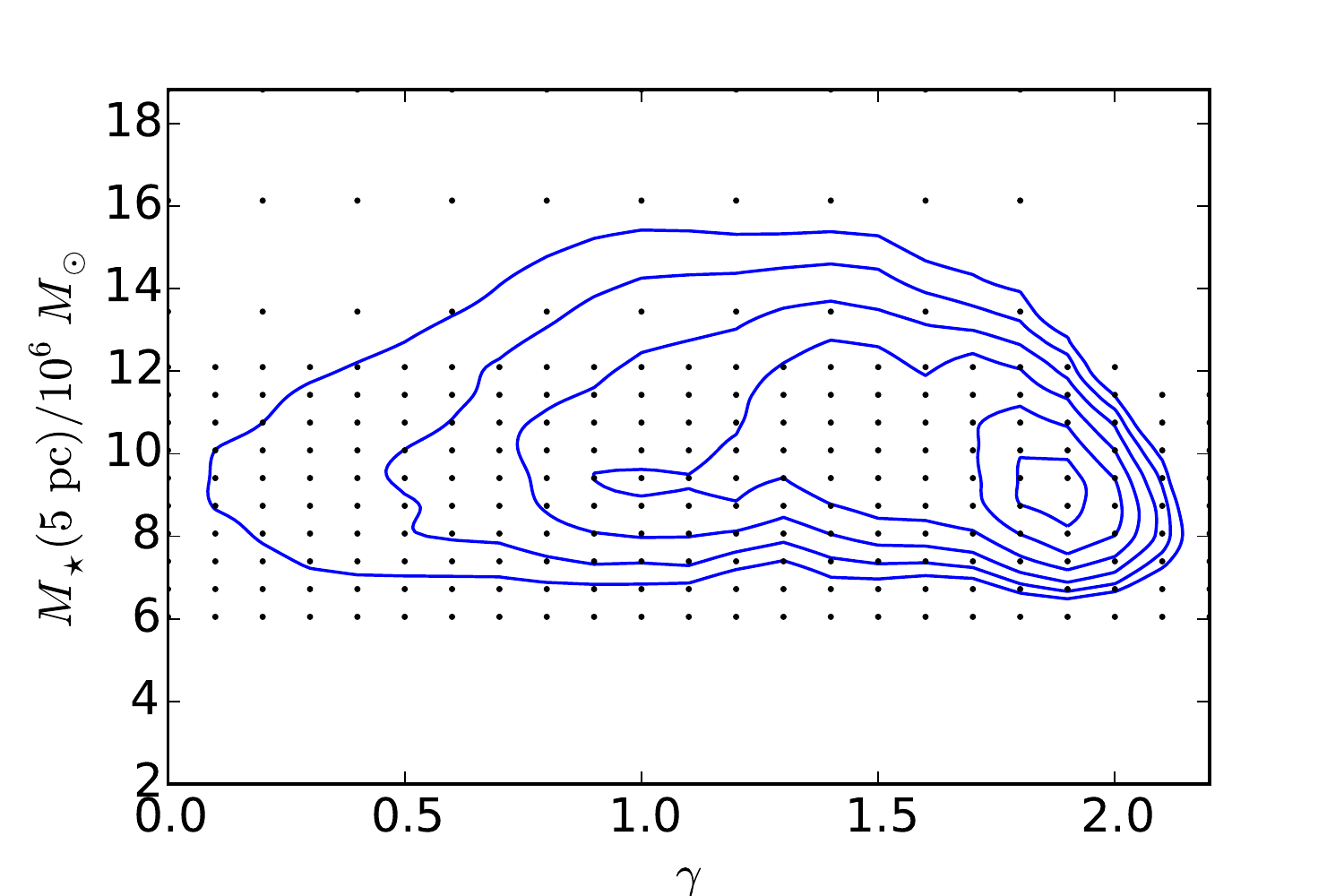}
    \caption{Dependence of log-likelihood of on $\gamma$ and
    $M_\star(5\pc)$ for models fit to all three components of velocity
    in our simulated dataset.  The models use the correct value of $M_\bullet$, but
    assume a scale radius $r_{\rm s}=100\,\pc$ that is a factor 10 too large.}
  \label{fig:gammadep}
\end{figure}

\subsubsection{How well is the radial mass profile constrained?}

\label{sec:testspherical2}
We have just seen that our modelling machinery can successfully
constrain the mass normalisation~$M_\star$ of the stellar cluster,
provided the functional form and scale radius of the mass density
profile were somehow known in advance.  To investigate how well the
models behave in more realistic situations we begin by fitting models that
adopt the correct $\gamma=1$ for the mass-density slope, but take the
scale radius of the mass distribution to be $r_{\rm s}=100\,\pc$ instead of the
correct 10~pc.

The resulting likelihood contours are plotted in
Figure~\ref{fig:discretebinbigr0}.  They show that the models predict
the correct black hole mass, but underestimate the extended mass
enclosed within 1 pc by a factor of at least~2.
Our motivation for choosing this $1\,\pc$ reference radius was that it
is approximately equal to the maximum projected radius of the discrete
kinematics ($R=19''\simeq0.74\,\pc$), but the Figure demonstrates that
this is actually a very poor choice.
It turns out that it is much better to choose a reference radius that
accounts for the three-dimensional distribution of the sampled stars.
The selection function for our discrete stellar sample defines a very
long, narrow cylinder of radius 19'' whose axis includes the observer
and the BH at the centre of the cluster.  The RMS three-dimensional radius
of stars within this cylinder is approximately $5\,\pc$: in three
dimensions the stellar sample is strongly elongated along the line of
sight.  The $r_{\rm s}=10\,\pc$ simulated cluster from which we drew
the kinematics has an extended mass of $13.4\times10^6\,M_\odot$
within this $5\,\pc$.

Figure~\ref{fig:gammadep} shows the effects on the likelihood of
varying the density slope $\gamma$ in a model that assumes the
correct~$M_\bullet$, but takes $r_{\rm s}=100\,\pc$, a factor $10$ too
large.
Although the overall peak occurs for $\gamma\sim1.8$ -- presumably
because that squeezes more mass close to the BH -- the best-fitting
models all have $M_\star(5\pc)\sim10\times10^6\,M_\odot$.  This is
systematically lower than the true
$M_\star(5\,\pc)=13.4\times10^6\,M_\odot$, but significantly less
biased than the models' estimate of $M_\star(1\,\pc)$ and, remarkably,
almost independent of~$\gamma$.  
The log likelihood of this best-fitting $\gamma=1.8$ model is only 0.5
less than that of the model that uses the correct mass distribution.
This is perhaps not surprising given the limited radial extent of the
discrete data we use.
Changing the reference radius much from $5\,\pc$ increases the bias on
the mass estimate and breaks the independence from~$\gamma$.

\begin{figure*}
  \null\hskip10pt
  \hbox to 0.3\hsize{\hfill all 3 components of $\vv$\hfill}
  \hbox to 0.3\hsize{\hfill proper motions only\hfill}
  \hbox to 0.3\hsize{\hfill line-of-sight only\hfill}

  % \vspace{-0.3cm}
  \null\leavevmode\raise35pt\hbox{\rotatebox{90}{\hbox{\small 40 disc stars}}}
  \includegraphics[width=0.3\hsize]{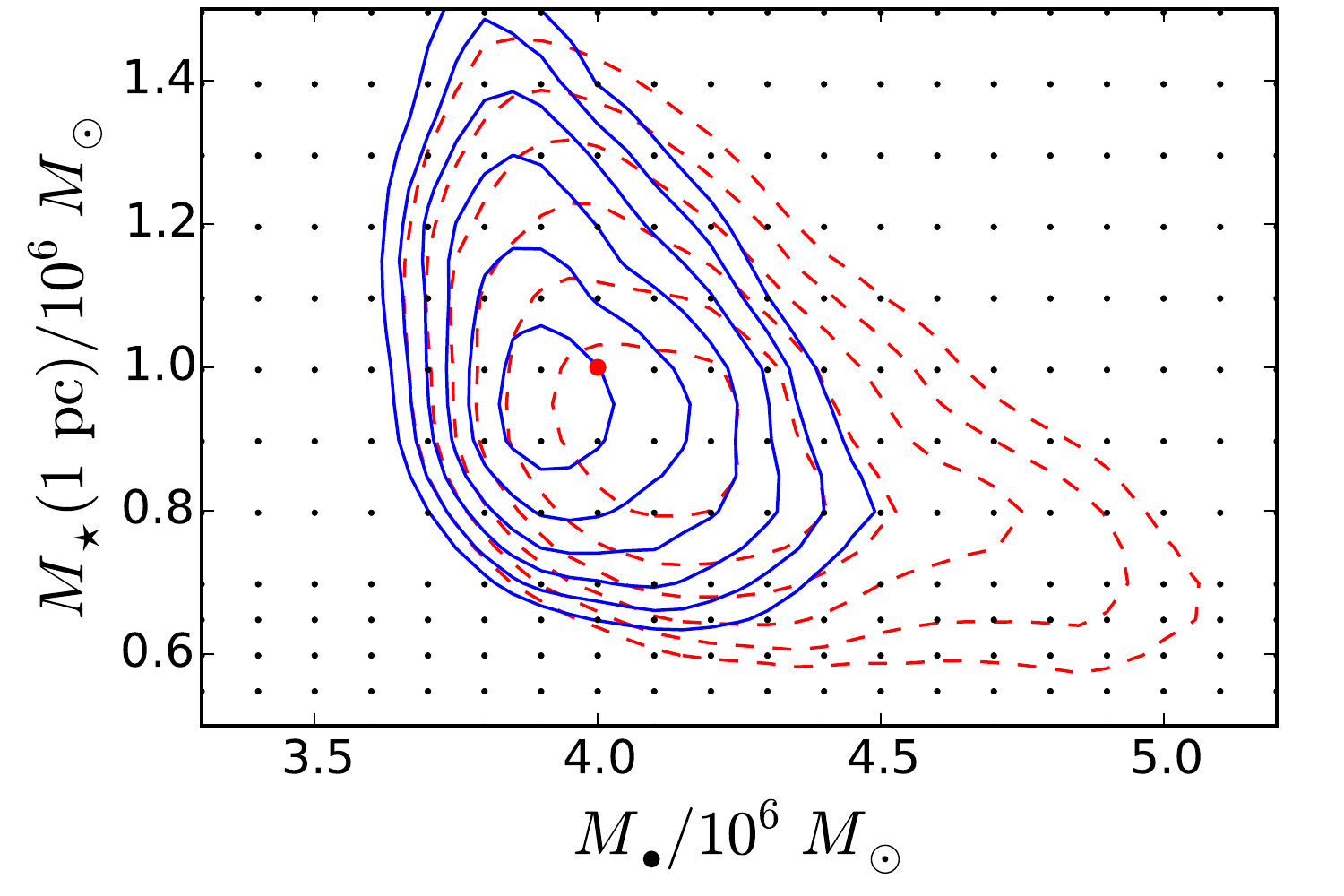}
  \includegraphics[width=0.3\hsize]{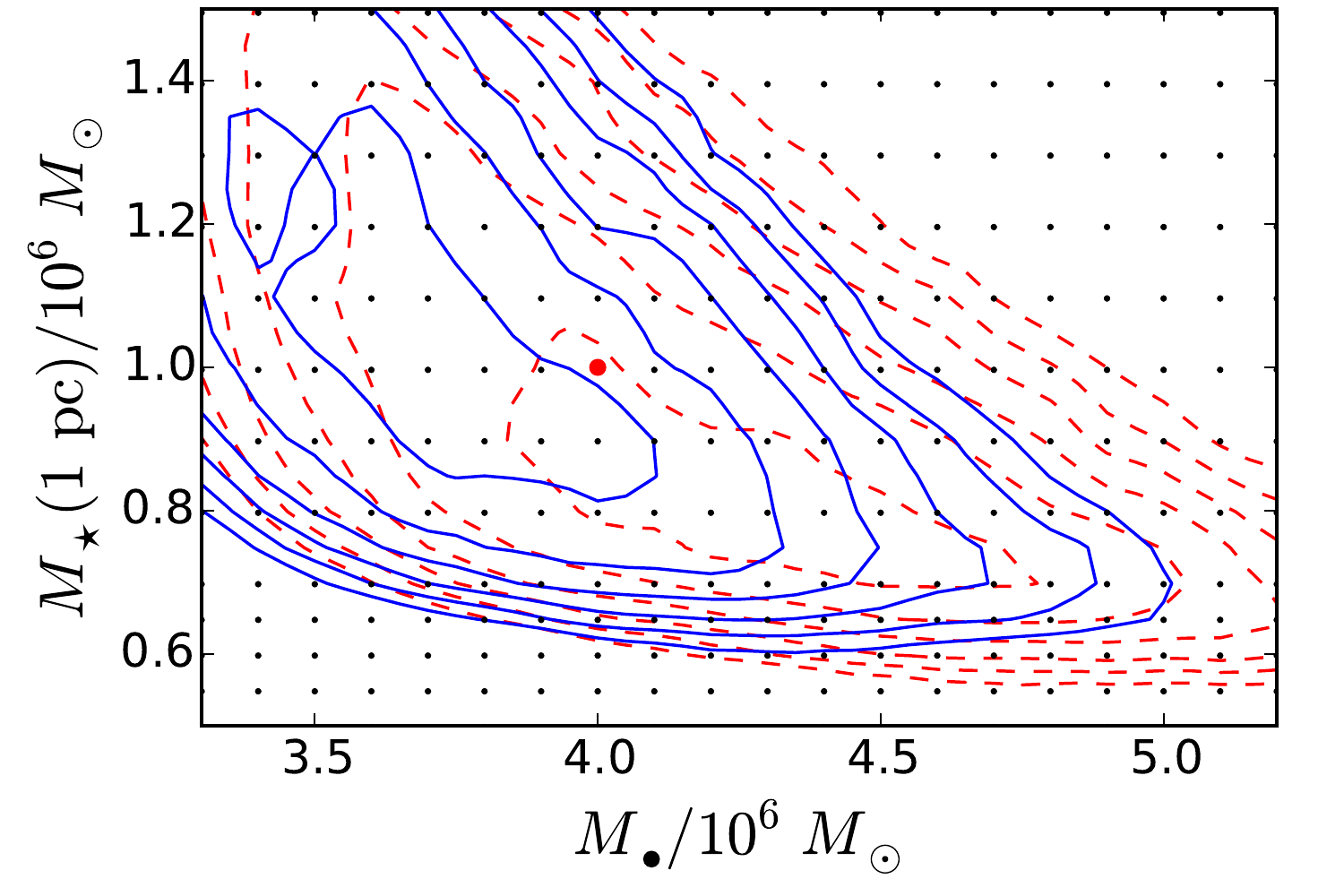}
  \includegraphics[width=0.3\hsize]{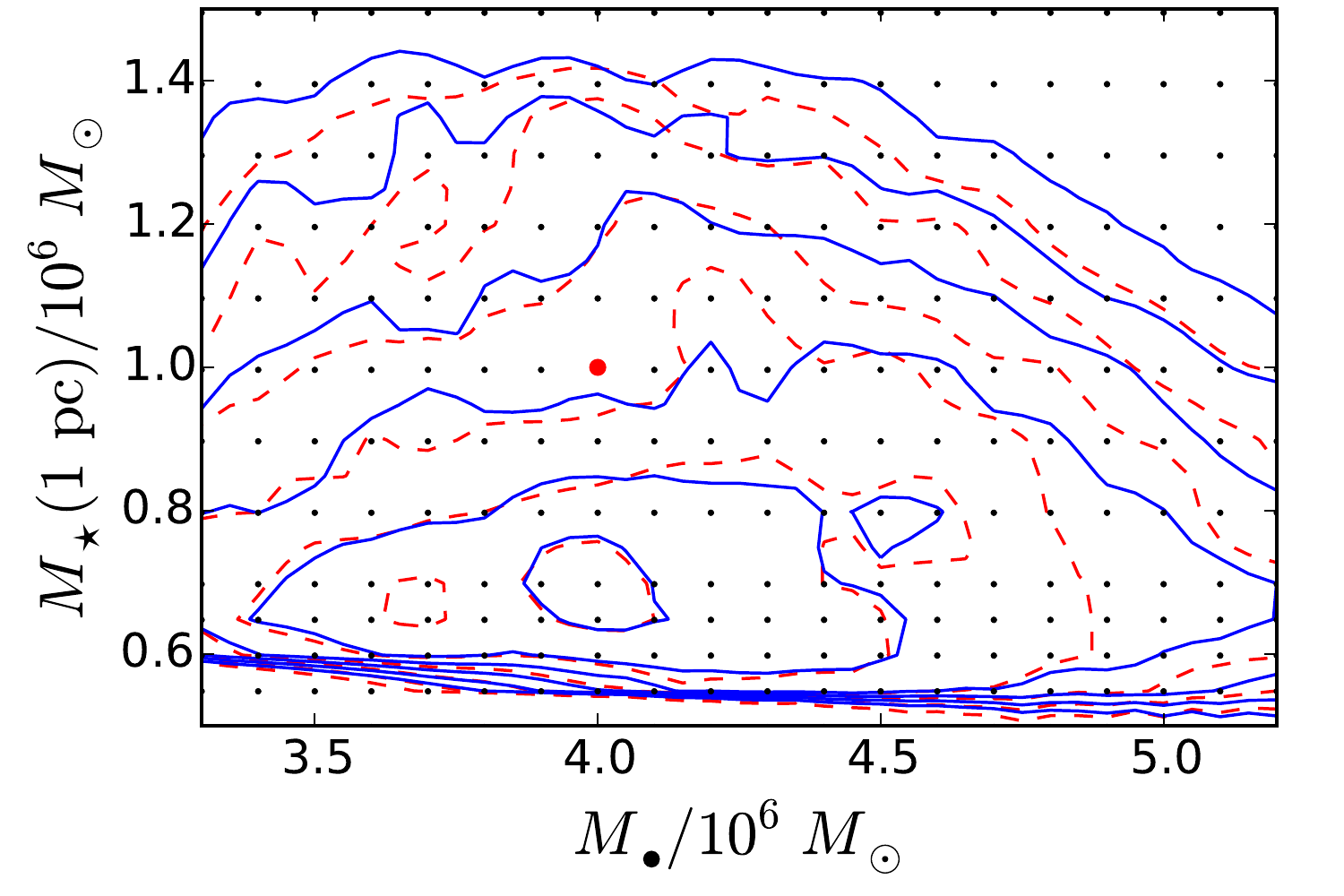}

  \vspace{-0.3cm}
  \null\leavevmode\raise35pt\hbox{\rotatebox{90}{\hbox{\small 100 disc stars}}}
  \includegraphics[width=0.3\hsize]{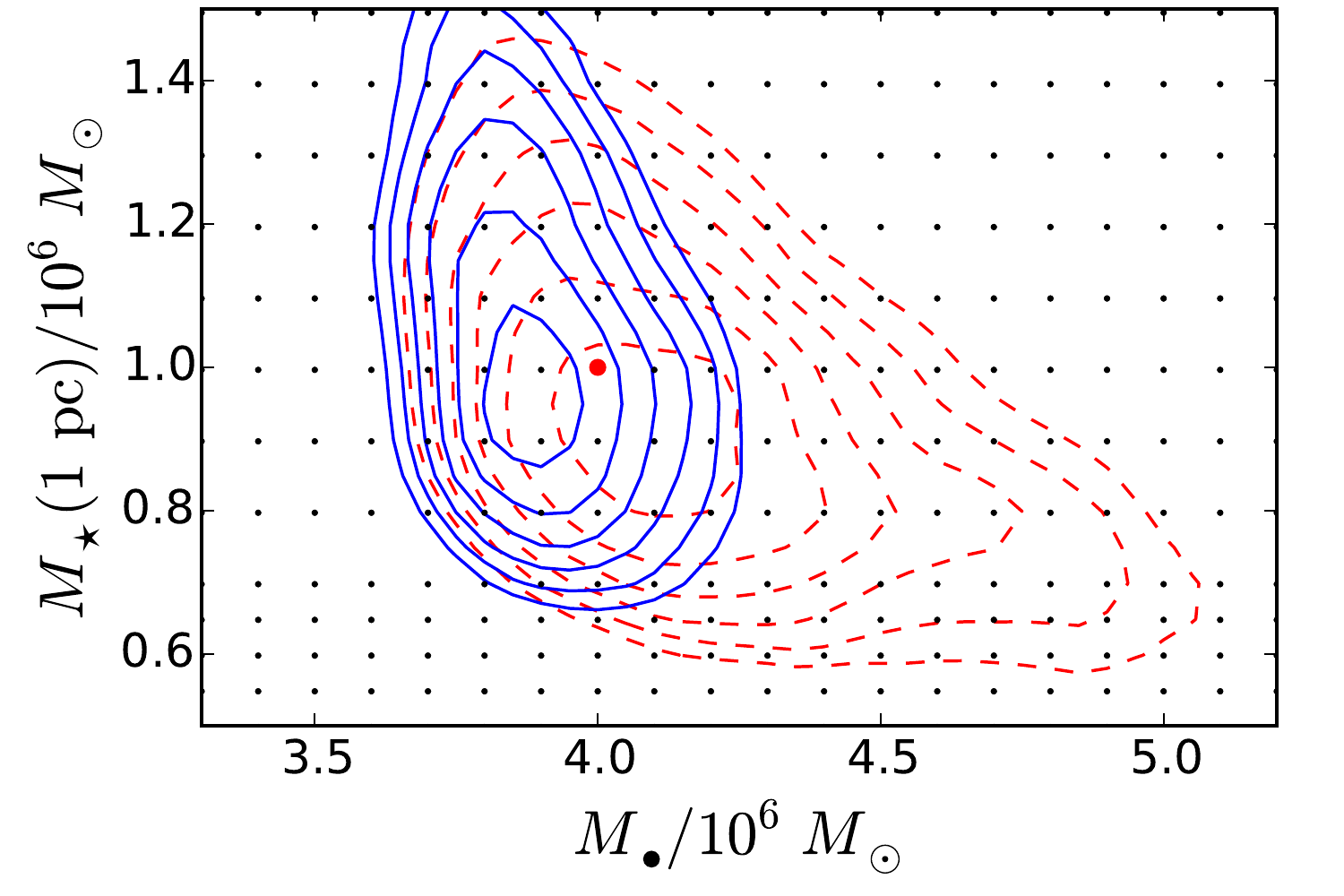}
  \includegraphics[width=0.3\hsize]{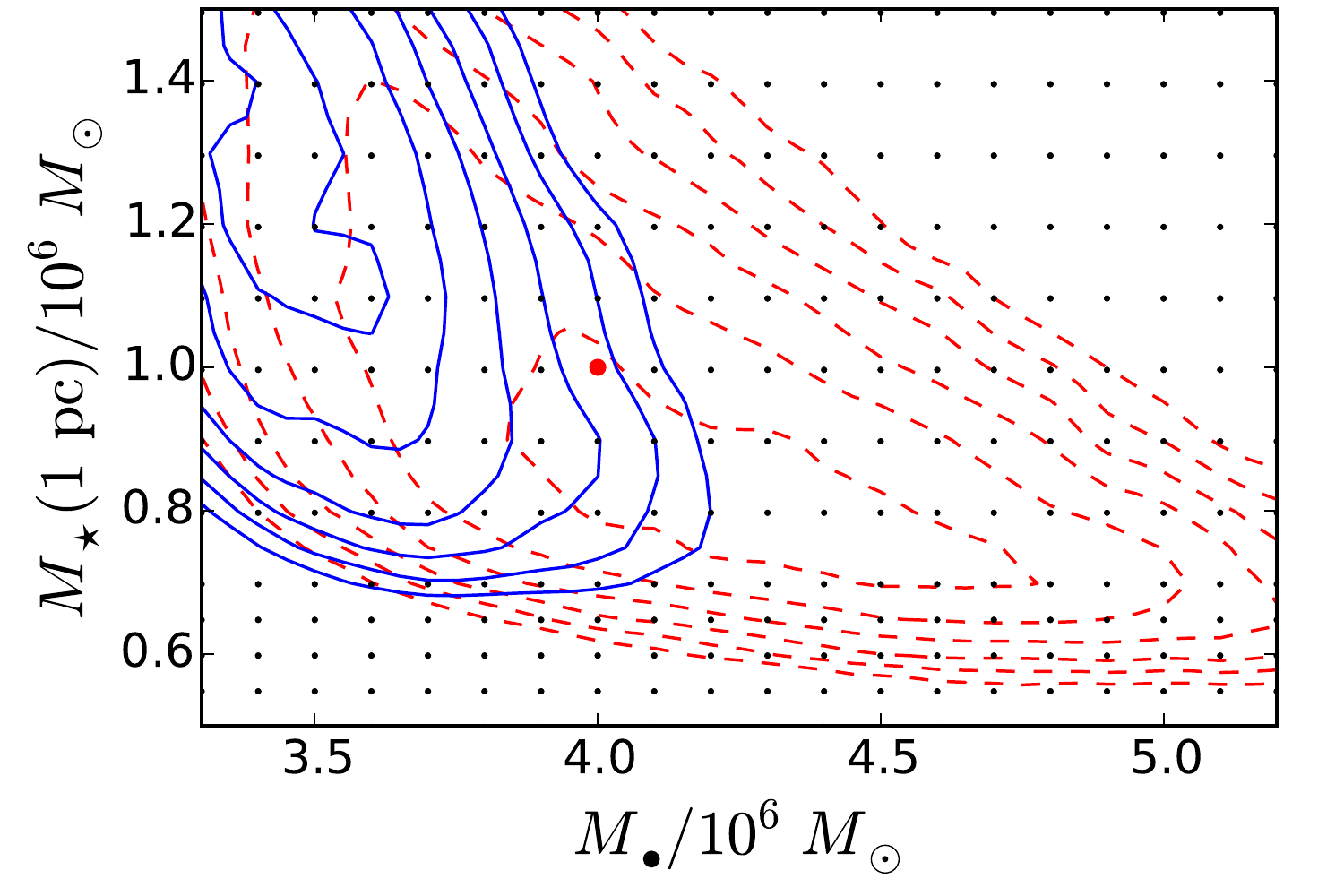}
  \includegraphics[width=0.3\hsize]{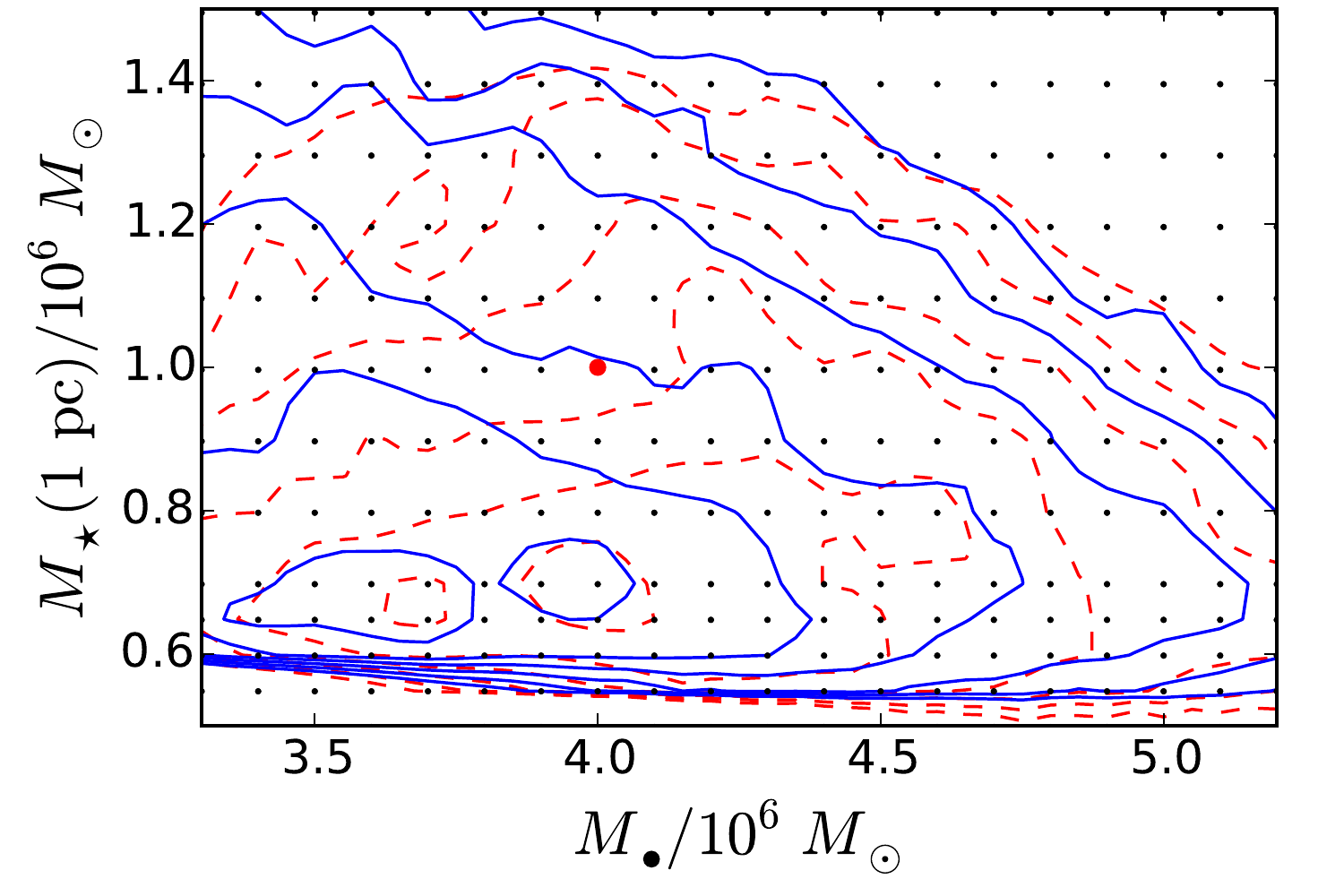}

  \includegraphics[width=0.3\hsize]{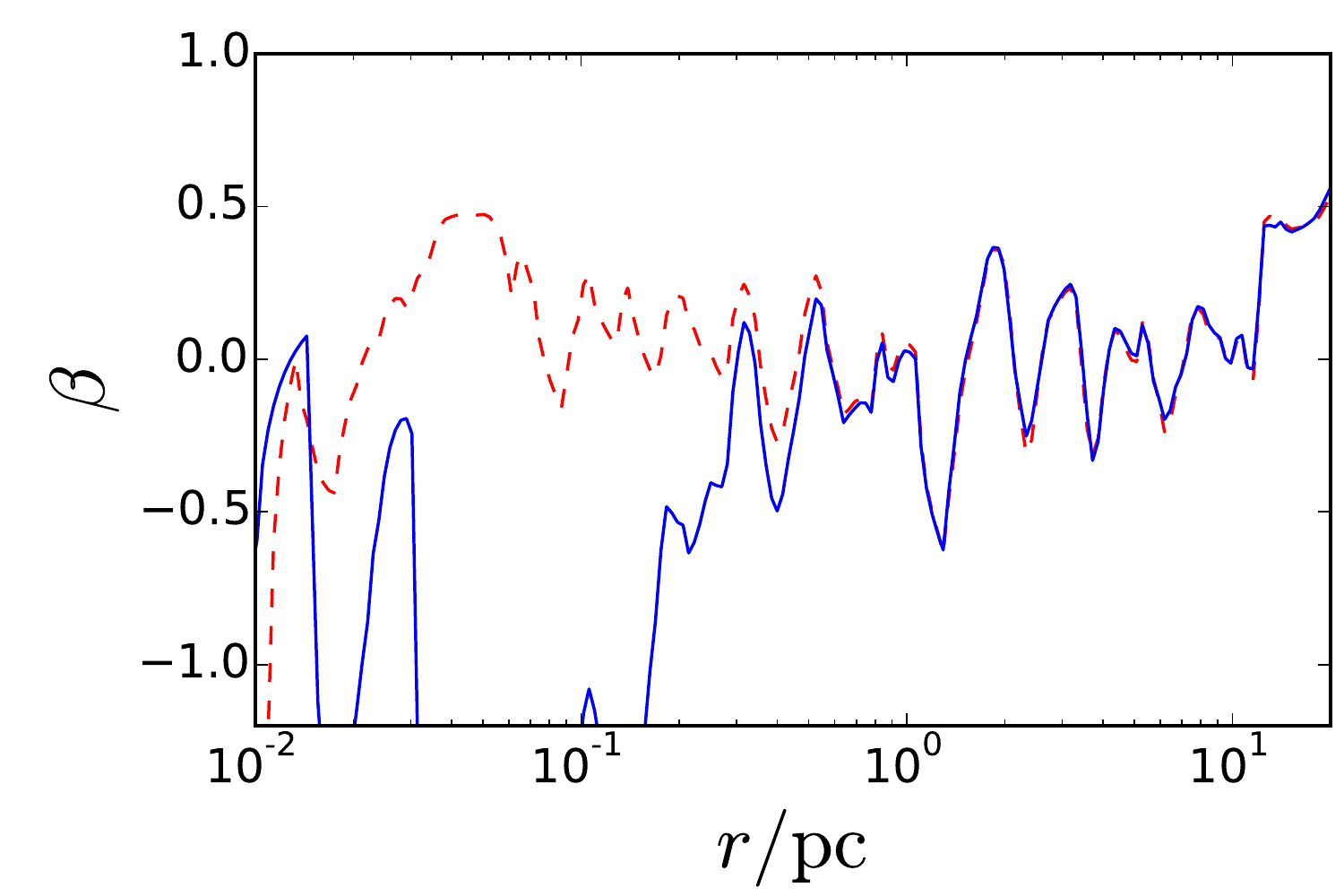}
  \includegraphics[width=0.3\hsize]{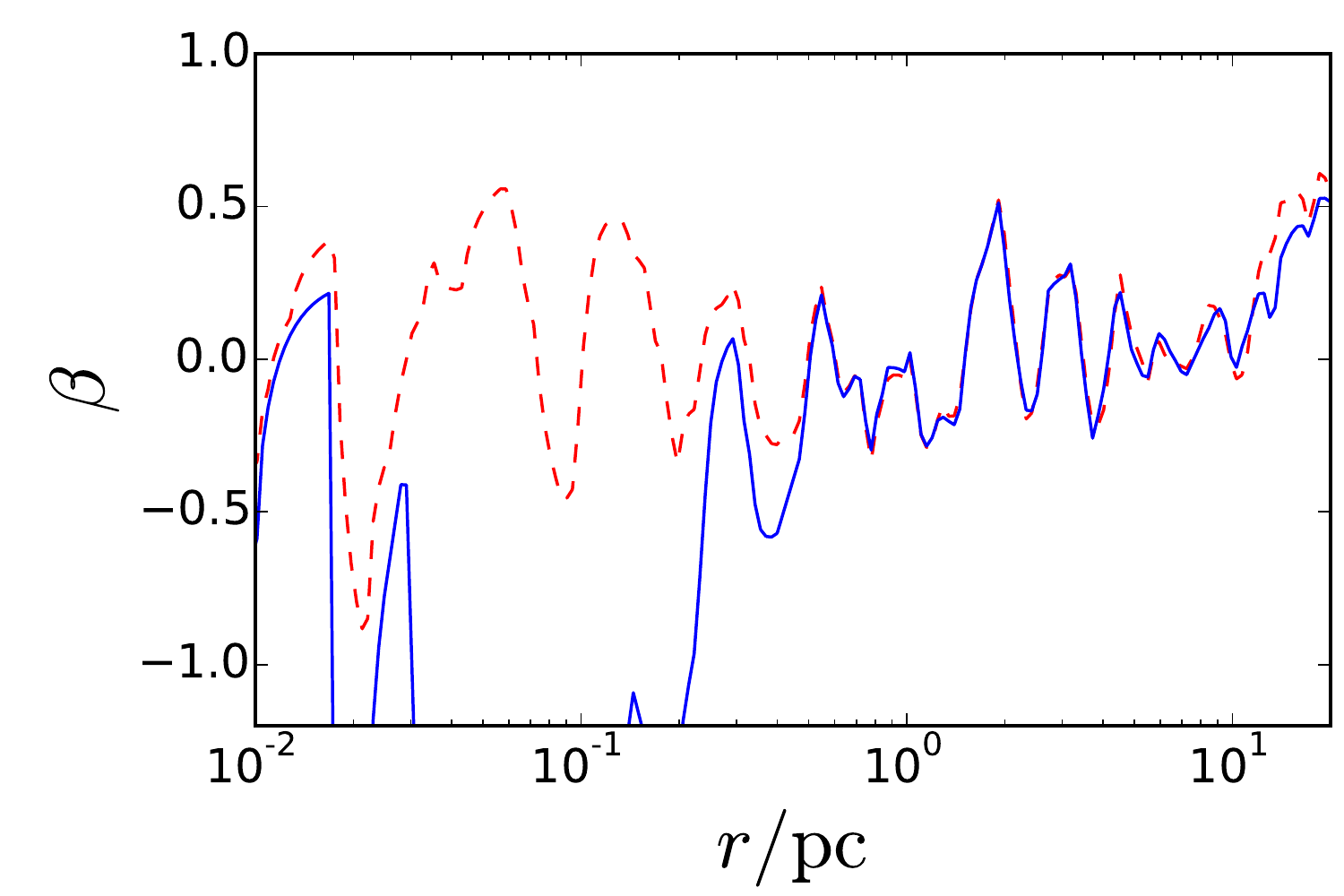}
  \includegraphics[width=0.3\hsize]{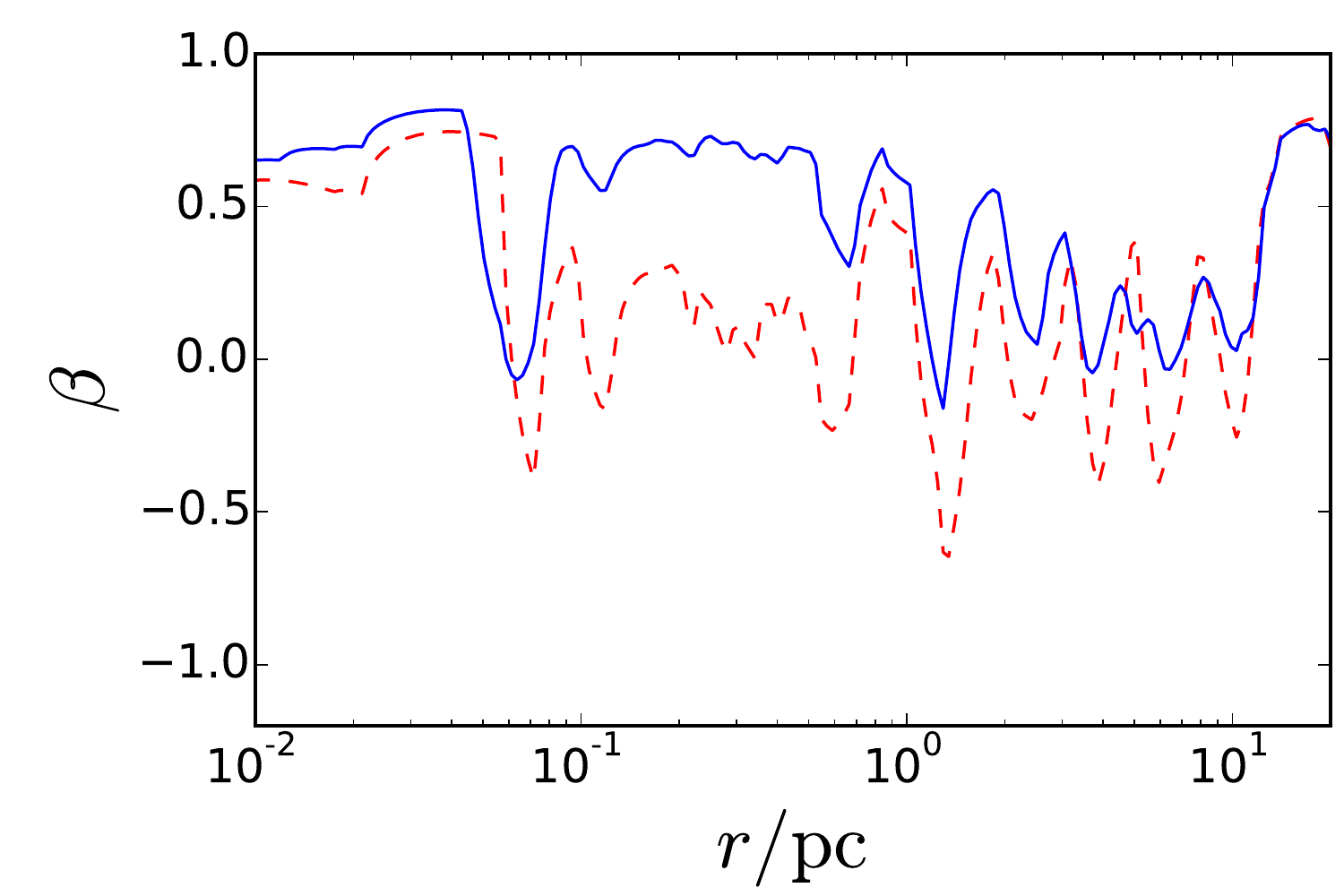}

  \caption{As for Figure~\ref{fig:discretebincorrectprof}, but
    illustrating the effects of a nonspherical contaminant population
    on the masses returned by spherical, anisotropic models.  The top
    row shows the result of adding 40 stars from the tilted ring
    distribution described in the text, the middle row the result of
    adding 100.  For reference, the dashed red contours plot the
    likelihoods (repeated from the middle row of
    Figure~\ref{fig:discretebincorrectprof}) of models fit to the
    uncontaminated spherical population.  The bottom panel compares
    the anisotropy parameters of models that assume that correct
    potential, but are fit to the uncontaminated sample (dashed red
    curves) versus those contaminated by 100 disc stars (solid blue
    curves).  }
  \label{fig:testpotring}
\end{figure*}

\subsection{Nonspherical clusters}

All of our tests so far have involved fitting spherical and
axisymmetric models to simulated observations of spherical clusters.
Having shown that our axisymmetric models tend to overfit the data, we
have confined our attention to spherical models.  Now we investigate
how well these spherical models perform when applied to simulated
observations of nonspherical clusters.

\subsubsection{Effects of a tilted ring of young stars}
\label{sec:ring}
Although the old population of our Galaxy's stellar cluster is roughly
spherical in its inner parsec or so, there is a substantial population
of young stars distributed in a ring-like structure in the innermost
few tenths of a parsec.  Ideally one would like to remove any such
young stars from the sample, but that relies on spectroscopic
identification, which is not always available.  To investigate the
effects of such nonspherical contaminants, we adopt a very simple
model of the Milky Way's young star population based on the fits in
\cite{PaumardTwoYoungStar2006} and \cite{YeldaPropertiesRemnantClockwise2014}.
Our model is a razor-thin ring of stars moving on circular orbits with
radii $1''<r<13''$.  Their face-on number density distribution goes as
$r^{-2}$.  Referred to our $(x,y,z)$ coordinate system
(\S\ref{sec:coords}) the disc's normal vector is taken to be
$\vn=(-\cos I,\sin I\cos\Omega,-\sin I\sin\Omega)$ with
$(I,\Omega)=(130^\circ,96^\circ)$.
We add stars from this tilted ring population to the sample of 1000
stars drawn from the simulated spherical cluster introduced
in~\S\ref{sec:testspherical}.  There is no change to the potential of
the simulated cluster: the ring stars are treated as massless test particles.

As in \S\ref{sec:testspherical1} we test our DF-superposition
modelling procedure by giving it the correct form for the radial mass
profile, leaving only the mass normalisation parameters $M_\bullet$
and $M_\star(1\pc)$ as parameters to be determined.
Figure~\ref{fig:testpotring} plots the likelihood as functions of
these two parameters for spherical anisotropic models fit to all three
components of velocity of the population of 1000 ``old'' stars used
for Figure~\ref{fig:discretebincorrectprof}, to which we have added a
further 40 (4\% contamination fraction, top row) or 100 (very high
10\% contamination fraction, bottom row) stars from the ``young''
tilted ring population.  We note first that the models that fit only
to line-of-sight velocities are largely unaffected by this
contamination: they are just as wrong afterwards as they were before.
The models that use proper motion information are more interesting:
increasing the fraction of disc contaminants biases the inferred orbit
distribution towards tangential orbits ($\beta<0$) for
$r\lesssim13''$, which in turn depresses the estimate of~$M_\bullet$.
This depression is less pronounced when all three components of
velocity are available.

We have investigated whether our simple model for outliers
(\S\ref{sec:outliers}) can correct this bias, but find that setting
$f_{\rm c}>0$ just depresses the estimate of~$M_\bullet$ further.
This failure is not too suprising though: the outlier model is
intended to account for stars whose velocities are simply mismeasured,
not significant populations of stars that break the underlying
symmetry assumption.

\begin{figure}
  \includegraphics[width=\hsize]{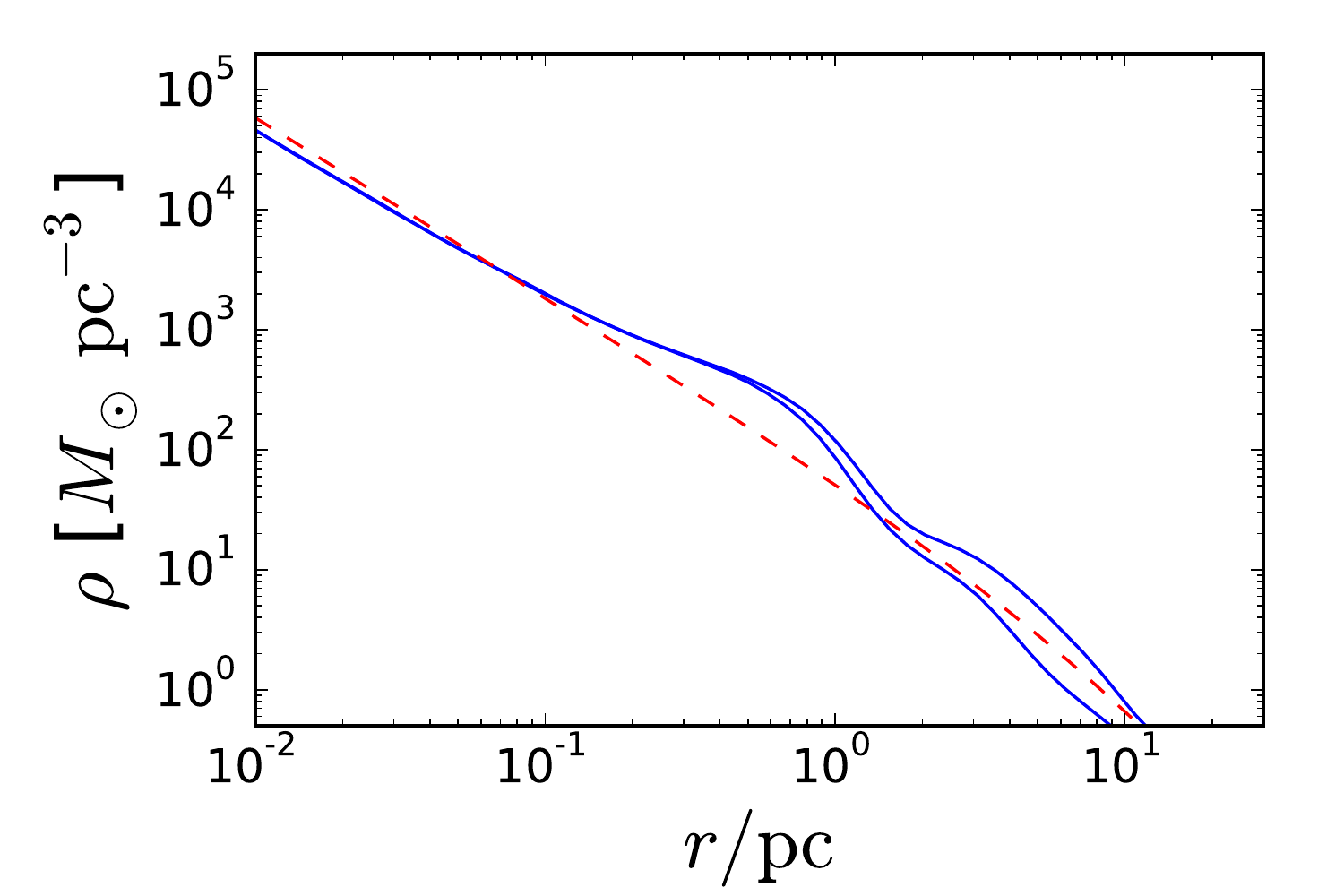}

  \caption{Density profiles of the flattened simulated cluster
    introduced in \S\ref{sec:flat}.  The blue curves plot the
    cluster's major-
    and minor-axis mass density profiles.
    For comparison, the
    dashed curve shows the form~\eqref{eq:rhoaxiprof} of the mass density profile assumed by our
    spherical anisotropic DF-superposition models.}
  \label{fig:axiaxes}
\end{figure}
\subsubsection{Fitting spherical models to a simulated flattened cluster}
\label{sec:flat}

Although the Milky Way's central star cluster is approximately round
in projection for projected radii $R<1\,\pc$, it does become flattened
further out.  Most of the stars that lie within the
$R<19''\simeq0.74\,\pc$ cylinder we use to sample the discrete
kinematics have three-dimensional radii $r\gg1\,\pc$, placing many of
them in the region where the cluster starts to become strongly
flattened.  To test to what extent this is likely to bias the masses
returned by our models we consider kinematics from a simulated
flattened cluster.

Our starting point for constructing this simulated cluster is the
multi-Gaussian fit to the Milky Way's NIR projected light distribution
between 10 and 2500 arcsec made by \citet[][their
Table~4]{SchodelSurfacebrightnessprofile2014}.  At smaller radii their
multi-Gaussian parametrization introduces a central core of almost
constant density.  To produce a more realistic, cusped density profile
we add to this a further component having surface brightness
\begin{equation}
I(R)=I_0\left(\frac R{R_0}\right)^{-0.5}\left(1+\frac R{R_0}\right)^{-3.5},
\end{equation}
with $R_0=20''$ and $I_0=10^7\,L_{\odot,4.5\,\mu{\rm m}}/\pc^2$.
We deproject the resulting surface brightness distribution under the
assumption that the cluster is axisymmetric and viewed edge on.  We
assume that mass follows light and normalise the resulting $\rho(R,z)$
mass density so that the stellar mass enclosed within a spherical
radius of $1\,\pc$ is $10^6\,M_\odot$.  
Figure~\ref{fig:axiaxes} plots the major- and minor-axis mass density
profiles of the deprojected cluster.  The $\nu(R,z)$ number-density
profile of our simulated cluster is directly proportional to this $\rho(R,z)$.

\begin{figure*}
  \null\hskip10pt
  \hbox to 0.32\hsize{\hfill all 3 components of $\vv$\hfill}
  \hbox to 0.32\hsize{\hfill proper motions only\hfill}
  \hbox to 0.32\hsize{\hfill line-of-sight only\hfill}

  \includegraphics[width=0.32\hsize]{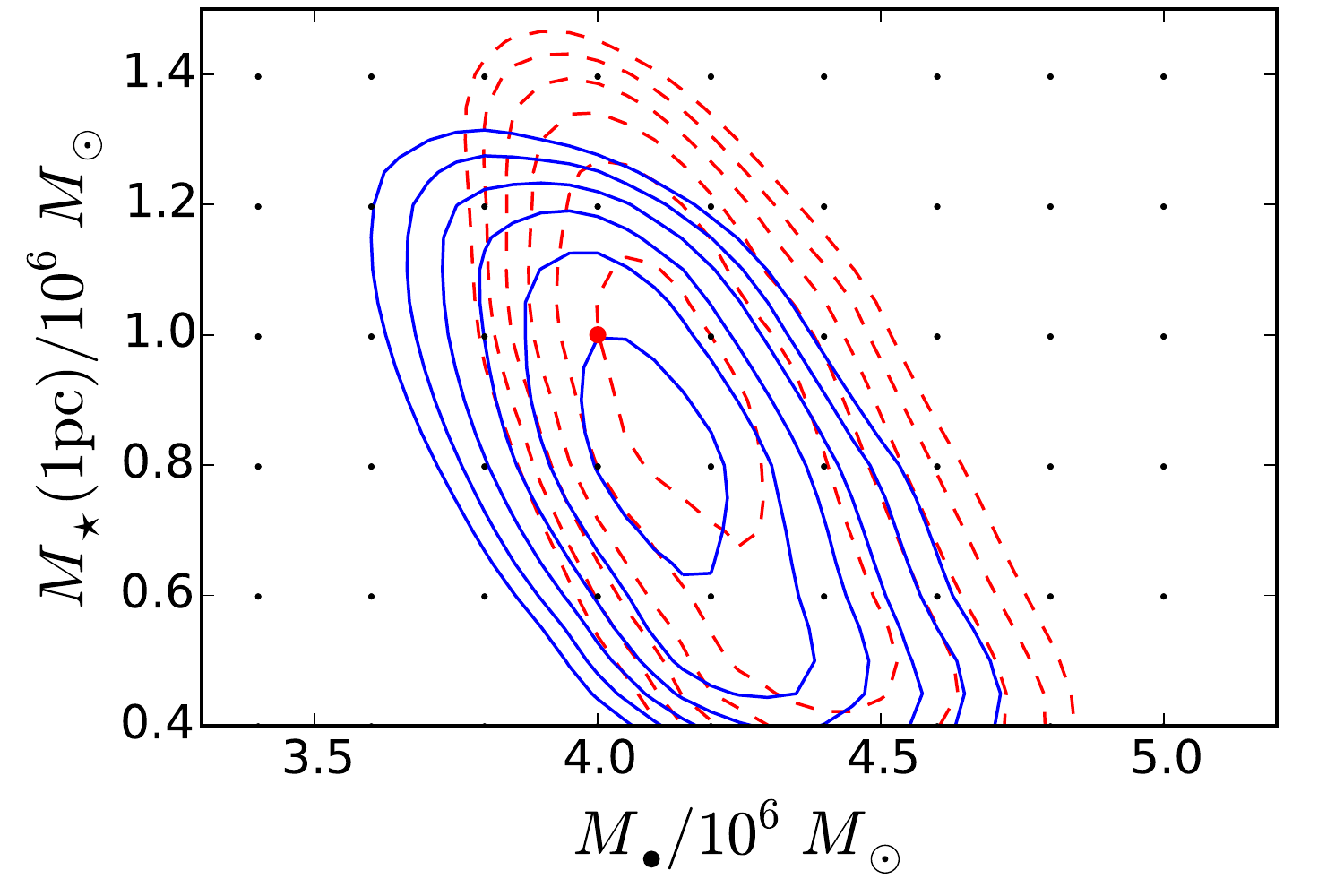}
  \includegraphics[width=0.32\hsize]{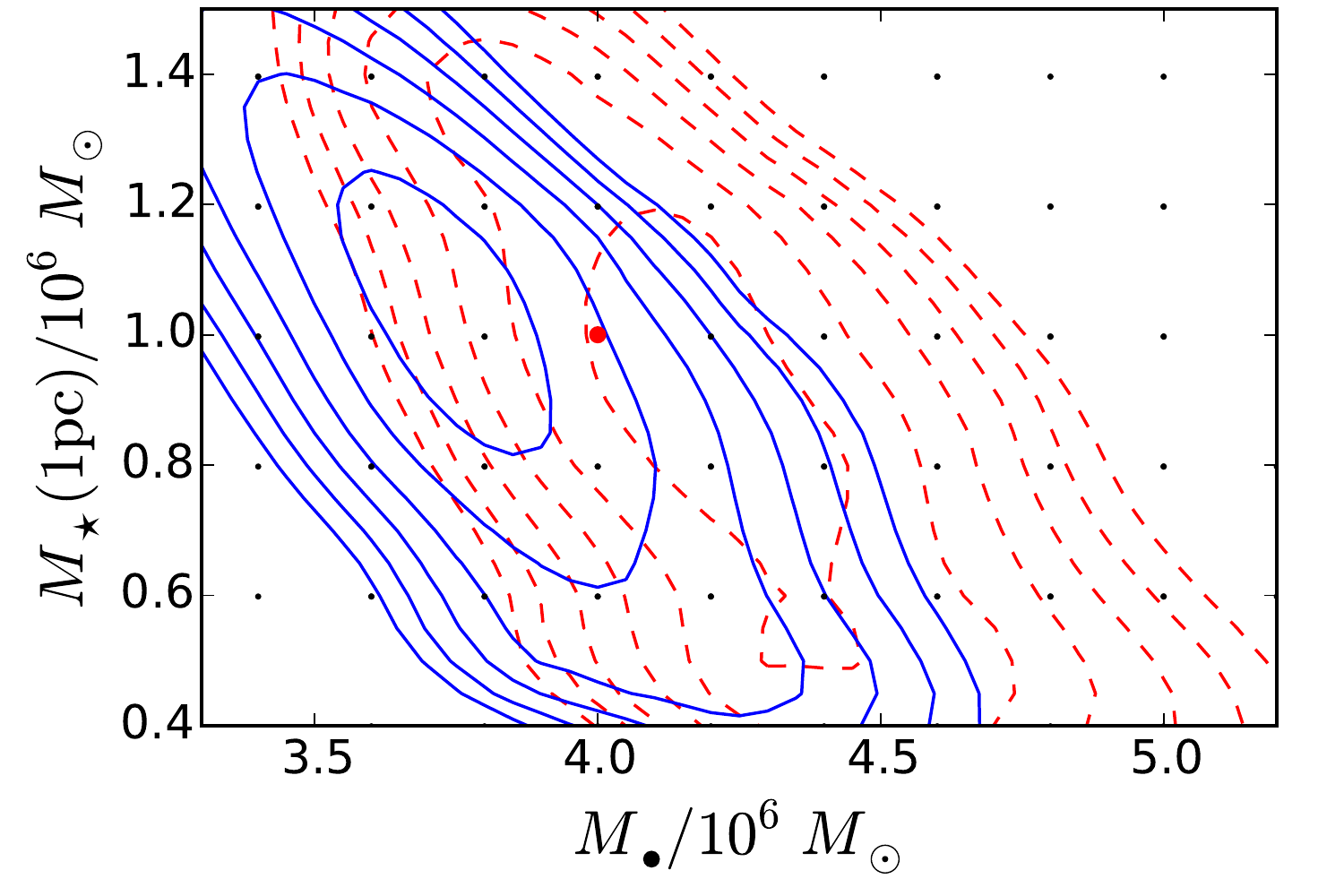}
  \includegraphics[width=0.32\hsize]{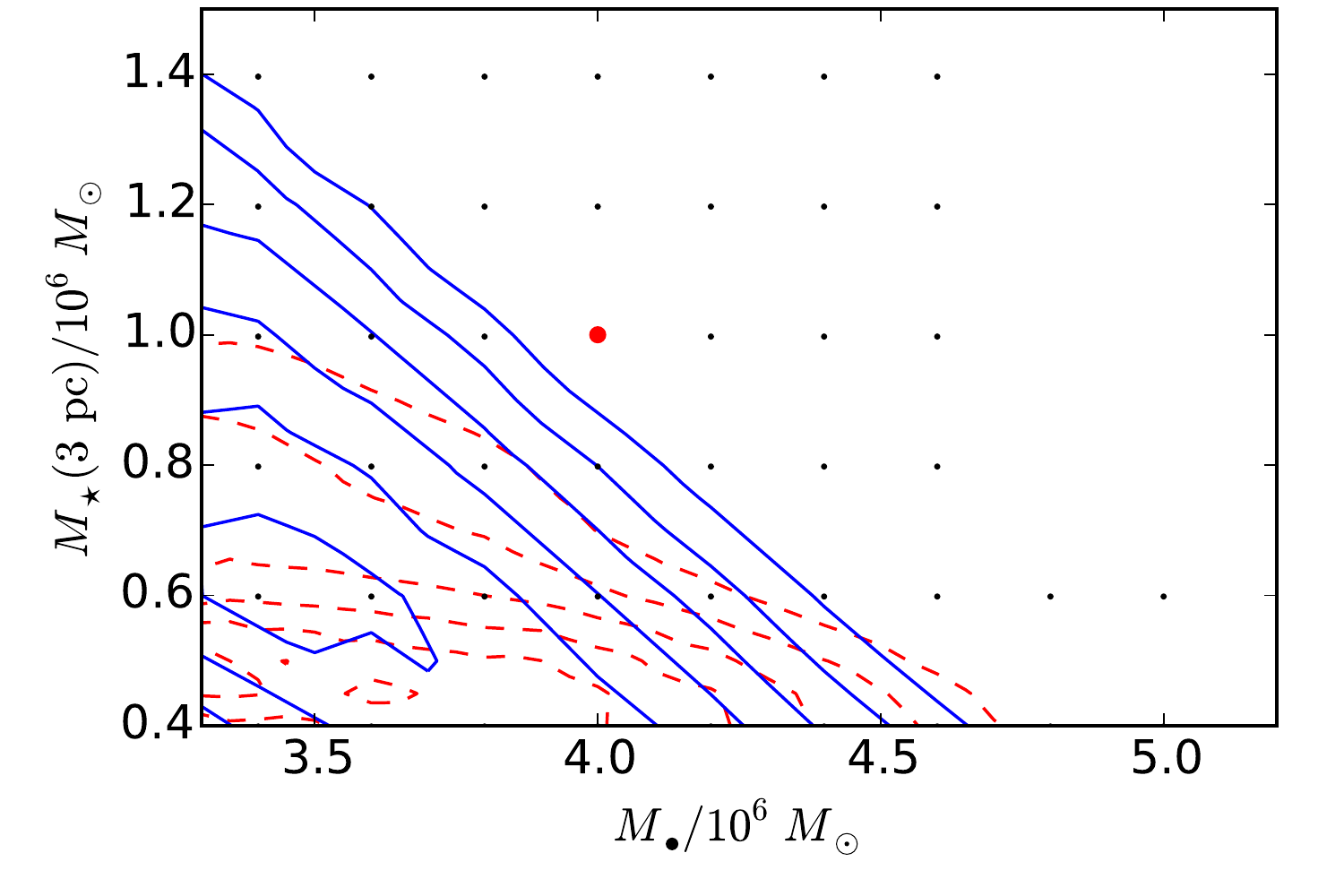}

  \includegraphics[width=0.32\hsize]{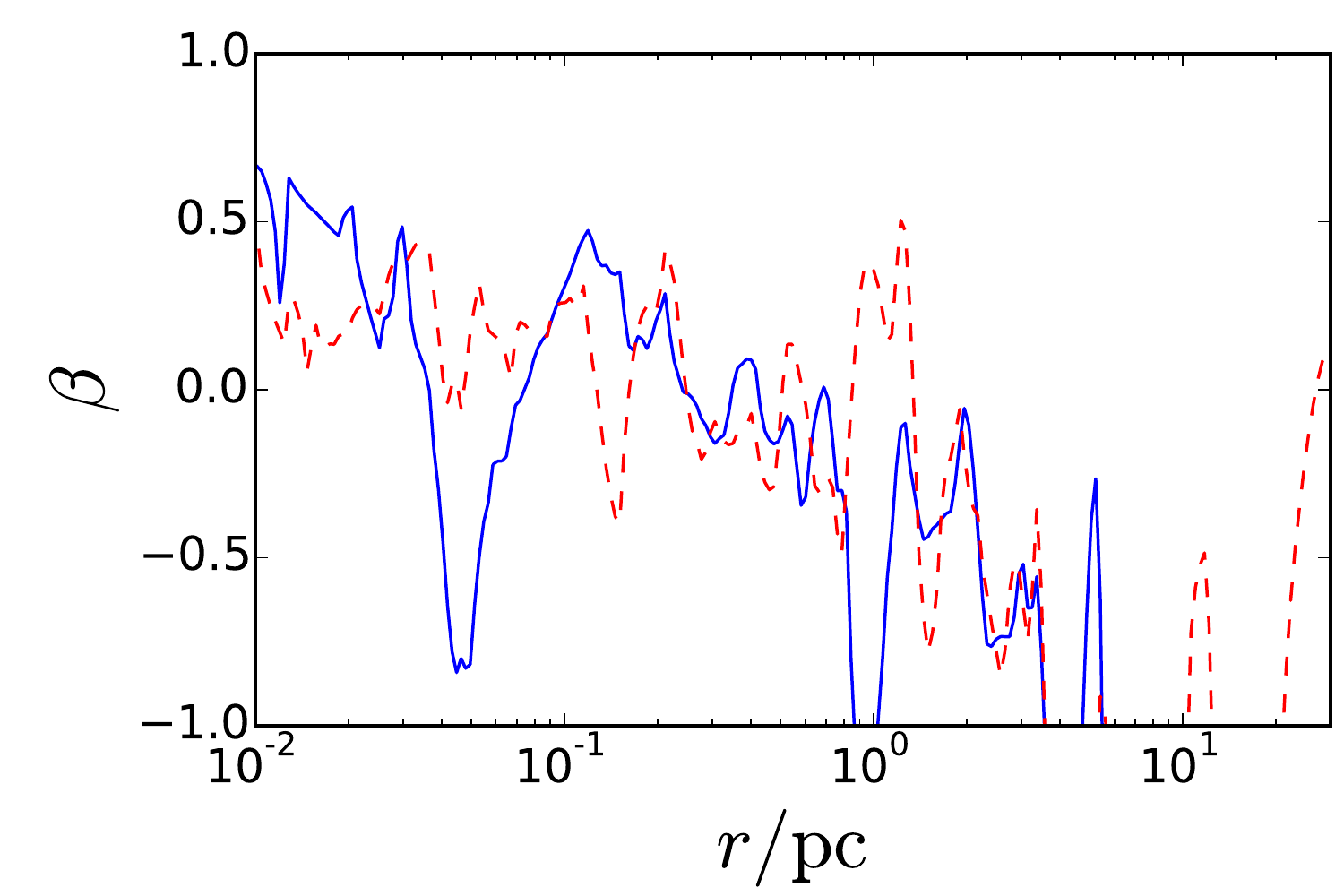}
  \includegraphics[width=0.32\hsize]{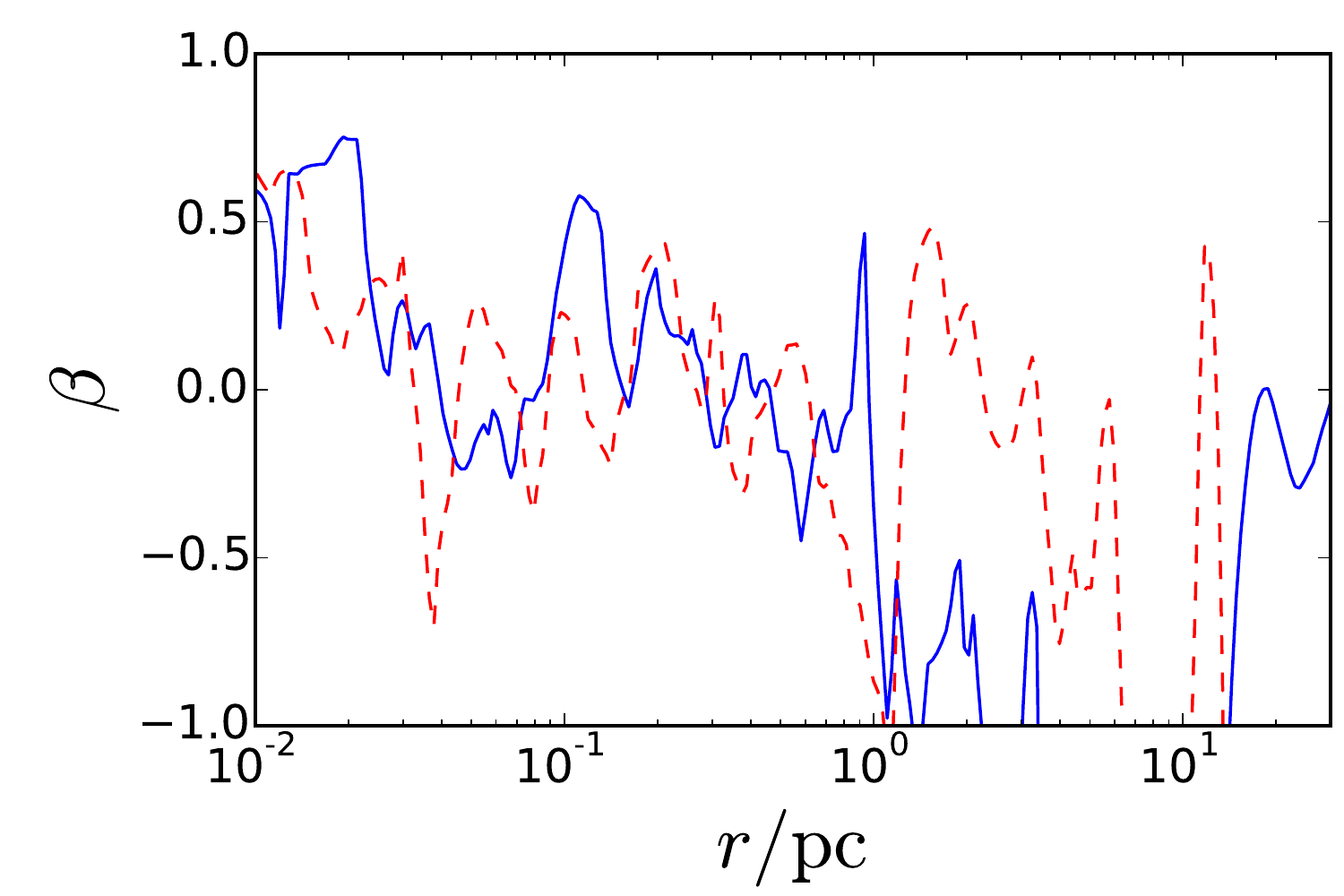}
  \includegraphics[width=0.32\hsize]{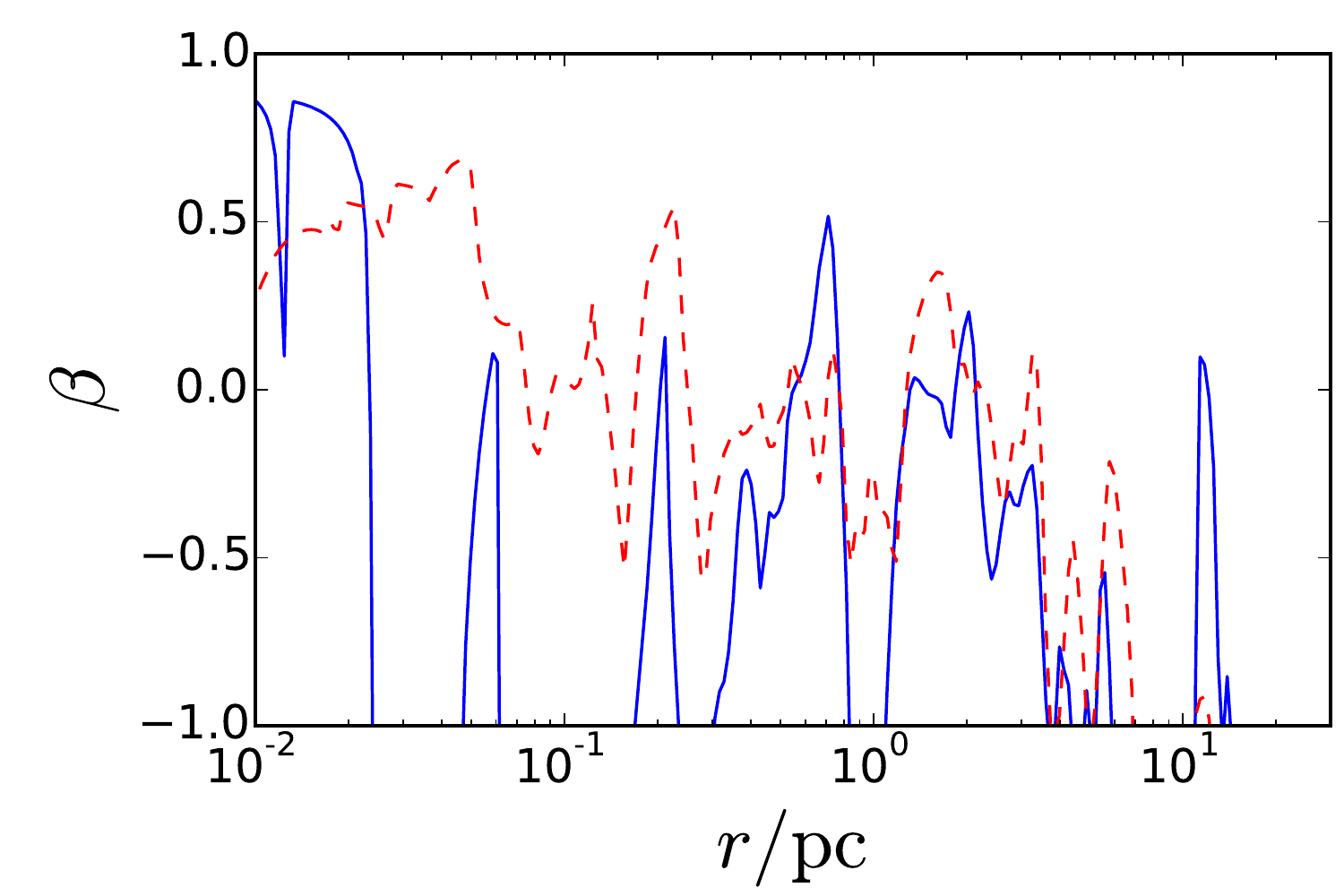}

  \caption{Top row: likelihood contours (blue curves) of spherical anisotropic models fit
    to the kinematics of the simulated flattened cluster described in
    \S\ref{sec:flat}.  For comparison, the dashed red contours show
    likelihoods of spherical models fit to the corresponding simulated spherical clusters.
    The red dot indicates the correct $(M_\bullet,M_\star)=(4,1)\times10^6\,M_\odot$.
    Bottom row: anisotropy profiles of the best-fitting models that
    assume this correct potential.
  }
  \label{fig:flat}
\end{figure*}
We use a multipole expansion
to calculate the stellar contribution to the potential, and add a
central BH of mass $M_\bullet=4\times10^6\,M_\odot$.
Our simulated cluster has a two-integral DF $f(\E,L_z)$, the
even-in-$L_z$ part of which, $f_+$, is completely determined by the known
$\nu(R,z)$ and $\Phi(R,z)$ through
\begin{equation}
  \label{eq:2iDF}
  \nu(R,z) =
  \frac{4\pi}R\int_0^{\psi(R,z)}\,\d\E\int_0^{\sqrt{2R^2(\psi(R,z)-\E)}}
  \d L_z\,f_+(\E,L_z).
\end{equation}
We ignore the odd-in-$L_z$ part of the DF because the
nonrotating spherical models that we will use to fit
the simulated kinematics are blind to it.
Our procedure for inverting~\eqref{eq:2iDF} to find $f_+$ borrows the
same techniques we use to construct our DF-superposition models: we
represent the DF $f_+(\E,L_z)$ as a sum of blocks~\eqref{eq:dfdefn} in $(\E,L_z)$,
calculate the contribution that each block makes to the densities
$\nu(R,z)$ at a set of points in the $(R,z)$ plane and then find the
set of block weights that best reproduces the deprojected
$\nu(R,z)$.  Having found $f_+$ we draw stellar positions and
velocities from it.  Our simulated discrete catalogue then consists of
1000 stars having projected radius $R<19''$ drawn from this $f_+$ and
scattered by $20\kms$ simulated observational uncertainty.  We also
calculate the zeroth- and second-order velocity moments of $f_+$
averaged over the annuli given in Table~\ref{tab:rad}.

For comparison purposes, we also construct a spherical, isotropic
reference cluster that matches the density and potential of the
flattened cluster along its symmetry axis.  That is, the DF $f(\E)$ of
this spherical cluster is set equal to $f_+(\E,L_z=0)$ and samples of
stars drawn from it to construct simulated catalogues in the same way
as the axisymmetric models.

We use the modelling procedure of Section~\ref{sec:models} to fit
spherical anisotropic models to the simulated kinematics from both
axisymmetric and spherical clusters.  Although it would be possible to
take, say, the deprojected minor-axis profile plotted in
Figure~\ref{fig:axiaxes} for the DF-superposition models' $\rho(r)$
profile, we have opted for convenience to take
\begin{equation}
  \label{eq:rhoaxiprof}
  \rho(r)\propto r^{-1.5}\left(1+\frac r{r_{\rm s}}\right)^{-1.5},
\end{equation}
with $r_{\rm s}=10\,\pc$ instead.  This was obtained using a crude
``by-eye'' fit to the deprojected $\rho(0,z)$ profile.

Figure~\ref{fig:flat} shows how the likelihoods of the models varies
with the assumed $(M_\bullet,M_\star)$ for this $\rho(r)$.  Models
that fit only the line-of-sight components of velocity do not
reproduce the correct potential or orbit distribution.  For the other
two cases it is clear that any systematic error in the inferred masses
caused by fitting these spherical models to the data from this
flattened simulated cluster is no larger than that induced by a
smattering of ``young'' stars on tilted ring orbits.  Internally, the
main systematic difference caused by the neglect of flattening is a
slight bias towards tangential orbits at radii $r>1\,\pc$; the effect on the
inferred masses is small.

%%%%%%%%%%%%%%%%%%%%%%%%%%%%%%%%%%%%%%%%%%%%%%%%%%
\begin{figure*}
  \null\hskip10pt
  \hbox to 0.24\hsize{\hfill (a) Omit inner 2''\hfill}
  \hbox to 0.24\hsize{\hfill (b) Omit inner 8''\hfill}
  \hbox to 0.24\hsize{\hfill (c) Dust\hfill}
  \hbox to 0.24\hsize{\hfill (d) Sch\"odel+2009 sample\hfill}

  \includegraphics[width=0.24\hsize]{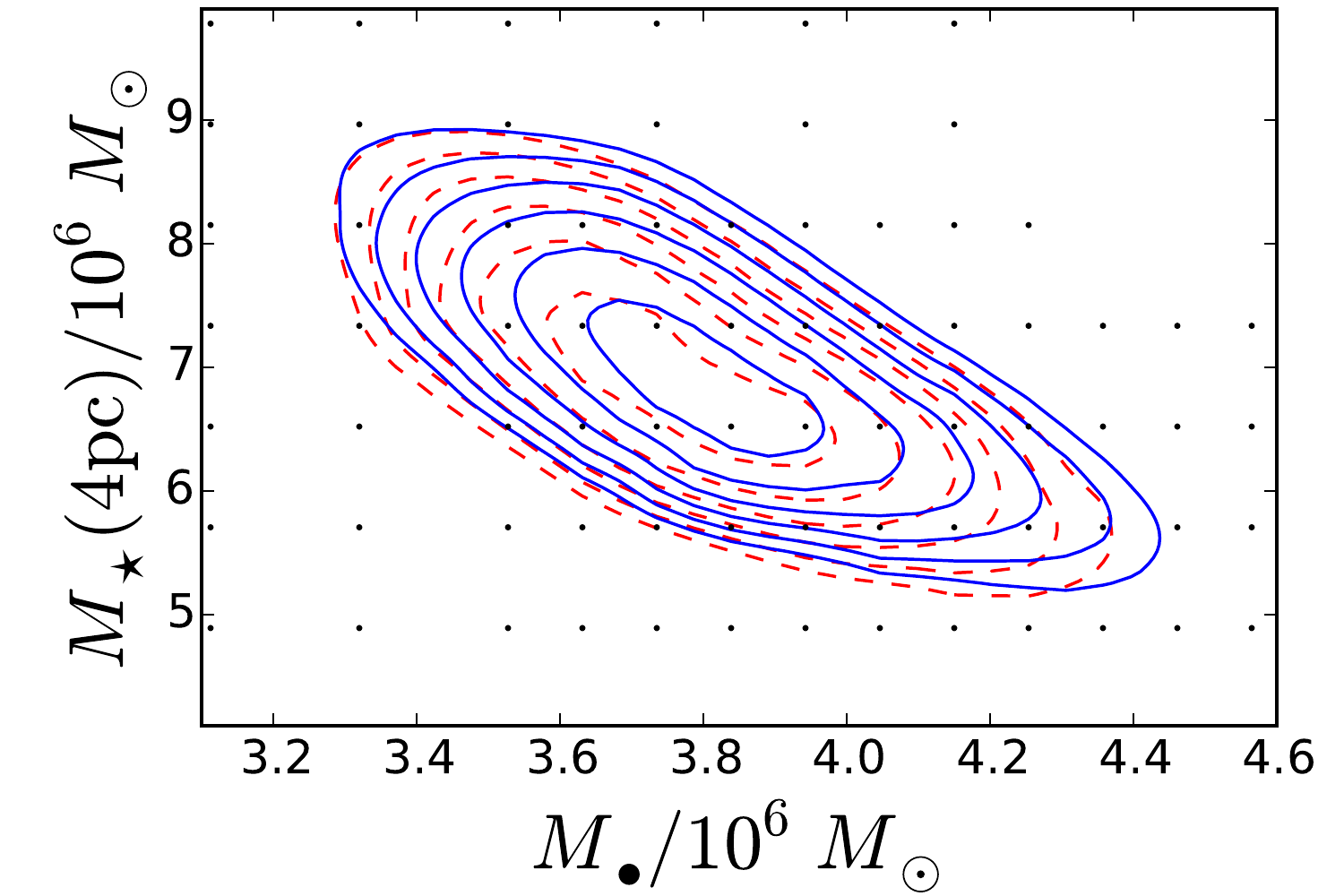}
  \includegraphics[width=0.24\hsize]{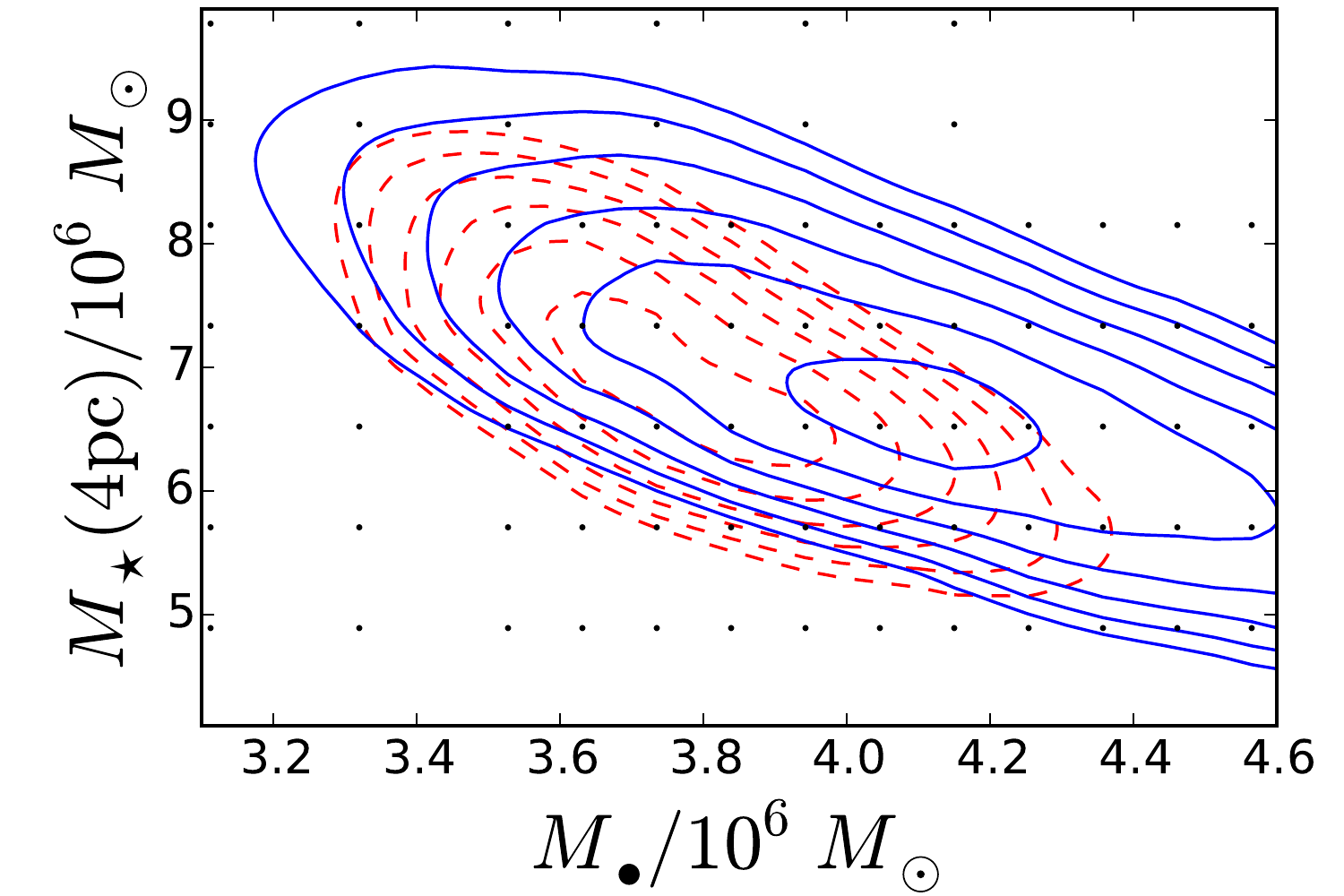}
  \includegraphics[width=0.24\hsize]{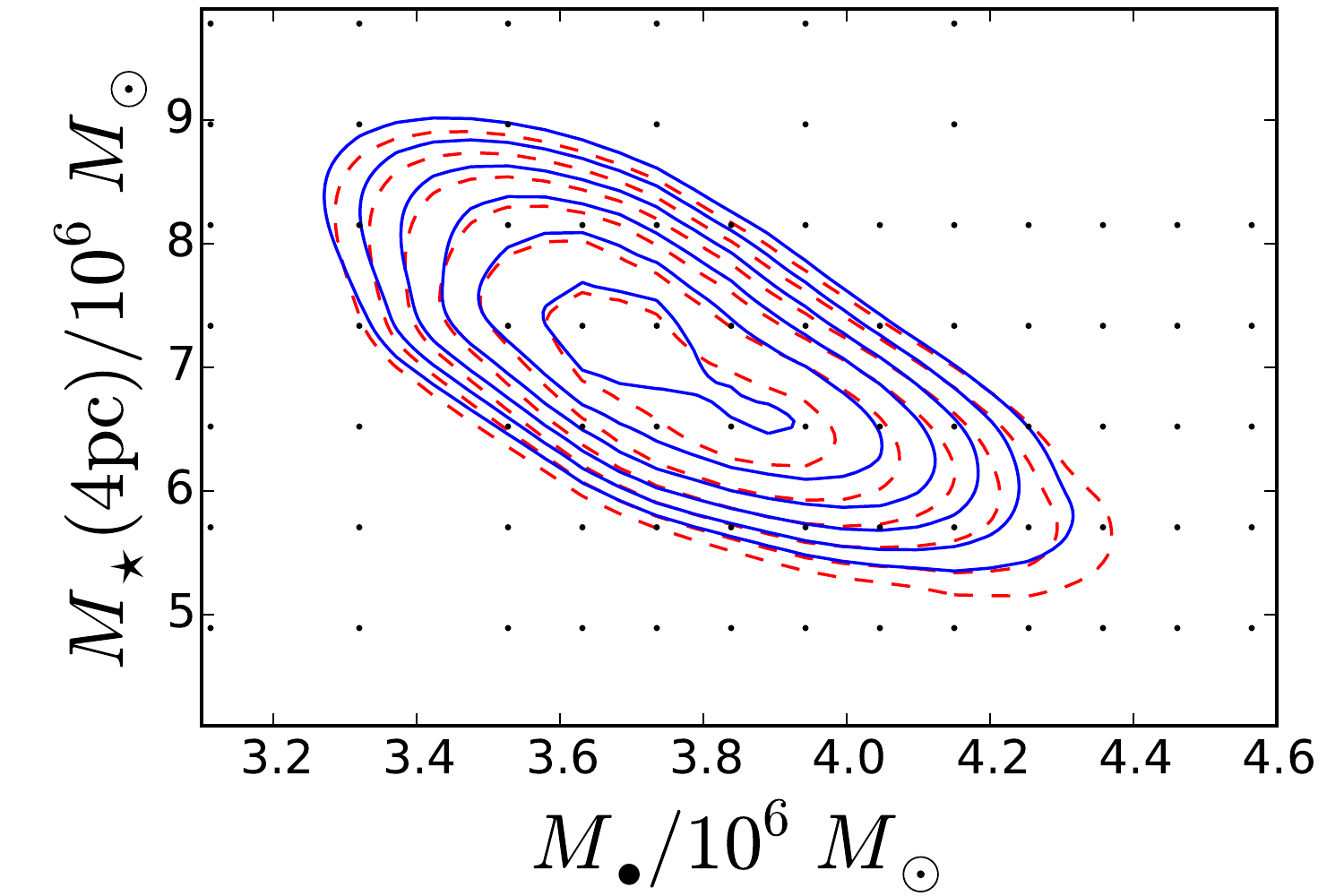}
  \includegraphics[width=0.24\hsize]{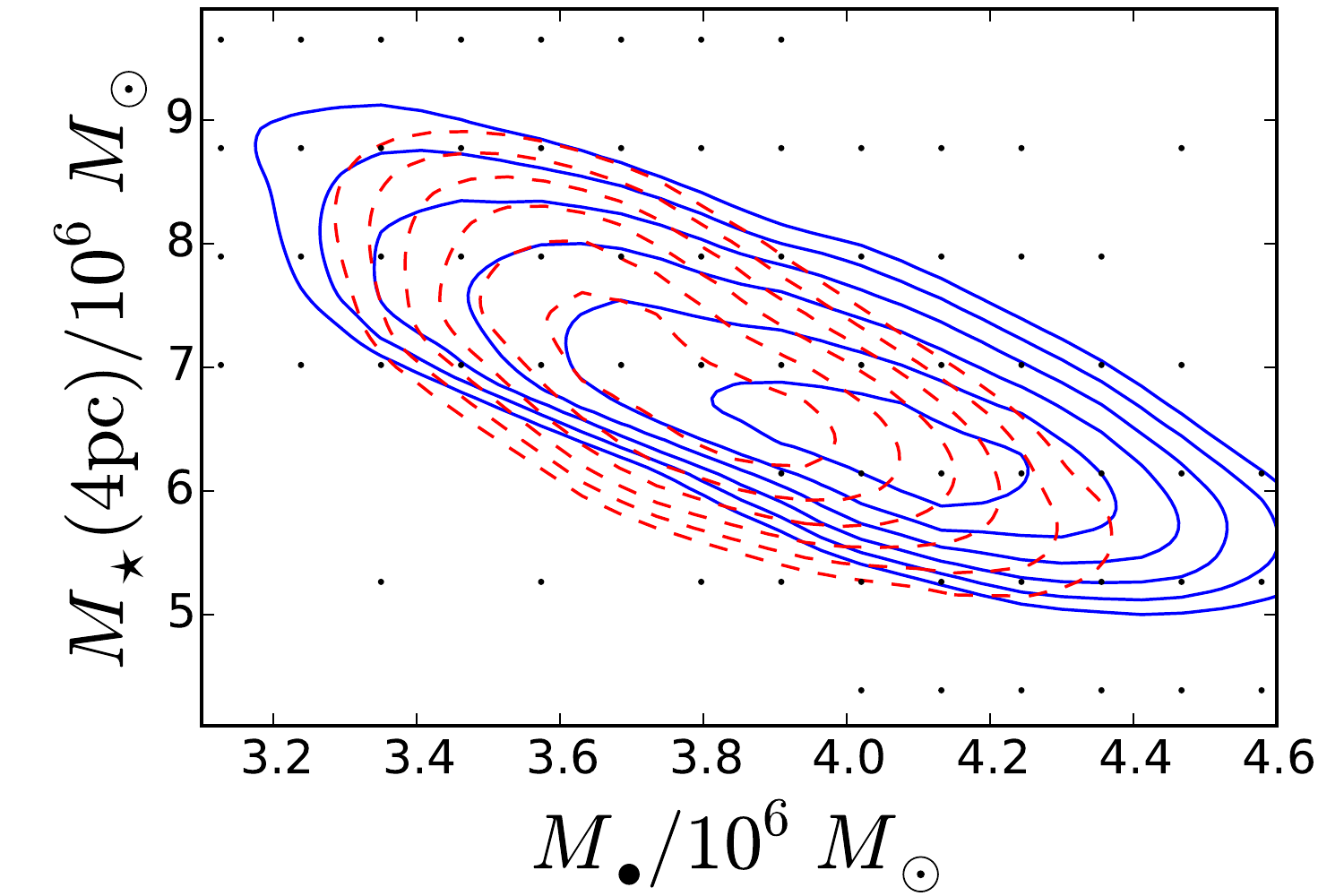}

  \caption{Likelihoods of models fit to real data, assuming mass
    density of the form~\eqref{eq:rhoprof} with an inner power-law slope
    $\gamma=1.5$ and a scale radius $r=300\,\pc$.  For reference, the dashed red
    contours in each panel show the likelihoods of models fit to the
    stellar kinematics of \citet{FritzNuclearClusterMilky2016} within
    $R=19''$; these contours are the same in all four panels.
    From left to right, the blue contours
    show models that: (a) fit only to stars having $2''<R<19''$; (b)
    fit only to stars that have $8''<R<19''$; 
    (c) use all stars having $R<19''$, but modifying the
    selection function to account for a simple two-dimenstional
    exintinction model; (d) fit to the $R<19''$ subset of the \citet{SchodelNuclearstarcluster2009} catalogue
    instead of \citet{FritzNuclearClusterMilky2016}.}
  \label{fig:realdata}
\end{figure*}

%\clearpage
\section{Application to the Galactic centre}
\label{sec:app}

Having investigated some of the most obvious potential sources of
systematic error introduced by our implementation of the
DF-superposition procedure with the spherical symmetry assumption,
we now apply it to the real Galactic centre.

\subsection{Power-law mass density profiles}

An important feature of our models is that we do not assume that the
cluster's unknown mass density profile is directly related to the
number-density distribution of the stars used as kinematical tracers.
Both observations \citep{SchodelDistributionstarsMilky2018} and
theoretical arguments \citep{BaumgardtDistributionstarsMilky2018}
suggest that the underlying mass density of the Milky Way's
nuclear cluster is approximately a power
law, $\rho\sim r^{-\gamma}$,
with $\gamma\simeq1.5$, at least in the innermost parsec or
so.
Our first models adopt a mass density of the form~\eqref{eq:rhoprof}
with $r_{\rm s}=300\,\pc$ and $\gamma=1.5$, so that the assumed
$\rho(r)$ is very close to this pure power law over the volume of
space probed by the observed stars.  That leaves the BH mass
$M_\bullet$ and the normalisation $M_\star$ of the surrounding
extended mass distribution as the free parameters in our potential.
For comparison with the results of \cite{ChatzopoulosOldnuclearstar2015}
we use $4\pc\simeq100''$ as the reference radius for
calculating~$M_\star$.

The dashed red contours in each panel of Figure~\ref{fig:realdata}
show the likelihood contours of $(M_\bullet,M_\star)$ obtained by
fitting the kinematics of all 4105 of the stars from \citet{FritzNuclearClusterMilky2016}
that have $R<19''$ together with the binned zeroth- and second-order
moments from Table~\ref{tab:rad}.  The best-fit $M_\bullet=(3.88\pm0.21)\times10^6\,M_\odot$ and
$M_\star(4\,\pc)=(6.76\pm0.54)\times10^6\,M_\odot$, only slightly lower than the
measurements of $4.0\times10^6\,M_\odot$ to $4.3\times10^6\,M_\odot$
obtained from the S stars
\citep{BoehleImprovedDistanceMass2016,GillessenUpdateMonitoringStellar2017}.

We now show that our result is independent of these S stars.  We can
excise the ``central field'' ($R<2''$) of the
\cite{FritzNuclearClusterMilky2016} sample by setting the selection
function $S(y,z)$ to zero outside the range $2''<R<19''$,
recalculating the likelihood normalisations $I_k$
(equation~\ref{eq:I}) and refitting the models ignoring the 59 stars
from the innermost 2''.  Doing this has almost no effect on
$(M_\bullet,M_\star)$ (blue contours on left-most panel of
Figure~\ref{fig:realdata}).

A more striking result is obtained by dropping the 793 stars that have
$R<8''$, within which most of the population of young stars live.
Omitting this region results in a significant increase in the BH mass
estimate to $M_\bullet=(4.08\pm0.32)\times10^6\,M_\odot$ (second panel
of Figure), but leaves
$M_\star(4\,\pc)=(6.75\pm0.75)\times10^6\,M_\odot$ almost unchanged.

We can also use the selection function to test the effects of dust
extinction.
\cite{SchodelPeeringveilnearinfrared2010} have used stellar
$H-K_{\rm s}$ colours to construct an two-dimensional extinction map
of the Galactic centre region: given a position $(y,z)$, they provide
an estimate of the $K_{\rm s}$-band extinction to the Galactic centre
along that line of sight.  Following their Figure~9 we assume that the
luminosity function of the stars that comprise the kinematical
catalogue scales as $L(m)\propto 10^{0.27m}$, where $m$ is the star's
absolute $K_{\rm s}$-band luminosity.  Then the probability that a
star is bright enough to be included in the catalogue scales as
$S(y,z)\propto10^{-0.27A_K(y,z)}$.  The results of fitting models in
which the normalisation integrals $I_k$ have been adjusted to use this
selection function are shown on the third panel of
Figure~\ref{fig:realdata}.  This suggests that our results are
unlikely to be strongly affected by extinction.

There are a couple of caveats associated with this reassuring result,
however.  The first is that the
\cite{SchodelPeeringveilnearinfrared2010} exinction map is available
for a slightly offset rectangular region that covers most, but not
all, of the 19''-radius cylinder that defines our kinematical sample:
for lines of sight $(y,z)$ that are unavailable in their map we assume
that the extinction is equal to the average of the available $A_K$
values that have the same projected radius.  The second is that this
dust model assumes that the extinction is caused by a two-dimensional
screen of dust that lies between us and the Galactic centre: it
ignores any variation in the dust density along lines of sight within
the Galactic centre region.  \cite{ChatzopoulosDustnuclearstar2015}
show how such variations can be constrained by looking for asymmetries
in velocity histograms, but note that they are not likely to strongly
affect the even-in-$L_z$ part of their $f(\E,L_z)$ DF (which is
effectively what our nonrotating spherical anisotropic models use).

Our models produce similar results when fit to the independently
obtained proper catalogue of \citet{SchodelNuclearstarcluster2009}:
the rightmost panel of Figure~\ref{fig:realdata} plots the
$(M_\bullet,M_\star)$ likelihood contours obtained by fitting all 5005
stars within $R=19''$ from this sample.  The best-fit
$M_\bullet=(3.98\pm0.52)\times10^6\,M_\odot$ with $M_\star(4\,\pc)=(6.62\pm0.98)\times10^6\,M_\odot$.
Even though we have adjusted the quoted uncertainties on both
kinematical datasets (Section~\ref{sec:data}) to be consistent with
each other, it is nevertheless reassuring that the two datasets
produce comparable mass estimates.

%%%%%%%%%%%%%%%%%%%%%%%%%%%%%%%%%%%%%%%%%%%%%%%%%%%%%%%%%%%%%%%%%%%%%%%%%%%%%%%%

\begin{figure*}
  \null\hskip10pt
  %\hbox to 0.24\hsize{\hfill Sch\"odel, $\gamma=1.5$\hfill}
  \hbox to 0.32\hsize{\hfill (a) $\gamma=1.5$\hfill}
  \hbox to 0.32\hsize{\hfill (b) Chatzopoulos+2015\hfill}
  \hbox to 0.32\hsize{\hfill (c) $\gamma=1.5$ extended\hfill}
  
  \vspace{-0.3cm}\null\hskip10pt%
  \includegraphics[width=0.32\hsize]{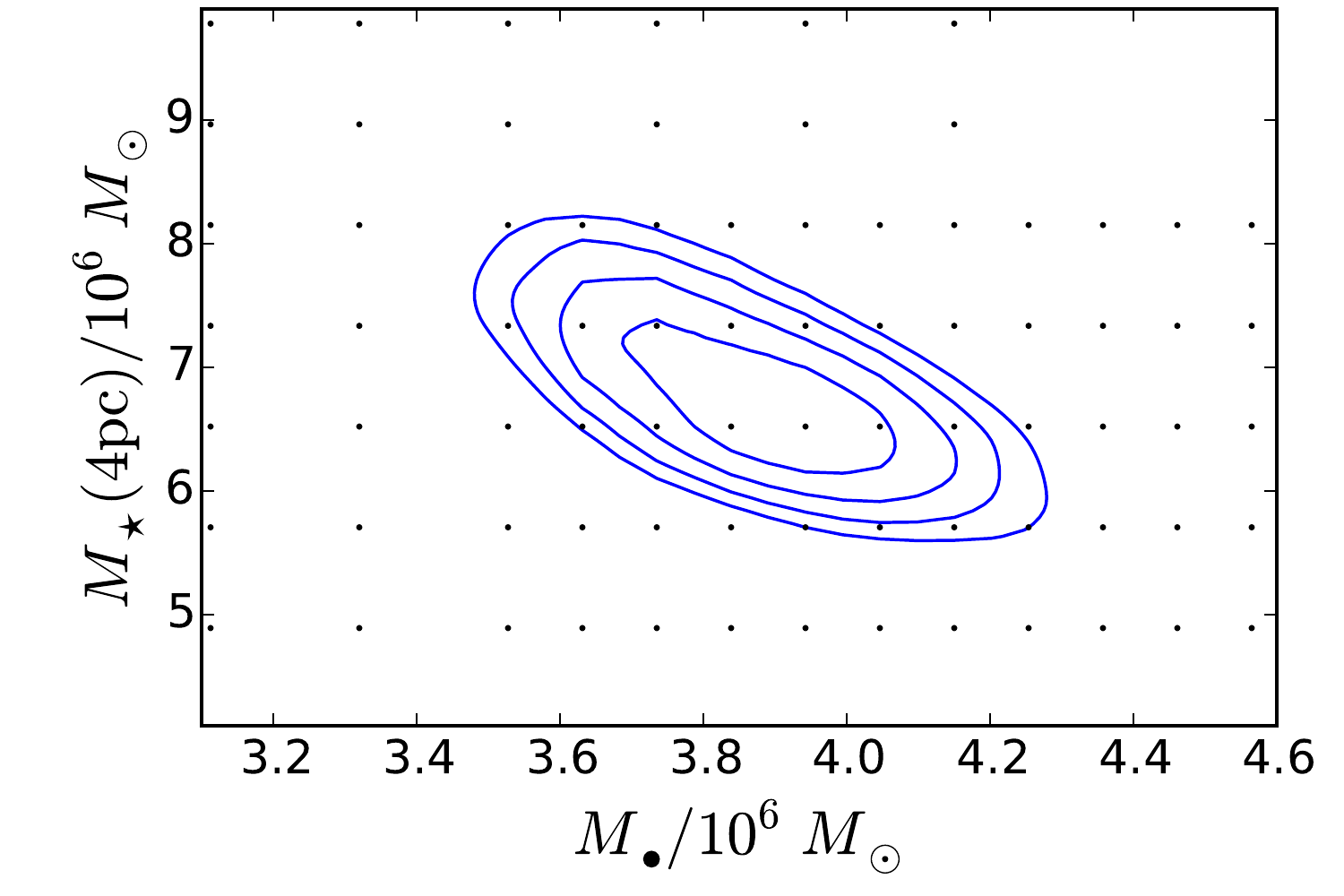}
  \includegraphics[width=0.32\hsize]{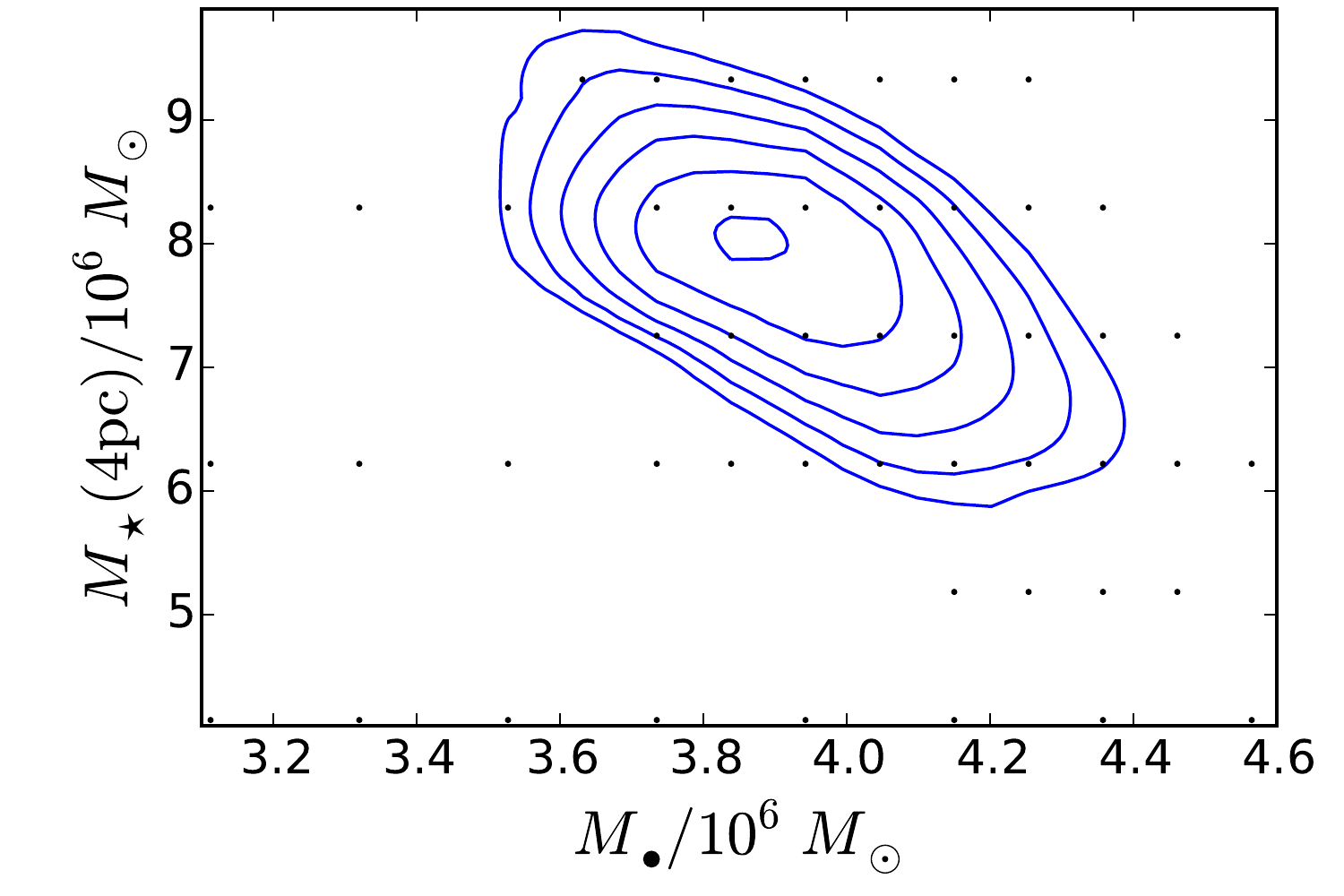}
  \includegraphics[width=0.32\hsize]{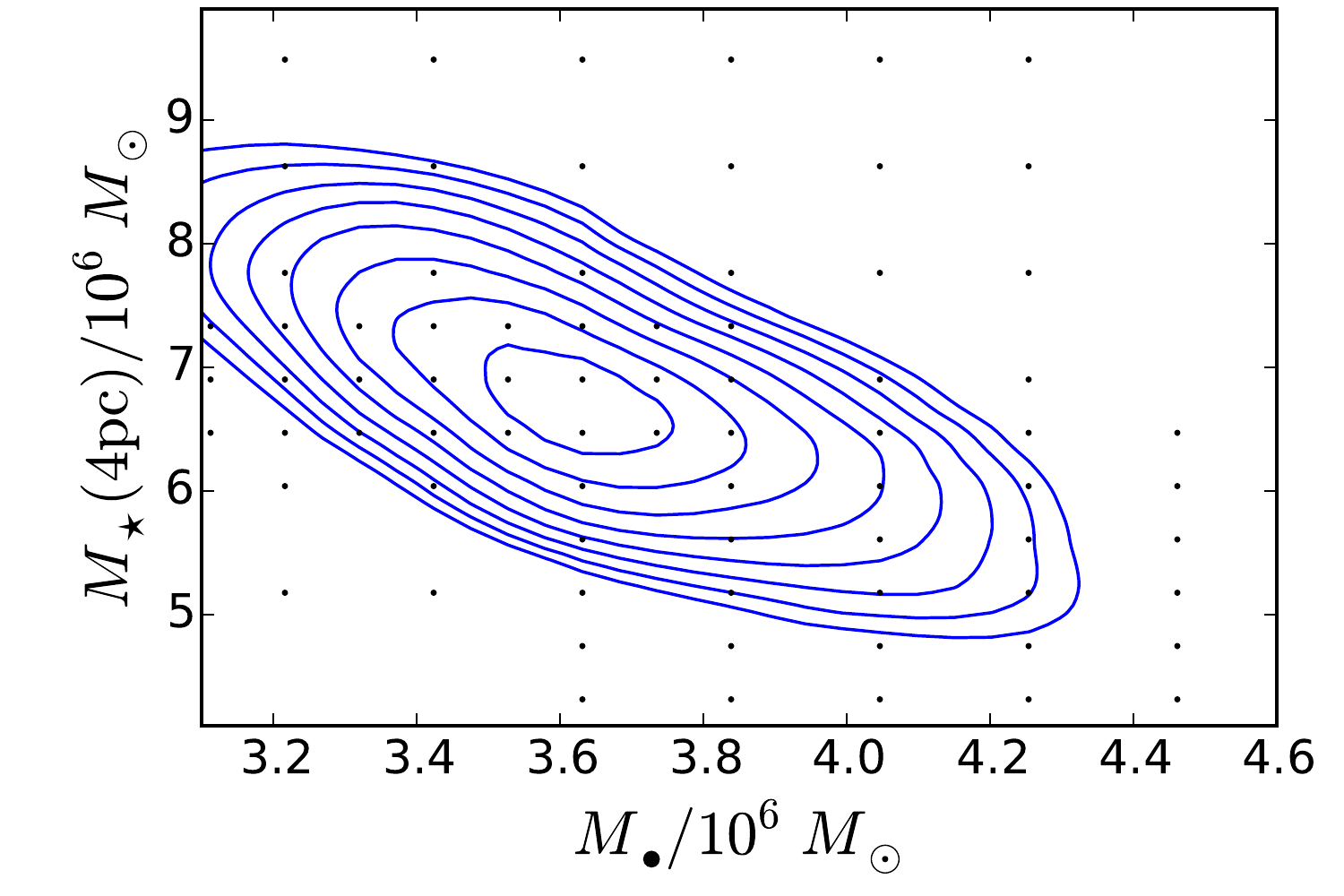}

  \includegraphics[width=0.32\hsize]{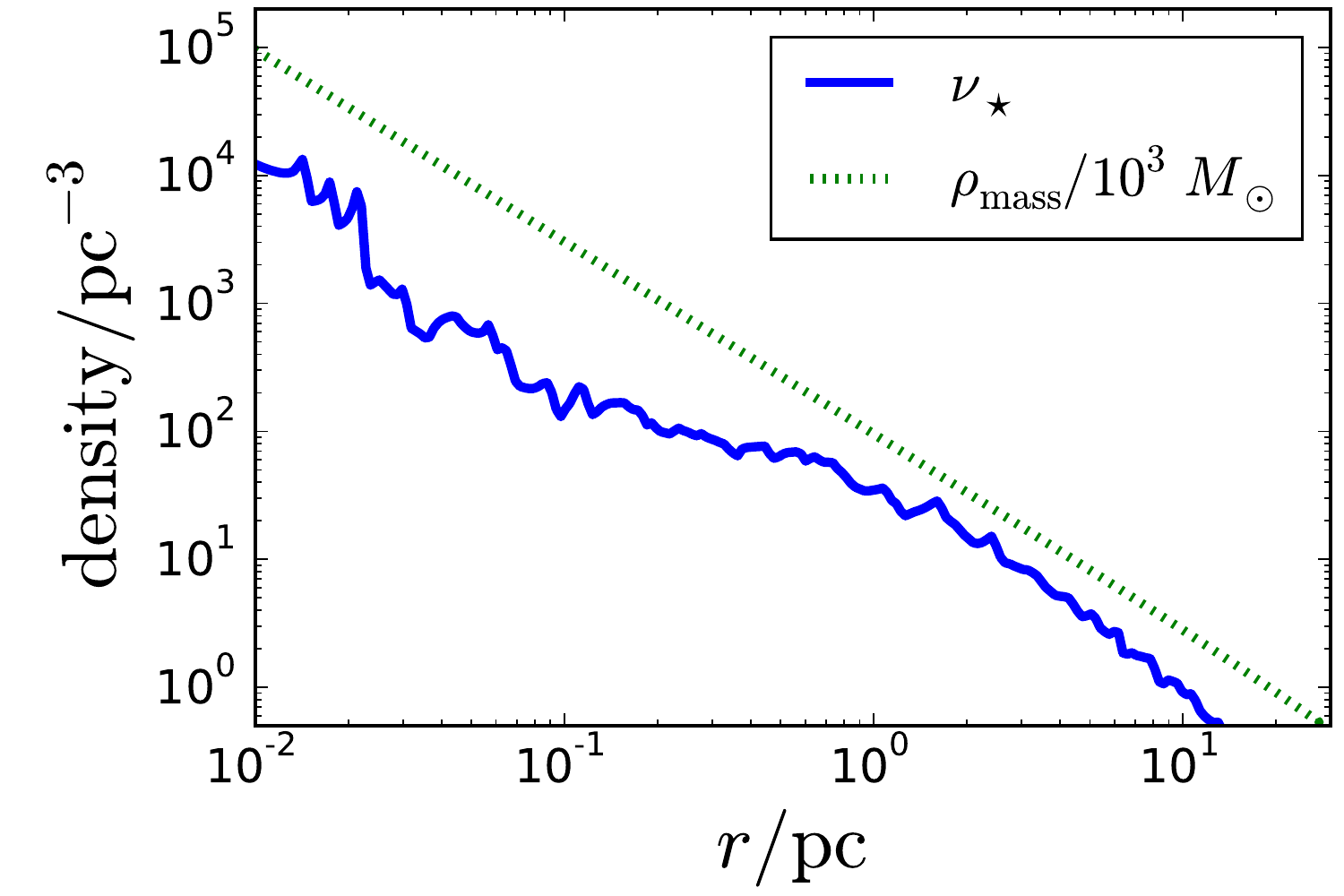}
  \includegraphics[width=0.32\hsize]{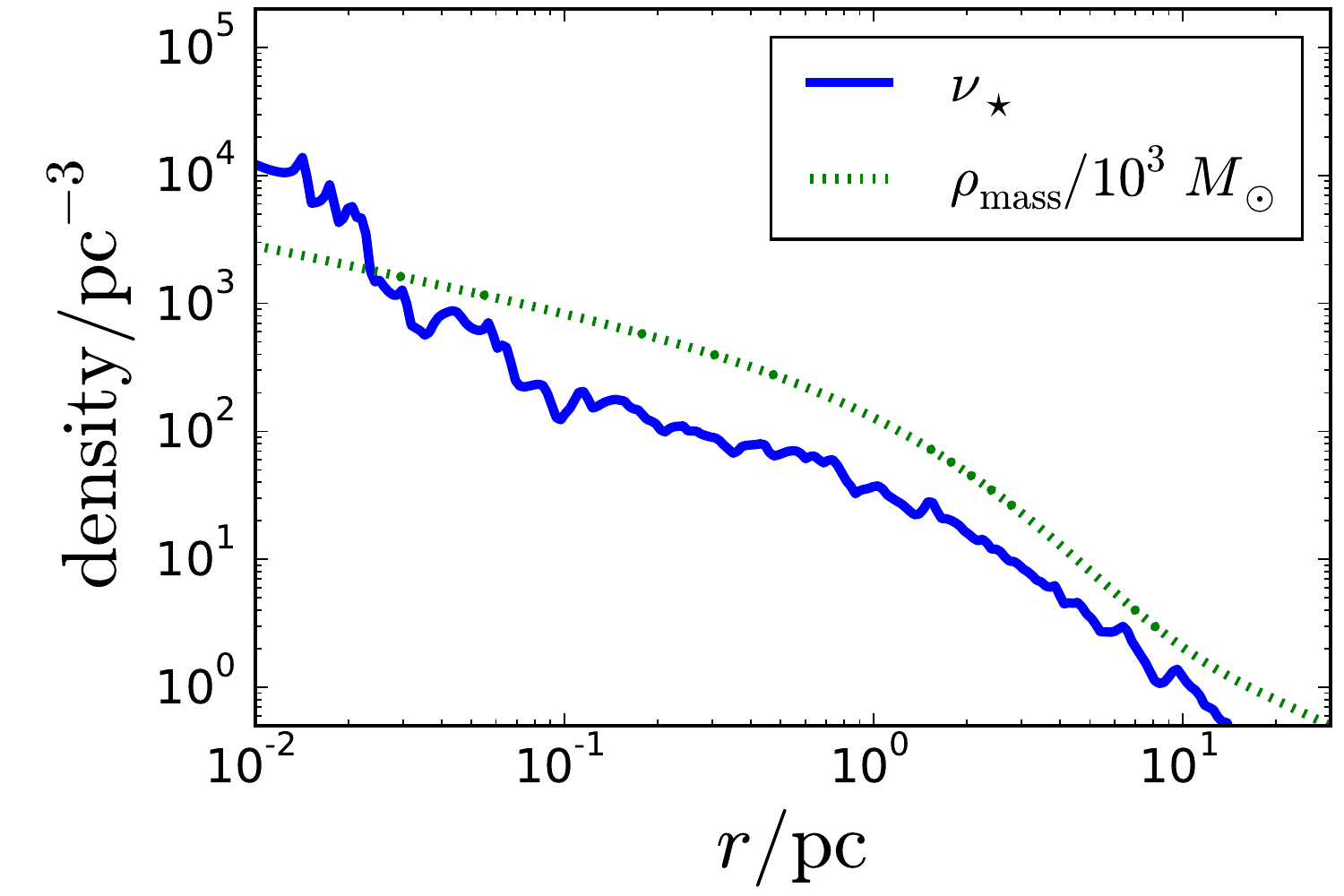}
  \includegraphics[width=0.32\hsize]{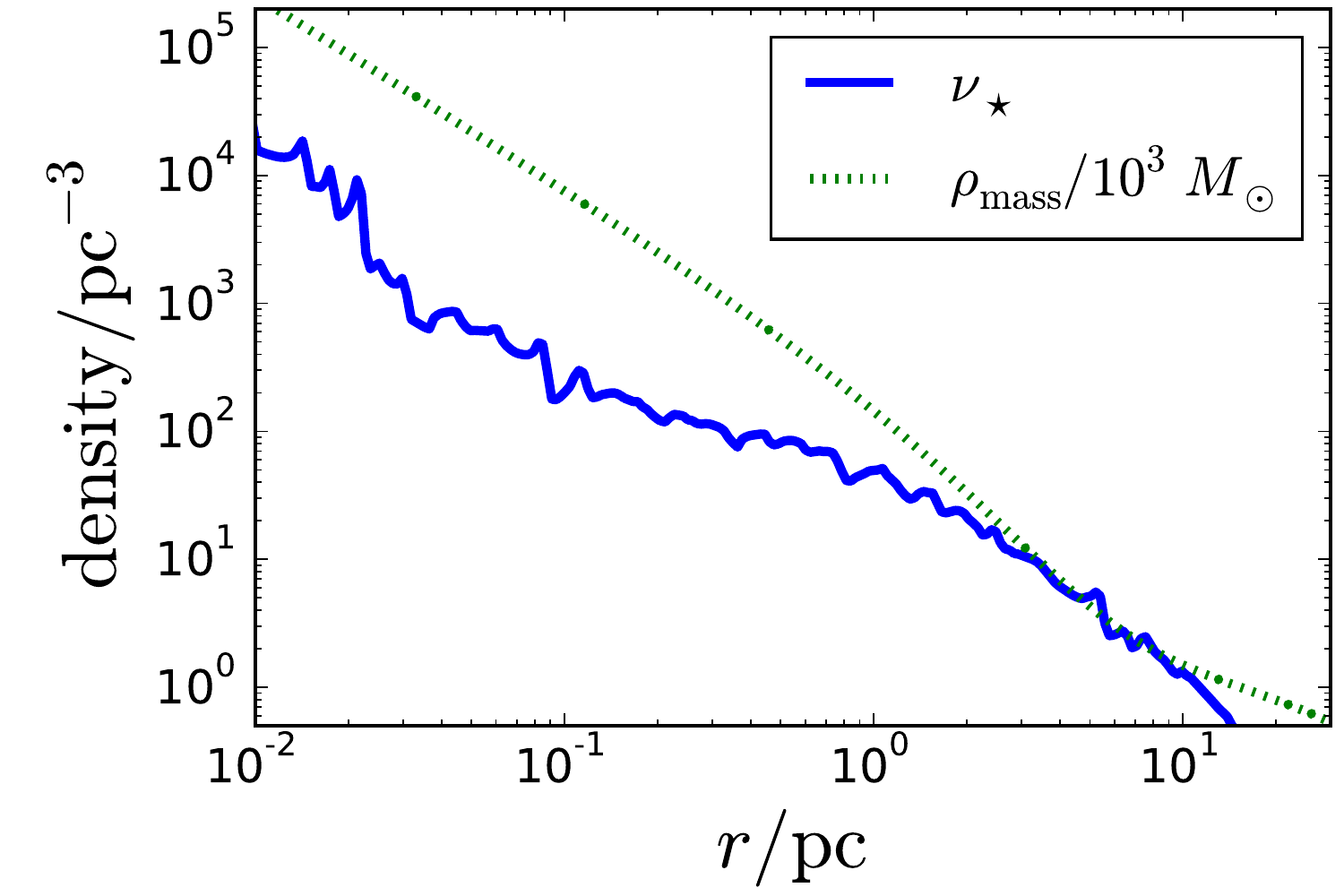}

  \includegraphics[width=0.32\hsize]{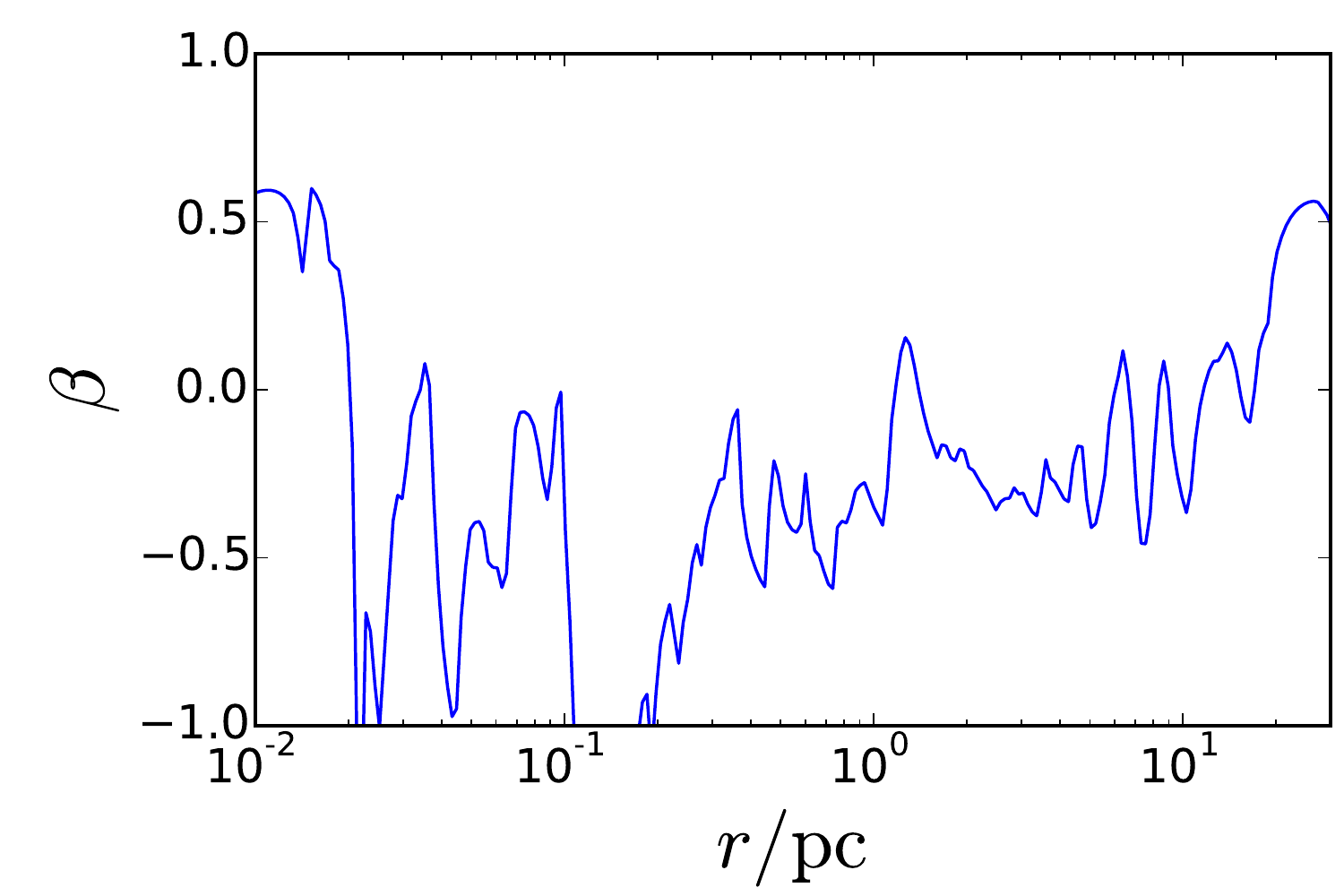}
  \includegraphics[width=0.32\hsize]{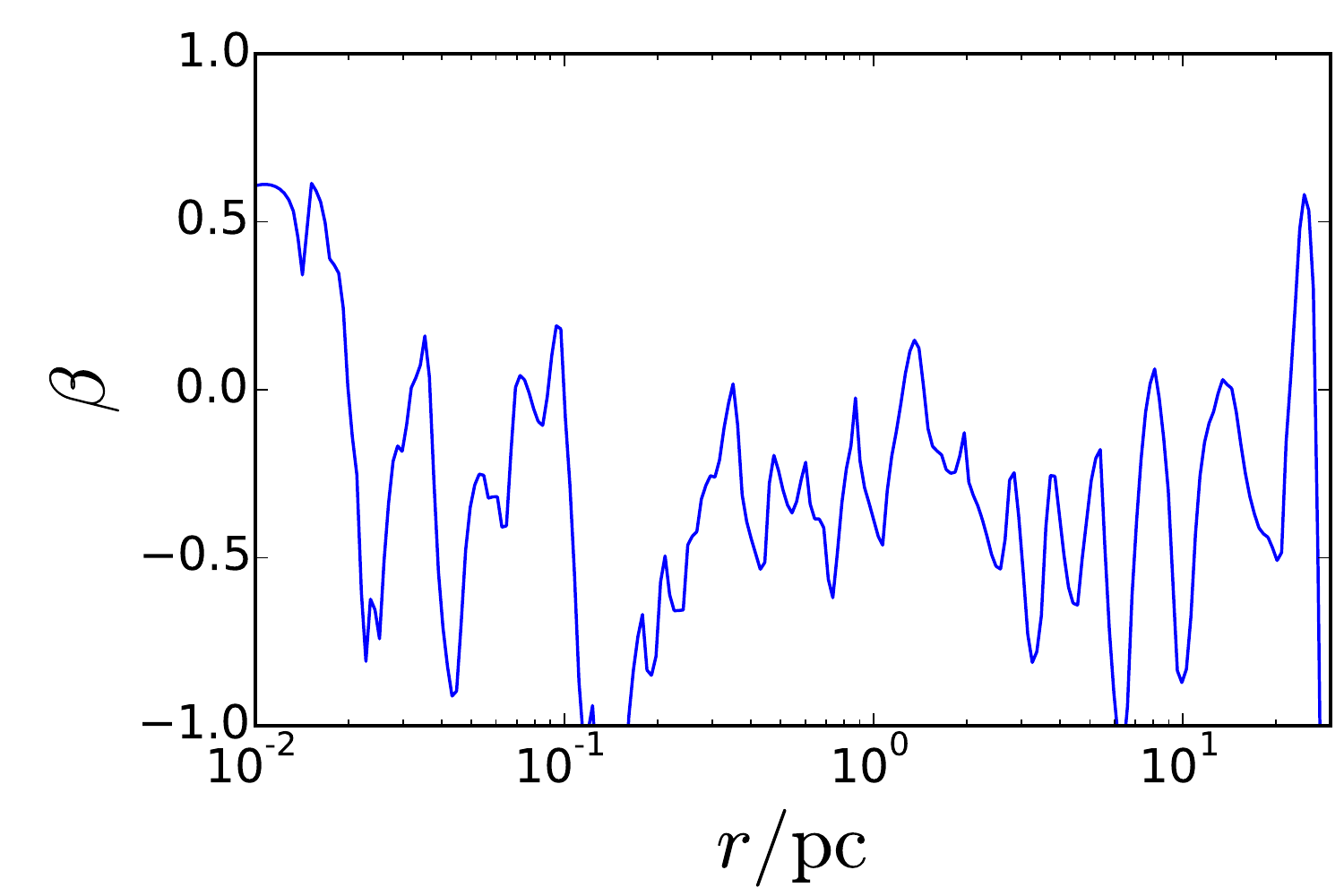}
  \includegraphics[width=0.32\hsize]{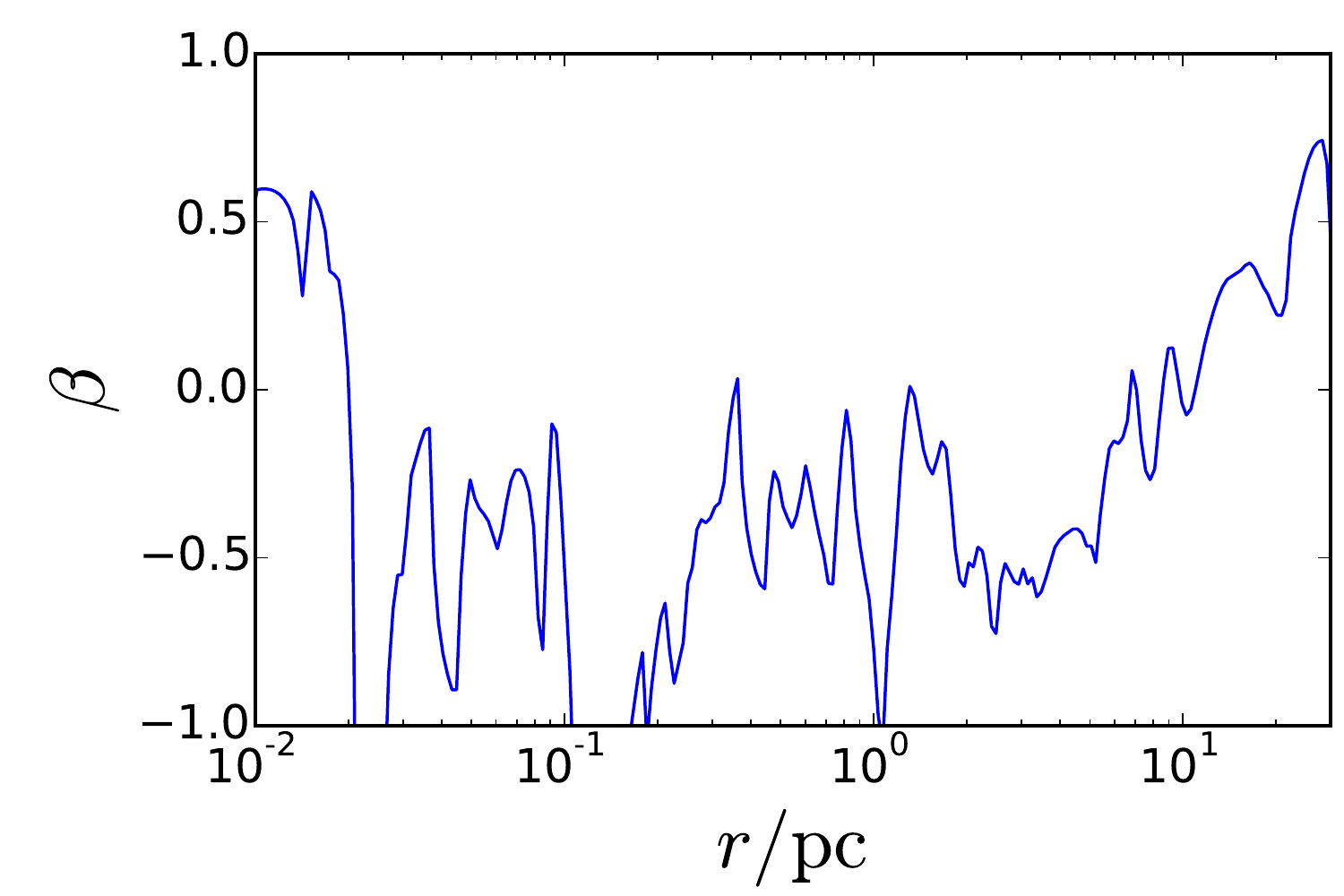}

  \caption{
    A comparison of different mass models fit to the
    \citet{FritzNuclearClusterMilky2016} kinematics.
    From left to right, the columns show:
    (a) the reference model from
    Figure~\ref{fig:realdata}, which has a mass density
    profile of the form~\eqref{eq:rhoprof} with $\gamma=1.5$ and
    $r_{\rm s}=300\,\pc$;
    (b) models that adopt the form of the mass density profile used by
    \citet{ChatzopoulosOldnuclearstar2015};
    (c) models having mass-density profile~\eqref{eq:extendedrhoprof}.
    The top panel in each column shows how the model's likelihoods depend on the assumed
    $(M_\bullet,M_\star)$.  The same contours levels are used in all
    three columns: the more contours shown, the more likely the model.
    The middle panel that shows the mass-density profile of the best-fitting
    model (dotted curves), together with the inferred number density profile
    (solid curve).  The lower panel plots the best-fit model's
    anisotropy profile.}
  \label{fig:realdata2}
\end{figure*}

Our models make no assumptions about the underlying number density or
anisotropy of the stars, apart from spherical symmetry.  The first
column of Figure~\ref{fig:realdata2} shows the number-density and
anisotropy profiles for our $\gamma=1.5$, $r_{\rm s}=300\,\pc$ fit to
the $R<19''$ subset of the \cite{FritzNuclearClusterMilky2016} sample in more
detail.  Notice that this number density profile becomes shallower
within about $1\,\pc$, only to steepen again inside $0.1\,\pc$.  In
this model there is a significant ``unseen'' mass component
within about 1 parsec of the centre that is not sampled by the
observed kinematical tracers.
The model is broadly isotropic, except for a bias towards circular orbits
for $0.1<r/\pc<0.2$ (approximately 2.5'' to 5'') and is consistent
with expectations from our simulated clusters (Figure~\ref{fig:testpotring}).

At radii $r<0.02\,\pc$ (approximately 0.5'') the model becomes
radially biased, with $\beta=0.5$.  The
\citet{FritzNuclearClusterMilky2016} sample contains only 7 stars
having {\it projected} radius
$R<0.5''$.  Therefore we do not attach any significance to the
$\beta(r)$ profile at these small radii.

\subsection{More realistic mass-density profiles}

\citet{ChatzopoulosOldnuclearstar2015} construct their spherical and
axisymmetric isotropic models assuming that number and mass densities are both
of the form
\begin{equation}
  \label{eq:extendedrhoprof}
  \begin{split}
  \rho(r)&=
  \frac{(3-\gamma)M_1}{4\pi}%%
  \frac{r_{\rm s}}{r^\gamma(r+r_{\rm s})^{4-\gamma}}\\%%
  &\quad+
  \frac{(3-\gamma_2)M_2}{4\pi}%%
  \frac{r_{\rm s,2}}{r^{\gamma_2}(r+r_{\rm s,2})^{4-\gamma_2}},%%
  \end{split}
\end{equation}
which includes an additional component of mass $M_2$ added to the
profile~\eqref{eq:rhoprof}.  They fit this profile to the projected
number density profile measured by
\citet{FritzNuclearClusterMilky2016}, obtaining best-fit parameters
$\gamma=0.51$, $r_{\rm s}=99''$, $\gamma_2=0.07$, $r_{\rm s,2}=2376''$
and mass ratio $M_2/M_1=105.45$ when they assume spherical symmetry.  Notice that the inner scale radius $r_{\rm
  s}=99''\simeq4\,\pc$ is much smaller than the $r_{\rm
  s}=300\,\pc$ we use for the models we have just considered.
The second column of Figure~\ref{fig:realdata2} shows our spherical
anisotropic models fit assuming this mass profile.  In these models
the assumed underlying mass-density profile is a closer fit to
the number-density profile of the kinematical tracers.
The BH mass returned by the models is 
$M_\bullet=(3.91\pm0.20)\times10^6\,M_\odot$, which 
is in good agreement with the $M_\bullet=(3.86\pm0.14)\times10^6\,M_\odot$
produced by the isotropic axisymmetric models of
\citet{ChatzopoulosOldnuclearstar2015}.  Our $M_\star$ lies between
the $M_\star(4\pc)=(8.94\pm0.31)\times10^6\,M_\odot$ they found using
their axisymmetric isotropic models and the
$M_\star(4\pc)\simeq 5\times10^6\,M_\odot$ from their spherical
isotropic models.

Our best-fitting model with this mass density profile has a log
likelihood that is 1.1 % 40821.7-40820
higher than the best-fitting $\rho\sim r^{-1.5}$ model shown in the
left column of the Figure.
Both assumed forms for the mass distribution yield similar
anisotropy profiles and best-fit values of $M_\bullet$, with only the
best-fit value of $M_\star(4\,\pc)$ changing slightly between the two.

We can produce yet better fits to the data by tweaking the parameters
used in the mass density~\eqref{eq:extendedrhoprof}. For example, the
right-most column of Figure~\ref{fig:realdata2} shows models that have
$M_2/M_1=300$ and $\gamma=1.5$, but keep all other parameters the same
those used by \citet{ChatzopoulosOldnuclearstar2015}.  The ``unseen''
mass fraction in the innermost parsec is higher even than the first
models and the BH mass is reduced to
$M_\bullet=(3.62\pm0.20)\times10^6\,M_\odot$ with
$M_\star(4\,\pc)=(6.77\pm0.54)\times10^6\,M_\odot$.  This has a log
likelihood that is 3.6 larger than our best $\rho\sim r^{-1.5}$,
$M_2=0$ model plotted in the left-most column.  Again, the anisotropy
profile is largely unchanged in the innermost few parsecs.

\begin{figure}
  \includegraphics[width=\hsize]{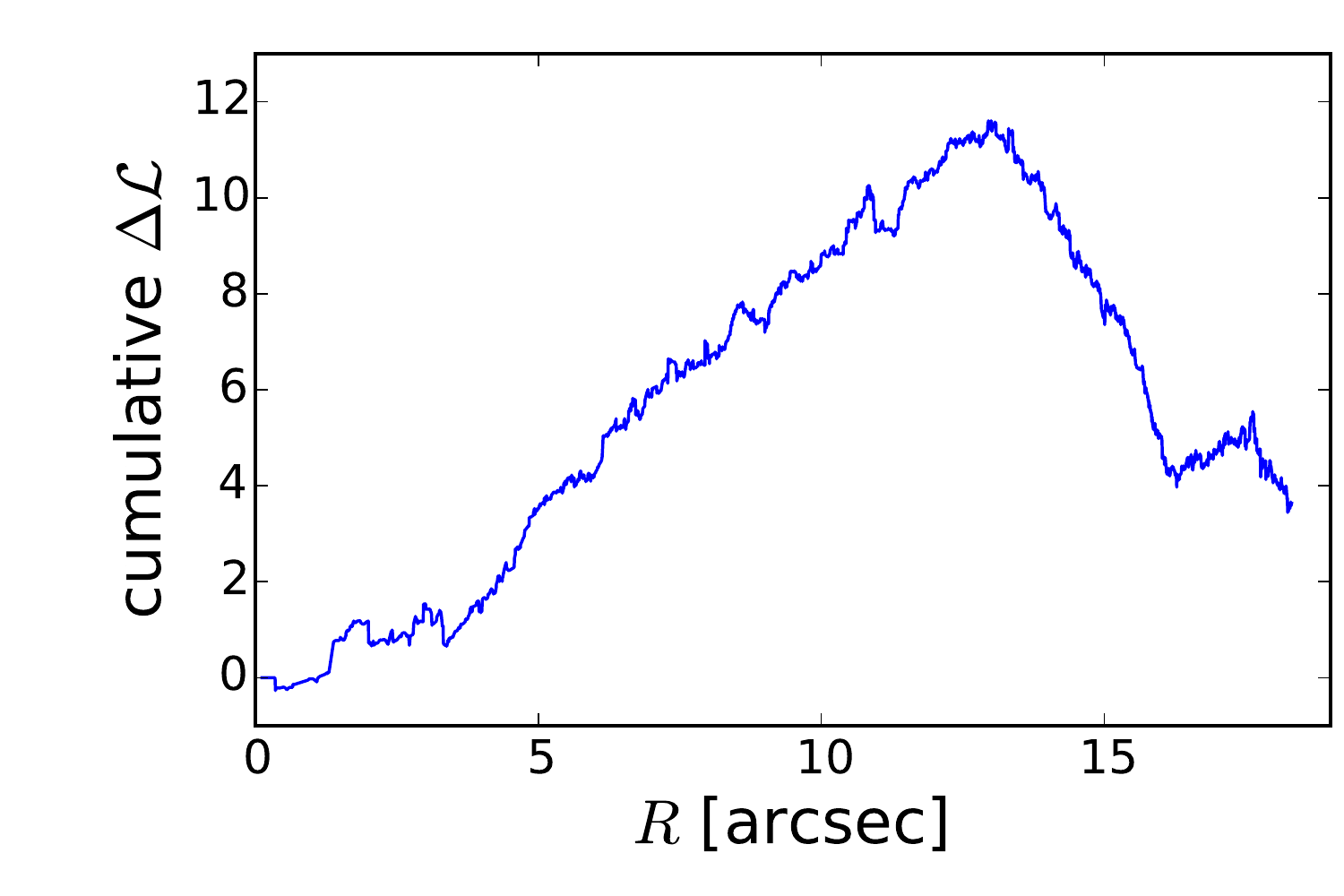}
  \caption{Cumulative change in the log-likelihood $\Delta{\cal
      L}_n=\log\pr(D_n|\Phi_2)-\log\pr(D_n|\Phi_1)$ of individual stars
    $n=1,...,4105$ ordered by their projected radius~$R_n$.  Here
    $\Phi_1$ is the potential of the almost pure ``power-law''
    $\rho\sim r^{-1.5}$ mass model used in the left panel of
    Figure~\ref{fig:realdata}, while $\Phi_2$ is the more realistic
    potential implied by~\eqref{eq:extendedrhoprof} shown in the right panel of the
    same figure.  In both cases
    $M_\bullet=3.7\times10^6\,M_\odot$ and $M_\star(4\pc)=6.2\times10^6\,M_\odot$.
  }
  \label{fig:cumlik}
\end{figure}
An immediate question is whether it is possible to identify the cause
of this increase of 3.6 in the log likelihood between the two
$\gamma=1.5$ models.  We have examined the change in the log
likelihoods of individual stars for models with
$(M_\bullet,M_\star)=(3.7,6.2)\times10^6\,M_\odot$ when one replaces
the almost-pure $r_{\rm s}=300\,\pc$ power-law cusp~\eqref{eq:rhoprof}
(left column of Figure~\ref{fig:realdata2}) with the varying slope on
superparsec scales implied by the density
profile~\eqref{eq:extendedrhoprof} (right column of same Figure).
Figure~\ref{fig:cumlik} shows that breaking the pure power-law in the
density tends to improve the likelihoods of stars between about 4 and
13 arcsec, albeit at the cost of those outside that.  This suggests
that it might be possible to ``tune'' the form of the mass density by
examining how the likelihoods of individual stars are affected as the
assumed mass profile changes.  We do not pursue it further in this
paper, however, because we expect that the systematic errors
induced in the likelihoods by our neglect of flattening are likely to
be comparable in size.

\begin{figure*}
  \includegraphics[width=0.45\hsize]{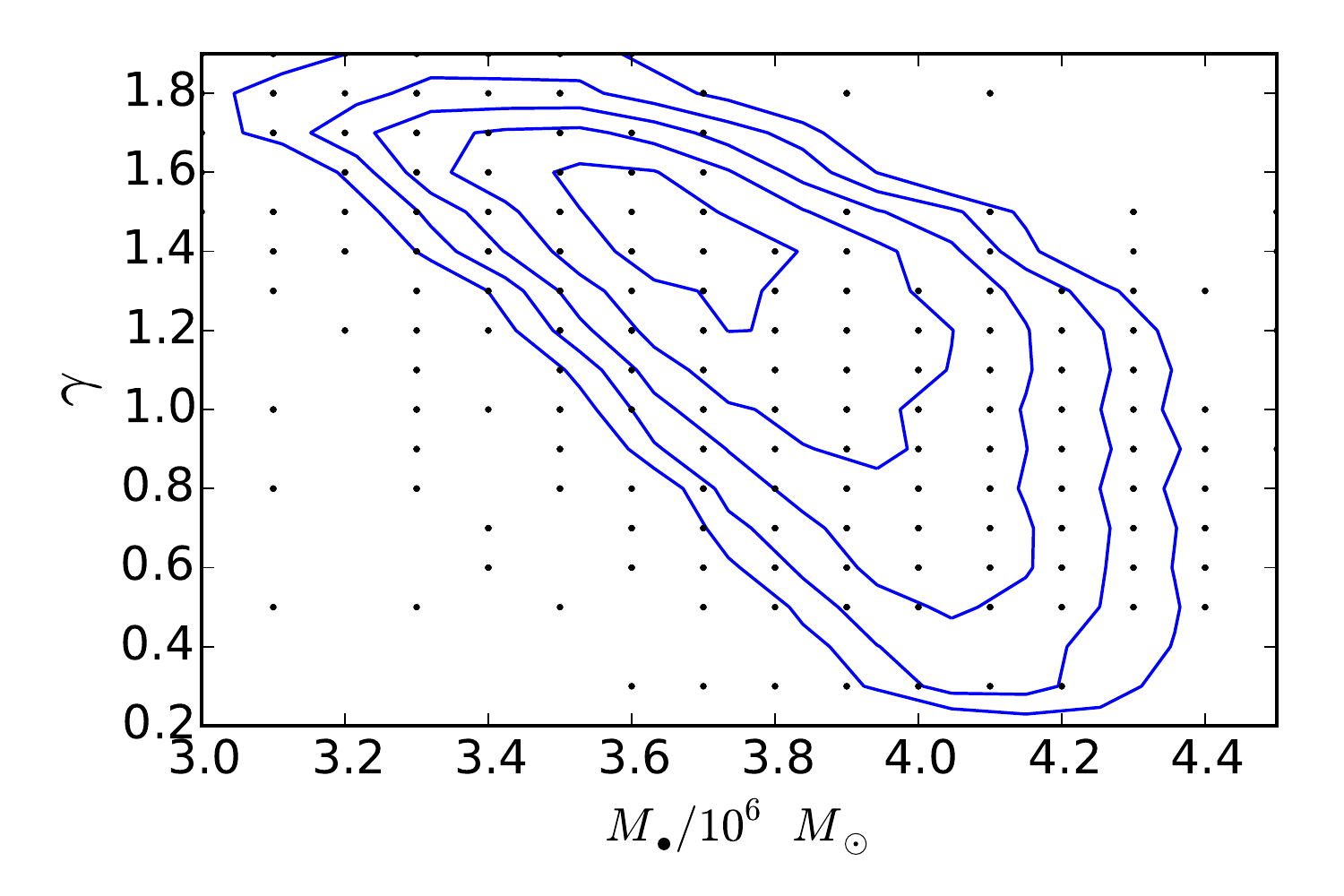}
  \includegraphics[width=0.45\hsize]{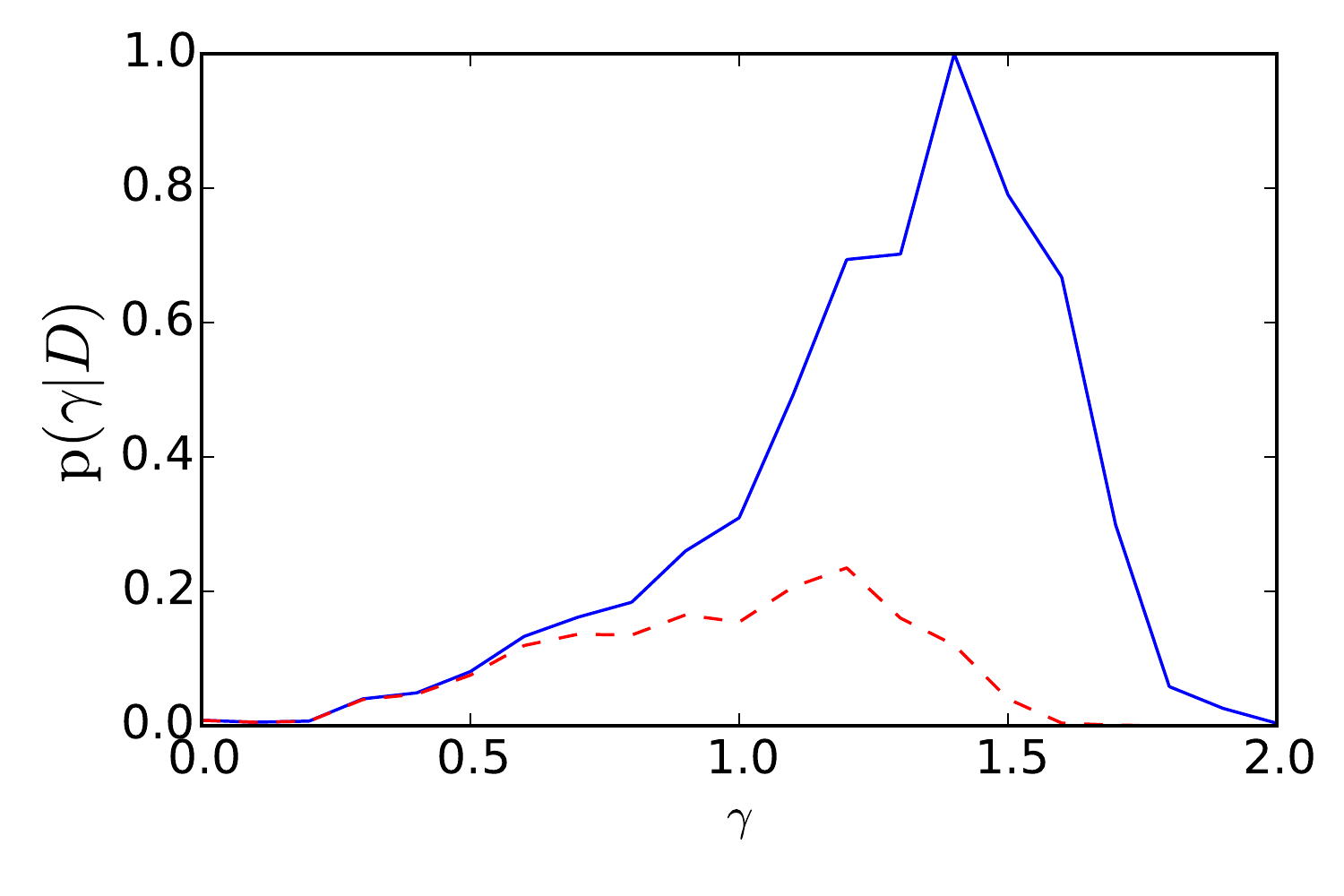}
  \caption{Constraints on the inner power-law slope, $\gamma$, of the
    mass density profile~\eqref{eq:extendedrhoprof} with
    $M_2/M_1=300$ and all other parameters equal to those used by \citet{ChatzopoulosOldnuclearstar2015}.
    The panel on the
    left shows the posterior distribution of $(\gamma,M_\bullet)$
    obtained by marginalizing~$M_\star$ assuming flat priors on
    $\gamma$, $M_\bullet$ and $M_\star$.
    The panel on the right shows
    the result of a further marginalization of $M_\bullet$ if all
    $M_\bullet$ are allowed (solid curve) or with the restriction that
    $M_\bullet>4\times10^6\,M_\odot$ (dashed curve).
  }
  \label{fig:marggamma}
\end{figure*}
Figure~\ref{fig:marggamma} shows the effect of different choices of
the assumed inner power-law slope $\gamma$ of the mass
density~\eqref{eq:extendedrhoprof} with $M_2/M_1=300$.
Marginalising $\gamma$, the constraints on the BH mass are
$M_\bullet=(3.76\pm0.22)\times10^6\,M_\odot$.  Conversely,
marginalising $M_\bullet$, the formal constraints on $\gamma$ are
$\gamma=1.3\pm0.3$, although this would drop to $\gamma=1.0\pm0.3$ if
we were to insist that $M_\bullet>4\times10^6\,M_\odot$.

\section{Conclusions}
\label{sec:byebye}

Although the mass of the BH at the Galactic centre is now reasonably
well constrained, much less is known about the structure of
surrounding stellar cluster.
The motivation for this paper was to find out what can be
established about the mass and orbit distribution of this cluster by
using DF-superposition models to fit positions and velocities of
stellar catalogues \citep{SchodelNuclearstarcluster2009,FritzNuclearClusterMilky2016}
that probe the central few parsecs.
Our models assume spherical symmetry, but otherwise make no assumption
about the cluster's internal anisotropy or its radial number-count
distribution: the observed stellar catalogues are treated as a
(projected) sample of the cluster's underlying DF, subject to a known
selection function $S(\vw)$.

\subsection{Constraints on BH mass}

The mass~$M_\bullet$ of the central BH sets an important boundary
condition that any dynamical model of the Galactic centre should
match.
We have carried out extensive tests with data from simulated clusters
to probe how well our models reproduce this constraint.  Although the
models tend to overfit the data, they nevetheless do reproduce the
correct $M_\bullet$ provided the simulated cluster is spherical.
Fitting spherical models to simulated clusters in which this symmetry
has been broken (e.g., by introducing a tilted ring of stars or by
flattening the cluster at larger radii) results in estimates of
$M_\bullet$ that are biased 5--10\% low.

When applied to the kinematical dataset of
\citet{FritzNuclearClusterMilky2016} our models produce a BH of mass
$M_\bullet=(3.76\pm0.22)\times10^6\,M_\odot$, very similar to the
value found by \citet{ChatzopoulosOldnuclearstar2015} using isotropic
axisymmetric Jeans models on the same data, but slightly lower than
the values obtained by directly fitting the orbits of the S stars
\citep{BoehleImprovedDistanceMass2016,GillessenUpdateMonitoringStellar2017}.
Our best-fitting models are broadly isotropic, but they do show a bias
towards circular orbits at radii where the young stellar disc is
strongest.  Omitting all stars at such radii from our fitting
procedure pushes our estimate of~$M_\bullet$ up to values that are
comfortably within the range favoured by the analyses based on the S
stars (Figure~\ref{fig:realdata}).

Although we have not mentioned it explicitly yet, an even more
important source of systematic error on~$M_\bullet$ is the assumed
distance $D$ to the Galactic centre.  In the absence of any
line-of-sight velocity constraints, the mass estimates produced by our
models scale as $(D/8.3\,\kpc)^3$: a 5\% error in $D$ produces a 15\%
error in~$M_\bullet$.  This is unlikely to be a significant
contributor to the origin of the small discrepancy between our models
and the the S-star analyses though, because fundamentally the latter
have the same $M_\bullet\propto D^3$ dependence (in the absence of any
line-of-sight velocity constraints).

\subsection{Constraints on the distribution of matter around the BH}

The stars from the ``extended field'' of
\citet{FritzNuclearClusterMilky2016} are largely confined to the BH's
sphere of influence.  It is not surprising then that the constraints
our models place on the mass distribution around the BH are much
weaker than those on the BH mass itself.  In particular, we find that
proper motions are essential if the models are to provide reliable
virial-like characteristic mass estimates of the surrounding cluster:
the DF of models that fit only line-of-sight velocities (e.g.,
Figures~\ref{fig:discretebincorrectprof} and~\ref{fig:flat}) are
underconstrained and, lacking any treatment for the degeneracies in
the orbit weight distribution, return a biased result.

On the other hand, when proper motions (or, even better, all three
components of velocity) are available then the stars do act as useful
probes of the extended mass distribution within which they move, even
if one has to work hard to place interesting constraints on $\rho(r)$
(Section~\ref{sec:testspherical2}).  For example, our best-fit models
of the Galaxy centre have power-law density profiles
$\rho\sim r^{-\gamma}$ with $\gamma\simeq1.3\pm0.3$ in the inner
parsec (Figure~\ref{fig:marggamma}).  That is based on a small sample
of plausible $\rho(r)$ parametrizations, but careful scrutiny of how the
likelihoods of individual stars change depending on the the assumed
potential (Figure~\ref{fig:cumlik}) may offer a way to tune the form
of $\rho(r)$.

We have deliberately restricted our models to stars having $R<19''$,
all of which lie within the ``extended field'' of
\citet{FritzNuclearClusterMilky2016}.  Including stars at larger radii
would help anchor the form of $\rho(r)$.  Indeed, kinematics for such
stars already exist in the literature (e.g., in the ``large field''
and ``outer field'' regions of \citet{FritzNuclearClusterMilky2016}
and in the sample of \citet{FeldmeierLargescalekinematics2014}).
There is nothing fundamentally difficult about modelling stars from
many such catalogues simultaneously: the simplest way would be to
adjust the selection function $S(\vw)$ to account for the inevitable
variations in the sampling depth of each catalogue.  For such an
endeavour to be worthwhile, however, we would need to relax our
convenient assumption that the potential is spherically symmetric:
this is marginally acceptable for stars having $R<19''$, but becomes
increasingly implausible as one moves beyond the sphere of influence
of the BH into the region where the flattened cluster starts to
dominate the potential.  It remains to be seen by how much the
overfitting problem would affect the reliability of results from these
more general models applied to more extensive data.

\subsection{General}

Stepping back, it is remarkable that discrete, likelihood-based
schemes can be used to place {\it any} interesting constraints on the
potential: one is looking for changes in the log likelihood of order
unity, with the log likelihood itself being a sum of contributions
from thousands of stars; even the slightest bias in the calculation of
likelihoods of individual stars will be amplified hugely.  As a
slightly artificial example of this, consider the treatment of
individual stellar velocity measurements.  We have assumed that the
observed velocity is normally distributed about the star's true
underlying velocity.  Our model for interlopers (eq.~\ref{eq:Pnkc})
can be thought of as adding an extended halo to the Gaussian that
represents the likelihood of each star.  When we include such halos by
setting $f_{\rm c}\gtrsim10^{-6}$ we find that the estimates
of~$M_\bullet$ for both simulated and real data are depressed.  The
most immediate interpretation is that this interloper model is at best
incomplete.

Apart from symmetry, an even more fundamental assumption shared by the
vast majority of dynamical mass estimation methods is that the system
has relaxed to a steady state.  We do not test the consequences of
that assumption here, but \citet{KowalczykEffectnonsphericitymass2018}
have, albeit in a different context -- using orbit-superposition
models to fit the kinematics of simulated $N$-body dwarf spheroidal
galaxies formed by disc mergers.  The machinery for fitting discrete
kinematics presented in this paper could also be applied to
dwarf spheroidal galaxies, but in many ways the Galactic centre
problem is simpler (e.g., the potential is simpler; there are no
issues with identifying the location of the centre of the cluster; the
effects of contamination by stellar binaries are less likely to be
significant).

\section*{Acknowledgments}
It is a pleasure to thank the members of the Oxford dynamics group for
helpful suggestions throughout the production of the results presented
here, the anonymous referee for many insightful comments that helped
improve the quality of the paper, and Christophe Pichon and the
Institut d'Astrophysique de Paris for their hospitality during the
early stages of this work.  This work was supported by the European
Research Council under the European Union's Seventh Framework
Programme (FP7/2007-2013)/ERC grant agreement no.~321067 and by the
``Research in Paris'' programme of Ville de Paris.

\def\aap{A\&A}\def\aj{AJ}\def\apj{ApJ}\def\apjl{ApJL}\def\mnras{MNRAS}\def\araa{ARA\&A}\def\aapr{Astronomy \&
  Astrophysics Review}

\bibliographystyle{mn2e}
%\bibliography{/home/magog/a/papers/biblio}

\end{document}